\documentclass[preprints,article,accept,moreauthors,pdftex,10pt,a4paper]{mdpi} 

\usepackage{ amsmath,amsthm, mathtools, amssymb, amsfonts, amsbsy,fixmath}
\usepackage{graphicx}
\usepackage{epsfig}
\usepackage{booktabs}
\usepackage{boldline,multirow}
\usepackage{array}
\usepackage{float}
\usepackage{kantlipsum}
\usepackage{pifont}
\usepackage{caption}
\usepackage{hyperref}
\usepackage[ruled, lined, longend, linesnumbered]{algorithm2e}
\usepackage{algpseudocode}
\usepackage{tikz}
\usepackage{subcaption}
\usepackage{tabu}
\usepackage{subcaption}
\usepackage{cleveref}
\usepackage[utf8]{inputenc}
\usepackage[english]{babel}
\usepackage{romannum}
\usepackage{alphabeta}
\usepackage{gensymb}
\usepackage{mathptmx}
\usepackage{mleftright}

\newtheorem{lemma}[theorem]{Lemma}
\def\bf #1{\mathbf{#1}}
\def\cbr#1{\left\lbrace #1 \right\rbrace} 
\def\sbr#1{\left[ #1\right]} 
\def\nbr#1{\left( #1\right)}
\def\bs #1{\boldsymbol{#1}}
\def\B #1{\textbf{#1}}
\def\E {\mathbb{E}}
\newcommand{\overbar}[1]{\mkern 1.5mu\overline{\mkern-1.5mu#1\mkern-1.5mu}\mkern 1.5mu}
\usepackage[colorinlistoftodos]{todonotes}

\definecolor{cgreen}{rgb}{0,0.4,0}

\DeclareMathAlphabet{\mathcal}{OMS}{cmsy}{m}{n}
\usepackage{nomencl}
\makenomenclature
\firstpage{1} 
\pubvolume{xx}
\issuenum{1}
\articlenumber{1}
\pubyear{2018}
\copyrightyear{2018}
\history{Received: date; Accepted: date; Published: date}

\newcommand{\norm}[1]{\left\lVert#1\right\rVert}
\def\w{\boldsymbol{\omega}} 
\def\l{\lambda} 
\def\k{\mathbf{k}} 
\def\E {\mathbb{E}}
\def\n{\mathbf{n}} 
\def\x{\mathbf{X}}
\def\sumk#1{\sum_{\mathbf{#1}\in{G}}}
\def\sumw{\sum_{\boldsymbol{\omega}\in{W}}}
\def\expp#1{e^{+j\langle\boldsymbol{\omega},#1\rangle}}
\def\expn#1{e^{-j\langle\boldsymbol{\omega},#1\rangle}}

\def\a{\overline{ \mathbf{a}}} 
\def\f{\overline{f}_\n(\bf z)}

\newcommand\LR[1]{\multicolumn{1}{|c|}{#1}}
\usepackage[colorinlistoftodos]{todonotes}

\Title{SURE-fuse WFF: A Multi-resolution Windowed Fourier Analysis for Interferometric Phase Denoising}

\Author{Joshin P. Krishnan $^{1*}$,  M\'ario A. T. Figueiredo$^{1}$ and Jos\'e M.  Bioucas-Dias $^{1}$}

\AuthorNames{Firstname Lastname, Firstname Lastname and Firstname Lastname}

\address{%
$^{1}$ \quad Instituto de Telecomunica\c c\~oes, Instituto Superior T\'ecnico, Universidade de Lisboa, Portugal; mtf@lx.it.pt, bioucas@lx.it.pt}

\corres{Correspondence: joshin.krishnan@lx.it.pt}

\abstract{Interferometric phase (InPhase) imaging is an important part of many present-day coherent imaging technologies. Often in such imaging techniques, the acquired images, known as interferograms, suffer from two major degradations: 1) phase wrapping caused by the fact that the sensing mechanism can only measure  sinusoidal $2\pi$-periodic functions of the actual phase, and 2) noise introduced by the acquisition process or the system. This work focusses on InPhase denoising which is a fundamental restoration step to many posterior applications of InPhase, namely to phase unwrapping. The presence of sharp fringes that arises from phase wrapping makes InPhase denoising a hard-inverse problem. Motivated by the fact that the InPhase images are often locally sparse in Fourier domain, we propose a multi-resolution windowed Fourier filtering (WFF) analysis that fuses  WFF estimates with different resolutions, thus overcoming the WFF fixed resolution  limitation. The proposed fusion relies on an unbiased estimate of the mean square error derived using the Stein's lemma adapted to complex-valued signals. This estimate, known as SURE, is minimized using an optimization framework to obtain the fusion weights. Strong experimental evidence, using synthetic and real (InSAR \& MRI) data, that the developed algorithm, termed as \textbf{SURE-fuse WFF}, outperforms the best hand-tuned fixed resolution WFF as well as  other state-of-the-art InPhase denoising algorithms, is provided.}

\keyword{Phase estimation, interferometric phase estimation, interferometry, phase unwrapping, Stein's Unbiases Risk Estimate (SURE), Short-time Fourier Transform (STFT), Multi-resolution STFT, windowed Fourier Transform (WFT), multi-resolution WFT, Synthetic aperture radar (SAR), InSAR, SURE-fuse.}

\begin{document}
\pagenumbering{arabic}

%

\section{Introduction}
\label{sec:intro}
In the recent decade, interferometry has strongly benefited from advancements in phase imaging techniques and has significantly contributed to many fields including remote sensing \cite{1974_Graham_Synthetic,2000_Rosen_Synthetic}, optical metrology \cite{1985_Hariharan_Optical_interferometry,1994_SPandit_Data}, astronomy \cite{2003_Monnier_astronomy,2002_John_Astronomy}, surveillance \cite{2017_Bonino_surveillance, 1978_Orr_Satellite}, medical diagnostic \cite{2008_Zhao_Molecular,2003_Vakhtin_Differential,2011_Hao_MRI}, weather forecasting \cite{1999_Hanssen_weather,2009_Liu_weather}, etc. In such coherent imaging modalities, the information related to the physical and geometric properties such as shape, deformation, movement, refractive index, structure, etc. of the illuminated objects is coded in the phase of the underlying signals.

\footnotetext[1]{The research leading to these results has received funding from the European Union’s H2020 Framework Programme (H2020-MSCA-ITN-2014) under grant agreement no 642685 MacSeNet. This work is also supported by the Portuguese Fundacão para a Ciência e Tecnologia (FCT) under grants UID/EEA/5008/2013.}

\begin{figure*}[h!]
	\centering
	\includegraphics[scale=0.7]{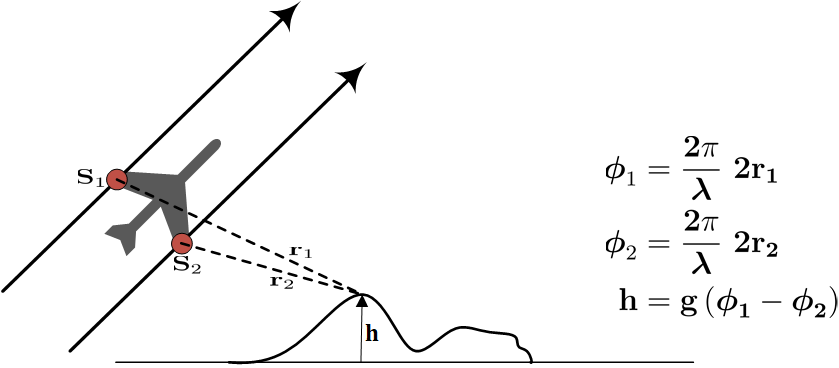}
	\caption{InSAR phase estimation problem.}
	\label{insar}
\end{figure*}

Radar and sonar interferometry \cite{2008_Garello_Radar,2000_Bonifant_sonar}, are  relevant technologies in the context of phase imaging, in particular Interferometric Synthetic Aperture Radar \& Sonar (InSAR/InSAS) \cite{1974_Graham_Synthetic,1986_Zebker_Topographic,1998_Ghiglia_Two,2000_Rosen_Synthetic,2002_Dias_Z}.  InSAR is based on the interference of the  electromagnetic field  scattered by an object or surface and measured by antennas (or sensors) located at different positions, as illustrated in the \cref{insar}. InSAS is similar to InSAR but operated with sound waves \cite{1997_Griffiths_Interferometric}.  Given the positions $s_1$ and $s_2$ of the sensors, the height $h$ of a given surface element, relative to a reference plane, is easily derived from   $r_1-r_2$, i.e., the difference between the distances of the surface element and the antennas.  Since the propagation phases are given by $\phi_1= \frac{4\pi}{\lambda} r_1$ and $\phi_2= \frac{4\pi}{\lambda} r_2$, where $\lambda$ is the carrier wavelength, we conclude that that $h = g(\phi_1-\phi_2)$, where the function $g$ is derived from the acquisition geometry. See \cite{1998_Ghiglia_Two} for further details. It happens, however, that  the measured signals depend only on the principal (wrapped) values of the original phase (absolute phase) $\phi$. Thus, the measured phase, which we term interferometric phase (InPhase) and takes values in $[-\pi,\pi)$, is a non-linear function of the actual phase. This process, termed as phase wrapping, makes the absolute phase inaccessible in the direct measurement. A second major challenge is the noise introduced by the acquisition systems/processes and this further complicates the task of estimating the absolute phase. Owing to these two degradation mechanisms, the estimation of the absolute phase is a challenging inverse problem. 

Though an InSAR scenario is adopted for the discussion, the underlying inverse problem pops ups in many technologies that incorporate phase imaging, namely, Magnetic Resonance Imaging (MRI) \cite{1973_Lauterbur_Image,1992_Hedley_new}, Optical Interferometry  \cite{1994_SPandit_Data} and High Dynamic Range (HDR) Photography \cite{2015_Zhao_Unbounded}. 

Most of the proposed approaches for absolute phase estimation follows a two-step procedure: In the first step, a clean phase is estimated in the interval $[-\pi,\pi)$. This step is termed as the InPhase estimation or phase denoising; while in the second step, the absolute phase is inferred from the InPhase obtained from first step and this stage is known as phase unwrapping. In this paper, our main focus is the InPhase estimation which is the first step of the two-step approach. The sharp interferometric fringes from phase wrapping make the InPhase images different from the natural images. InPhase image denoising is difficult since the wrapping discontinuities should be preserved carefully for the second stage of unwrapping.    

Despite these difficulties, many strategies have been proposed to tackle this problem. Local polynomial approximation(LPA) \citep{2006_katkovnik_local} is an early stage attempt in InPhase estimation in which phase is approximated by zero order polynomials in a small local area (in a rectangular window). The lack of an adaptive window size selection is a major drawback of this method. An \textit{oversized} window fails to estimate sharp discontinuities and non-smooth regions while, on the other hand, a  \textit{small} window may yield poor denoising performance. The selection of the  window size is addressed in \citep{2008_Bioucas_Absolute} in which zero- and first-order LPA of the phase are calculated in sliding windows of varying size. A criterion for adapting the window size is obtained from the zero-order approximations, whereas the filtering is performed using first-order approximations. The performance of the  proposed algorithm, termed PEARLS, is also limited owing the limited modelling power of first-order polynomials in the areas containing discontinuities.

The recent trend in image denoising exploits non-local self-similarity of the natural images \cite{2006_elad_image,2010_Elad_sparse}. The Block matching with 3D filtering (BM3D \citep{2007_Dabov_Image}) algorithm is a popular tool in this category in which collaborative filtering is applied to the groups of similar image patches. The  non-local self-similarity has been exploited in the context of InPhase estimation. The methods   NL-InSAR \citep{2011_Deledalle_NL},  SpInPhase  \citep{2015_Hongxing_Interferometric}, and MoGInPhase \cite{2017_Joshin_MoGInpahse} are three representative examples of this class. SpInPhase and MoGInPhase exploit non-local properties of the InPhase images in complex domain and yield state-of-the-art denoising performance. SpInPhase reformulates the phase denoising problem as a dictionary-based sparse regression problem in the complex domain, where each patch is modelled as a sparse linear combination of the atoms of the dictionary. The sparse regression implicitly projects the noise from a high dimensional space to a low dimensional subspace, which largely  attenuates the noise. In MoGInPhase, the patches of complex phase images are modelled using Mixture of Gaussian (MoG) densities in the complex domain. The phase patches, due to the non-local self-similarity, are well modelled by very few eigen-directions of the covariance matrices of the MoG components, which underlies the  MoGInPhase denoising performance.

Both SpInPhase and MoGInPhase are data adaptive in the sense that they learn representations from the image to be denoised. In spite of the power of the data adaptive representations, the fact that, in many applications, the phase is mostly locally smooth makes fixed representations in the Fourier domain still powerful tools in phase modelling, as they yield sparse representations of local complex phase patches. A state-of-the-art algorithm of this class is the  Windowed Fourier filtering (WFF) based algorithm \cite{2007_Kemao_Twodimensional}. Arguably, the major weakness of WFF is the lack of adaptiveness of the used windows in windowed Fourier transform (WFT).

Research on automatic window size selection has been carried out in time-frequency analysis of one dimensional signals. In \cite{1990_Jones_high}, a data-adaptive time-frequency representation is developed using Gaussian basis functions with varying time width and chirp rate; but this adaptive method has high computational complexity. A simple, computationally efficient and signal-dependent time-frequency representations is proposed in \cite{1992_Jones_Asimple}, in which the window is characterized by a free parameter, which is adapted over time. The temporal  estimation of the parameter is done by maximizing a short-term quality measure of the time-frequency representation. More recently, a fast version of adaptive WFT using B-splines that have near optimal time–frequency localization is proposed in \cite{2006_Liu_Adaptive}. A new adaptive short-time Fourier analysis and synthesis scheme, applied to speech enhancement, is proposed in \cite{2008_Rudoy_adaptive}. This scheme uses an efficient modified overlap-add procedure. It is to be noted that these adaptive time-frequency analysis methods are developed for real domain one dimensional signals, such as speech signal.

For the two-dimensional image analysis, the wavelet transform, due to its  multi-resolution ability, outperforms the Fourier based analyses. We conclude this section by specially mentioning a series of work presented by T. Blu \textit{et al.} \cite{2007_Blu_SURELET,2007_Luisier_New_sure,2008_Luisier_SURE_Multichannel,2010_SURE_Luisier} for natural images and videos in which a new strategy, termed as `SURE-LET', is developed for wavelet-based denoising. In SURE-LET approach, the denoising process is parametrized as a sum of elementary non-linear processes with unknown weights. The optimal weights are computed by minimizing an estimate of the mean square error (MSE) between the clean image and the denoised one. The key point of this strategy is that it makes use of a statistically unbiased estimate of MSE, namely, Stein’s unbiased risk estimate (SURE), that does not depend on the clean image. This \textit{a priori} avoids the random process modelization of wavelet coefficients which is the basis of most of the wavelet-based denoising schemes. A similar strategy is followed in \cite{2008_Gauthier_Two}. Although the wavelet transform and other multi-resolution analysis have been developed for natural two-dimensional images, they do not adapt well to InPhase estimation, where the underlying signals are complex-valued.

In this paper, we adopt a SURE-based strategy and propose a new multi-resolution Fourier analysis for complex-valued InPhase signals.


\subsection{Proposed approach}
As already mentioned, the research herein developed is motivated by the fact that, in many applications, the  phase $\upPhi$,  is locally (i.e., in 2D neighborhoods) smooth \cite{1998_Ghiglia_Two}. Therefore, the Fourier transform of a 2D patch tends to be sparse. These ideas have already been exploited in the literature. We refer to the WFF work \cite{2007_Kemao_Comparative,2007_Kemao_Twodimensional}, which exhibits competitive performance in InPhase denoising of locally smooth phase images. WFF works with fixed windows. This is acceptable if the smoothness of the image phase does not vary considerably across the image. If this is not the case, ie., the image smoothness varies widely across the image, the lack of adaptiveness  of the window size  greatly limits the WFF performance. The influence of window size was recently studied in windowed Fourier ridge algorithm \cite{2013_Windowseln_zhao}.  Without surprise, it was concluded that small windows are good for detecting sharp discontinuities whereas larger windows yield strong noise attenuation in  smooth phase regions and/or for high amplitude of noise.

Aiming at endowing WFF with window size adaptiveness, and inspired by the works \cite{2007_Blu_SURELET,2007_Luisier_New_sure,2008_Luisier_SURE_Multichannel}, this paper proposes a novel approach in which WFF estimates from different window sizes are fused in an optimal way. The fusion is pixel-wise, linear and is implemented in the complex domain such that the estimates from larger windows get higher weights in smooth regions and vice versa. The linear weights for the fusion is designed in such a way that the mean square error (MSE) between the clean and the estimated image is minimum. Since the oracle MSE is not available in a practical inverse problem, we resort to Stein's unbiased risk Estimate (SURE) adapted to the complex domain.

\subsection{Contributions}
The main contributions of this paper are summarized as follows:
\begin{itemize}	
	\item Reformulation of InPhase estimation as sparse regression in the frequency domain using Fourier analysis.
	
	\item Mathematical formulation of a SURE-based unbiased estimate of the MSE, which is derived in the complex domain for the specific InPhase sparse regression model.
	
	\item Design of multi-resolution WFF by pixel-wise linear fusion of WFF estimates with different resolutions. The optimal fusion is achieved through a quadratic programming, designed to minimize the SURE. 
\end{itemize} 

The paper is organized as follows: 
\Cref{sec:format} introduces the observation model and the underlying interferometric phase problem.
A brief revisit of WFT-based denoising strategy, in the perspective of sparse regression in frequency domain, is presented in \Cref{sec:WFT} to build the mathematical foundation for the main proposal. \Cref{sec:sure} discusses the resolution-related limitations of WFF and formulates the main research contribution, i.e., a multi-resolution WFF based on SURE-fusion. Strong experimental evidence, using synthetic and real (InSAR \& MRI) data, of the competitiveness of the proposed algorithm termed as \textbf{SURE-fuse WFF},  is given in \cref{sec:exp}. Finally, the paper concludes in \Cref{sec:concl}.

\section{Problem formulation}
\label{sec:format}

\noindent Let us assume that the images are defined in a 2D grid ${G}:=\lbrace 1,....m \rbrace \times \lbrace 1,....n \rbrace$ with total number of pixels $N=m \times n$. Also, let a $\bf{k} := \nbr{k_1, k_2}\in{G}$ be a point in the 2D grid ${G}$. Herein, for the observed image $\bf{z} := \cbr{\bf z_\bf{k} \in \mathbb{C},\bf{k}\in{G}}$, we assume the following model:
\begin{equation}
\bf{z}=\bf{x}+\bf{n},\ j=\sqrt{-1}, \label{obsmodel}
\end{equation}
where $\bf{x=a}e^{j\boldsymbol{\upPhi}}$, $\bf{a}  := \cbr{\bf a_\bf{k} \in \mathbb{R}_{\geq 0},\ \bf{k}\in{G}}$\footnote{ $\mathbb{R}_{\geq 0}$ denotes the set of non-negative real numbers.} is the amplitude image, $\bs{\upPhi}  := \cbr{ \bs \upPhi_\bf{k} \in \mathbb{R},\ \bf{k}\in{G}}$ is the absolute phase image and $\bf{n}=\bf n^\Re+j\bf n^\Im  := \cbr{\bf n_\bf{k}=\bf n_\k^\Re+j \bf n_\k^\Im \in \mathbb{C},\ \bf{k}\in{G}}$ is the complex-valued noise, having $\bf n^\Re$ and $\bf n^\Im$ as its real and imaginary parts respectively. The noise $\bf n_\k$ is assumed to be zero-mean circular Gaussian \cite{2006_Gallager_Circular} and white  with variance $\sigma^2$, i.e., $\bf n_\k^\Re$ and $\bf n_\k^\Im$ are zero-mean independent Gaussian random variables with variance $\sigma^2/2$. Model \eqref{obsmodel} captures the essential features of the interferometric phase estimation problem \cite{1996_Ghiglia_Minimum, 2011_Deledalle_NL, 2002_Dias_ZpiM, 2015_Hongxing_Interferometric} and it is also a good approximation for magnetic resonance imaging (MRI) \citep{1998_Ghiglia_Two}. Even in InSAR applications \cite{1996_Ghiglia_Minimum, 2011_Deledalle_NL,2002_Dias_ZpiM}, model \eqref{obsmodel} endowed with a suitable spatially variant noise, is a good replacement for the true observation model, as shown in \cite{2015_Hongxing_Interferometric}. The interferometric phase $\bs{\upPhi}_{2\pi} := \cbr{\bs \upPhi_{2\pi_\bf{k}} \in  \left[ -\pi, \pi\right),\ \bf{k}\in{G}}$ is defined as
\begin{eqnarray}
\bs \upPhi_{2\pi}:=\mathcal{W}( \boldsymbol{\bs \upPhi}),
\end{eqnarray}
where $\mathcal{W}(.)$ is the wrapping operator that performs component wise 2$\pi$-modulo wrapping operation defined by
\begin{alignat}{2}
\mathcal{W}:\mathbb{R} &\longrightarrow && \left[ -\pi, \pi\right), \nonumber \\ 
\boldsymbol \upPhi & \longrightarrow && ~\textrm{mod}(\boldsymbol  \upPhi+\pi, 2\pi)-\pi, \label{eqn:wrap}
\end{alignat}
where the function$\mod(. , 2\pi)$ denotes modulo-$2\pi$ operation. Let a clean complex image be denoted as $\bf{x}  := \cbr{\bf x_\bf{k} \in \mathbb{C},\ \bf{k}\in{G}}$ with $\bf x_\bf{k} = \bf a_\bf{k} e^{j\bs{\upPhi}_\bf{k}}$. We remark that $\boldsymbol \upPhi_{2\pi}$ = $\arg (\bf x)$. Hereafter, for the sake of mathematical convenience, we assume that the images $\bf {z, x, n, a,}\bs{\upPhi, \text{ and }\upPhi_{2\pi}}$ are stacked into a column vector of size $N = m \times n$ according to the lexicographic ordering of set $ G$. Throughout our discussion, the components of these column vectors are indexed using their original 2-D grid location $\k \in G$. Also, all the mathematical operations in model \eqref{obsmodel} is to be understood as component-wise. With these notations in hand, we define InPhase denoising as the estimation of $\boldsymbol{\upPhi}_{2\pi}$ from the noisy observation $\mathbf{z}$.

\section{InPhase Denoising via Sparse Regression in the Frequency Domain}
\label{sec:WFT}

We start by briefly reviewing the WFT and the inverse WFT (IWFT) in terms of analysis and synthesis frames. Then, we formulate a sparse regression framework in the frequency domain for InPhase denoising.

\subsection{Two-dimensional windowed Fourier transform (WFT)} 
The discrete WFT of $\bf z$, computed at a spatial point $\bf k \in  G$ and frequency  $\bs{\omega}:= \nbr{\omega_1, {\omega_2}} \in \left[ -\pi,\pi \right)^2 $, is defined as (see \cite{2007_Kemao_Twodimensional} for more details)
\begin{eqnarray}
\bf Z _{\k,\w} = \sumk{\k'}\bf z_{\bf {k'}}\bf h^s_{\k-\k'}\expn{\k'}, \label{eqn:wft}
\end{eqnarray}
where $\bf h^s_{\k}$  is a Gaussian weighting window centered at location $\bf k$, defined as $\bf h^s_{\k}=\bf h^s_{k_1,k_2}:=e^{-\frac{k_1^2+k_2^2}{s^2}}$ and $\langle\bs{\omega,\bf{k}} \rangle := \omega_1 k_1 +\omega_2 k_2$. Size of the window is controlled by the parameter $s$ termed as \textit{scale}.\footnote{In this paper, size of the Gaussian window $\bf h^s$ is fixed to be $\lceil 6s \rceil_{odd}\times \lceil 6s \rceil_{odd}$, where  $\lceil y \rceil_{odd}$ denotes the nearest odd integer greater than or equal to $y$. Please see \cref{sec:paramset} for the details.} We remark that the analysis function $\bf h^s_{\k-\k'}\expn{\k'}$ in \eqref{eqn:wft}, with $\bf h^s_{\k}$ being Gaussian, has the structure of a Gabor wavelet \cite{1996_lee_image}, which has optimal concentration properties in terms of time duration and frequency bandwidth \cite{1985_Daugman_Uncertainty}. Also, we assume that both $\bf z$ and $\bf h^s$ are periodic images of size $m \times n$.
An equivalent way to express $\bf Z _{\k,\w}$ is
\begin{eqnarray}
\bf Z_{\k,\w} = \sbr{\bf z_{\k}\expn{\k}}\circledast \bf h^s_{\bf k}, \label{eqn:wft2}
\end{eqnarray}
where $\circledast$ denotes-2D cyclic convolution. The interpretation of \eqref{eqn:wft2} is clear: the WFT of $\bf z$ computed at a given space-frequency point $\nbr{\k, \w}$ may be obtained by modulating $\bf z_\k$ with the complex exponential $\expn{\k}$ and then convolving the modulated signal with the weighting function $\bf h^s_\k$. In practice, we compute $\bf Z_{\k,\w}$ at frequencies $\nbr{\omega_i, \omega_j}=\nbr{2\pi i /m,2\pi j /n}$ for $(i,j) \in {I}
\subset  {G}$. This can efficiently be implemented with Discrete Fourier Transform (DFT) or its fast version Fast Fourier Transform (FFT).

The IWFT associated with  \eqref{eqn:wft}  is not uniquely defined. In this work, we adopt the following definition:
\begin{eqnarray}
\widehat{\bf z}_{\bf k}&=&\frac{1}{E_{\bf h^s}|W|}\sumk{\k''}\sumw \bf Z _{\k'',\w}\bf h^s_{\k''-\k} \expp{\k}, \label{eqn:iwft0}
\end{eqnarray}
where $E_{\bf h^s}=\sumk{\k}|\bf {h}^s_\k|^2 $ is the energy of $\bf h^s$ and
$$ 
  {W}: = \{(2\pi i/m,2\pi j/n): (i,j)\in {I}\}.
$$
is a set of 2D-frequencies,  and  $|{W}|$ denotes the cardinality of ${W}$. Following below is a derivation to obtain the perfect reconstruction condition (PRC) of the IWFT, i.e., the condition under which $\widehat{\bf{z}}_\k=\bf{z}_\k$. We start by substituting \eqref{eqn:wft} in \eqref{eqn:iwft0}:
\begin{align}
\widehat{\bf z}_{\bf k}&= \frac{1}{E_{\bf h^s}|W|}\sum_{\bf {k',k''} \in  G} \bf{z}_{\bf k'} \bf h^s_{\k''-\k'}\bf h^s_{\k''-\k} \sumw \expp{\k-\k'}.\label{eqn:iwft1}
\end{align}

Let  $D_1$ and $D_2$ to be  sub-multiples of $m$ and $n$ respectively and ${I}:=\{0,D_1,2D_1,\dots, m-D_1\}\times \{0,D_2,2D_2,\dots,n-D_2\}$. Thus the WFT has a 2D-frequency resolution $\nbr{\frac{2\pi}{m/D_1},\frac{2\pi}{n/D_2}}$. Then the summation in \eqref{eqn:iwft1} w.r.t. $\w$ yields:

\begin{align}
\sumw \expp{\k-\k'}&=\sum_{r_1,r_2\in \mathbb{Z}}|W|\delta\nbr{k_1-k'_1-{r}_1\frac{m}{D_1},k_2-k'_2-{r}_2\frac{n}{D_2}}, \label{eqn:iwft3}
\end{align}
where $r_1,r_2 \in \mathbb{Z} $ (integer) and 

 \[
\hspace{0.5cm}  \delta \nbr{y_1,y_2} =
\begin{cases}
1, & \nbr{y_1,y_2} = \nbr{0,0}, \\
0,              & \text{otherwise}.
\end{cases}
\]
Evaluating \eqref{eqn:iwft1} for the cases when $\delta\nbr{.,.}=1$ in \eqref{eqn:iwft3} (i.e.,when $k_1-k'_1=r_1\frac{m}{D_1}$ and $k_2-k'_2=r_2\frac{n}{D_2}$) gives:
\begin{align}
\widehat{\bf z}_{\bf k}&= \frac{1}{E_{\bf h^s}|W|}|W|\sum_{\bf {k''} \in  G} \sum_{~r_1,r_2\in \mathbb{Z}} \bf{z}_{\nbr{k_1-r_1m/D_1,k_1-r_2n/D_2}} \bf h^s_{\nbr{k_1''-k_1+r_1m/D_1,k_2''-k_2+r_2n/D_2}} \bf h^s_{\k''-\k}.\label{eqn:iwft7}
\end{align}
In order to obtain PRC, we assume that the support of $\bf h^s_{k_1,k_2}$ w.r.t. $k_1$ and $k_2$ are not greater than $\frac{m}{D_1}$ and $\frac{n}{D_2}$ respectively. With this assumption, all the terms in \eqref{eqn:iwft7} vanishes except the one corresponding to $r_1,r_1=0$, i.e,
\begin{align}
\widehat{\bf z}_{\bf k}&= \frac{1}{E_{\bf h^s}}\sum_{\bf {k''} \in  G}  \bf{z}_{\k} \bf h^s_{\k''-\k} \bf h^s_{\k''-\k}=\frac{1}{E_{\bf h^s}}\bf{z}_{\k}\sum_{\bf {k''} \in  G}   \nbr{\bf h_{\k''-\k}^{s}}^2=\bf{z}_{\k}.\label{eqn:iwft6}
\end{align}
We emphasis that the PRC is achieved only when the window function $\bf{h^s_\k}$ satisfies the aforementioned support constraints and when $D_1$ and $D_2$ are sub-multiple of $m$ and $n$ respectively.\footnote{The specific PRC settings adopted in our implementation, i.e., settings of $D_1$ and $D_2$ in relation to the window support, are explained in \cref{sec:paramset}.} In all the proceeding IWFT expressions, a default assumption of PRC is taken.
For the efficient FFT implementation, \eqref{eqn:iwft0} is rewritten as:
\begin{eqnarray}
{\bf z}_{\bf k}&=& \frac{1}{E_{\bf h^s}|W|}\sumw \sbr{\bf Z_{\k,\w}\expp{\k}} \circledast \sbr{\bf h^s_{-\k}\expp{\k}}. \label{eqn:iwft4}
\end{eqnarray}

\subsection{Denoising via sparse regression in the frequency domain}

Let $\bf A_s: \mathbb{C}^{N} \mapsto \mathbb{C}^{N \times |W|}$ be the linear operator representing the  WFT analysis associated with the definition \eqref{eqn:wft2} and $\bf S_s: \mathbb{C}^{N \times |W|} \mapsto \mathbb{C}^{N} $ be the linear operator  representing the IWFT synthesis associated with the definition \eqref{eqn:iwft4}. We recall that, according to the observation model \eqref{obsmodel}, our objective is to estimate $\bf x$ from $\bf z$.   We attack this inference problem by solving an optimization problem with two facts in mind:
\begin{enumerate}
	\item $\bf X: = \bf{A}_s\bf{x}$ is sparse or at least compressible \footnote{$\bf X$ is compressible if its magnitude decays rapidly.}
	\item Since the  $\bf{A}_s^H\bf{A}_s\propto \bf I$,\footnote{$H$ denotes Hermitian transpose operation} if the noise $\bf n$ is circular complex Gaussian, independent, and identically distributed (i.i.d.), then $\bf{A}_s \bf{n}$ is also circular complex Gaussian  and i.i.d. 
\end{enumerate}
Therefore, we propose to solve the following optimization 
\begin{align}
\widehat{\bf X} = \arg\min_{\bf X \in \mathbb{C}^{N \times |W|}} ||\bf X ||_0 + \frac{1}{\lambda^2}||\bf{X}-\bf{Z}||^2_2,  \label{opt2}
\end{align}
where $\bf{Z=A}_s\bf{z}$ and the final estimate of $\bf x$ is given by  $\widehat{\bf x}= \bf{S}_s \widehat{\bf X}$. The motivation for the formulation \eqref{opt2} stems from the above two facts: the $\ell_0$ term is a regularizer that promotes sparse WFT representations of the original data and the second term is the negative loglikelihood associated to the circular complex Gaussian  i.i.d. noise. The relative weight between the two terms is set by the parameter $\lambda > 0$.

Optimization $\eqref{opt2}$, although non-convex, has a simple closed-form solution: the hard-threshold given by
\begin{align}
\widehat{\bf X}&=   \Theta_H^\lambda \nbr{\bf{Z}} \label{hardt},
\shortintertext{where} 
\Theta_H^\lambda \nbr{y}   &:= \begin{cases} 
0, & \text{if } |y| \leq \lambda\\
y,              & \text{otherwise}.
\end{cases}  \label{ht}
\end{align}

The WFT-based denoising strategy, formulated under a sparse regression framework, is summarized in the following steps:
\begin{enumerate}
	\item Compute the WFT of $\bf z$, i.e., $\bf {Z=A}_s\bf{z}$ give by \eqref{eqn:wft2}.
	\item Compute a sparse solution in the frequency domain by hard-thesholding $\bf Z$ yielding $\widehat{\bf X}=   \Theta_H^\lambda \nbr{\bf{ Z}}$
	\item Compute the IWFT of the thresholded signals to get the spatial domain estimate, i.e., $\widehat{\bf x}= \bf{S}_s \widehat{\bf X}$.
\end{enumerate}
We term this denoising scheme, courtesy of \cite{2007_Kemao_Twodimensional}, as \textit{windowed Fourier filtering}(WFF). We remark that this paper does not claim the novelty WFF. Our proposal in \cref{sec:WFT} aims to present  WFF in a solid framework necessary to the following developments, in  addition to a computational efficient implementation. 
  
This section is concluded by formally defining WFF as follows: 
\begin{align}
\widehat{\bf x}_\k=f_\k\nbr{\bf z}&=\frac{1}{E_{\bf h^s}|W|}\sumk{\k''}\sumw  \Theta^\lambda_H \nbr{\bf Z _{\k'',\w}}\bf h^s_{\k''-\k} \expp{\k}. \label{eqn:wftestmn}
\intertext{This can also be rewritten as:}
\widehat{\bf x}_\k=f_\k\nbr{\bf z}&= \frac{1}{E_{\bf h^s}|W|}\sumw \sbr{\Theta_H^\lambda \nbr{\bf Z_{\k,\w}}\expp{\k}} \circledast \sbr{\bf h^s_{-\k}\expp{\k}}. \label{eqn:wftestmn1}
\end{align}

\section{SURE-fuse WFF: A multi-resolution WFF for improved InPhase denoising}
\label{sec:sure}
WFT and its variants are widely used in speech and natural image denoising. A significant related work, WFF \cite{2007_Kemao_Twodimensional} proposed by Kemao, shows that WFT is a powerful tool to efficiently represent interferometric fringes. Although WFT-based techniques are very popular, they are designed with fixed resolution,   which sets a bottleneck to their performance. The spatial and frequency resolution of the WFT are determined by the size of the windowing function. Often in a WFT-based signal analysis, a trade-off between frequency and spatial (or temporal) resolution is considered. Usually, this trade-off is application specific and is set using the Heisenberg uncertainty principle.   

\begin{figure}[h!]
	\centering
	\begin{subfigure}{0.49\textwidth}
		
		\includegraphics[width=\textwidth]{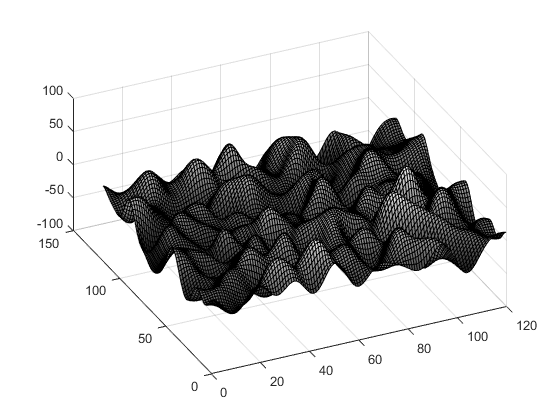}
		\caption{\centering Non-smooth  mountains. \hspace{3cm} Size:$120\times 120$, phase range:-64 to 60 radians}
		\label{pre_hf}
	\end{subfigure}
	\hfill
	\begin{subfigure}{0.49\textwidth}
		\includegraphics[width=\textwidth]{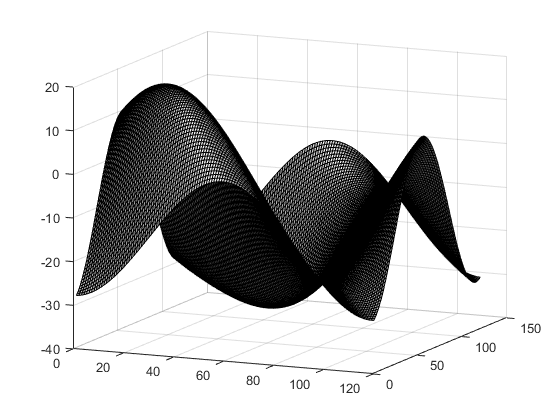}
		\caption{\centering Smooth  mountains. \hspace{3cm} Size:$120\times 120$, phase range:-35 to 18 radians}
		\label{pre_lf}
	\end{subfigure}
	\caption{Absolute phase surfaces}
	\label{predisc}
\end{figure}

We present an illustrative experiment, as a pre-discussion, to show the role of window size in WFF. Two different types of phase surfaces are simulated - \textbf{i}) a `non-smooth mountain' (\cref{pre_hf}), which has frequent ups and down (compared to the WFF window size), and  \textbf{ii}) a `smooth smooth mountain' (\cref{pre_lf}), which is a smooth surface. The complex-valued image $\bf a e^{j\upPhi}$is generated by using each of these simulated surfaces as the absolute phase $\bs \upPhi$ with amplitude $\bf a$ set to unity. Observations as per model \eqref{obsmodel} are generated for low to high level of noise ($\sigma \in \lbrace0.3, 0.5, 0.7, 0.9 \rbrace$). InPhase $\bs \upPhi_{2\pi}$ is estimated from these noisy observations using WFF \cite{2007_Kemao_Twodimensional}. The experiments are repeated using various windows whose sizes are determined by the scales $s \in S=\lbrace 1, 2, 3, ..., 10 \rbrace$. A small value of $s$ corresponds to a small window and vice versa.

 The performance measurements, in terms of \textit{peak signal to noise ratio} $PSNR$ \eqref{psnr}, are given in \cref{table0} from which the following observations are made: \textbf{i}) The non-smooth mountain possesses abrupt variations and hence  the spatial resolution is preferred over the frequency resolution. Here a small-window WFF yields better results compared to a large-window WFF. \textbf{ii}) The smooth mountain has a smooth terrain and here the frequency resolution is preferred over the spatial resolution. A larger window is preferred in such cases. However, in the latter case, since the surface is not entirely flat like a plane, there is an optimal window size ($s=7$), beyond which, if the window size is increased, the performance starts to decrease.

\begin{table}[h!]
	\begin{center}
		\scalebox{0.9}{
			\begin{tabular}{c||c||cccccccccc}
				\multicolumn{1}{r||}{\multirow{4}[1]{*}{Surf.}} & \multirow{3}[1]{*}{$\sigma$} & \multicolumn{10}{c}{{\normalsize PSNR (dB) }} \\
				\multicolumn{1}{r||}{\multirow{2}[1]{*}{}} & \multirow{2}[1]{*}{} & \multicolumn{10}{c}{\footnotesize WFF (by Kemao \cite{2007_Kemao_Twodimensional})}  \\
				\multicolumn{1}{r||}{} &       & \multicolumn{1}{c}{$s=1$} & \multicolumn{1}{c}{$s=2$} & \multicolumn{1}{c}{$s=3$} & \multicolumn{1}{c}{$s=4$} & \multicolumn{1}{c}{$s=5$} & \multicolumn{1}{c}{$s=6$}& \multicolumn{1}{c}{$s=7$}& \multicolumn{1}{c}{$s=8$}& \multicolumn{1}{c}{$s=9$} & \multicolumn{1}{c}{$s=10$}\\
				
				\midrule
				\midrule
				\clineB{3-3}{2.5}
				\multirow{2}[1]{*}{Non-smooth}     & 0.3   &\LR{31.27}	&29.53	&28.77	&28.49	&28.36	&28.29	&28.24	&28.20	&28.16	&28.12 \\
				                                 & 0.5   &\LR{24.69}	&21.98	&21.01	&20.71	&20.60	&20.45	&20.37	&20.35	&20.27	&20.18 \\
				\multirow{2}[1]{*}{mountain} & 0.7   &\LR{19.45}	&17.36	&16.59	&16.33	&16.12	&16.05	&15.98	&15.89	&15.84	&15.77 \\
				                                 & 0.9   &\LR{16.16}	&15.00	&14.40	&14.20	&14.12	&14.01	&13.93	&13.85	&13.80	&13.77 \\
				\clineB{3-3}{2.5}
				\midrule
				\midrule
				\clineB{8-8}{2.5}
                \multirow{2}[1]{*}{Mountain}     & 0.3   &36.15	&40.74	&42.95	&44.08	&44.59	&\LR{44.78}	&44.45	&44.40	&44.39	&44.14 \\
                                                 & 0.5   &32.68	&37.15	&39.41	&40.54	&41.08	&\LR{41.32}	&41.13	&41.09	&41.02	&40.78  \\
                \multirow{2}[1]{*}{(smooth)}     & 0.7   &30.29	&34.82	&37.13	&38.28	&38.81	&\LR{39.04}	&38.90	&38.85	&38.79	&38.55  \\
                                                 & 0.9   &28.19	&33.09	&35.44	&36.59	&37.13	&\LR{37.34}	&37.21	&37.15	&37.11	&36.86  \\
                \clineB{8-8}{2.5}
				\midrule
				\midrule
		\end{tabular}}%
		\caption{Performance indicators for surfaces shown in \cref{predisc}. s: scale of WFF. Best values are shown in boxes. PSNR: Peak signal-to-noise ratio as defined by \eqref{psnr}}
		\label{table0}
	\end{center}
\end{table}
This experiment motivates the necessity of a WFT-based algorithm that can adapt the window size depending on the data. Also, the real life InPhase images are often more challenging than the two examples just considered. In real images, we  often have different levels of smoothness across the image, abrupt peaks and valleys, sharp discontinuities, etc. As shown later, having adaptive window size selection yields large denoising gains.

Inspired by the works \cite{2007_Blu_SURELET,2007_Luisier_New_sure,2008_Luisier_SURE_Multichannel}, we address the adaptive window selection by proposing a pixel-wise fusion mechanism of WFF estimates having different resolutions. The fused estimate at a pixel $\k \in  G $ is given by
\begin{align}
\widehat{\bf x}_\k^{\textrm{fuse}}=\sum_{s\in S}a_\k^s\widehat{\bf x}_\k^s, \label{sure0}
\end{align}
where $\widehat{\bf x}_\k^s $ is the image estimate at the pixel $\k$ obtained from WFF of scale $s$ and $\bf a_\k^s$ is the corresponding fusion coefficient. The main challenge here is the design of $\bf a_\k^s$ to obtain a data-adaptive fusion. We propose a strategy in which the fusion coefficients are computed by minimizing SURE, an unbiased estimate of $\bf{mse}$. Before introducing the fusion framework, SURE, adapted to complex-valued signals, is derived below.

\subsection{SURE: Unbiased estimate of $\bf{mse}$}
\label{sec:SURE}

Let  $\widehat{\bf x}:=\bf{f}(\bf z)\in\mathbb{C}^N$ be the estimate of the true image $\bf{x}$, stacked in vector form. Also we use the notation $f_\k\nbr{\bf z}$ to represent the $\k^{th}$ component of the estimate $\widehat{\bf x}$. The $\bf{mse}$ of $\widehat{\bf x}$, denoted by $\bf{mse}(\widehat{\bf x})$,  is given by
\begin{align}
\label{eqn:mse_def}
  \bf{mse}(\widehat{\bf x}) := & \;\frac{1}{N}\|\bf{f}(\bf{z}) - {\bf x}\|^2 \\
               = & \;\frac{1}{N}\left[{\|\bf{f}(\bf{z})\|^2} - 2{\Re\left\{\bf{f}(\bf{z})^H\bf{x}\right\}} +
               {\|\bf{x}\|^2}\right].
\end{align}
Expression \eqref{eqn:mse_def}, the {\em oracle}   \textbf{mse},  cannot be evaluated in a real-life scenario, as we do not have access to the true signal ${\bf x}$. However, in the case of real-valued estimates,  the Stein’s lemma \cite{1981_Stein_Estimation} opens the door to replace \eqref{eqn:mse_def}  by an unbiased estimate, which is a function of only the observations. Stein's lemma \cite{1981_Stein_Estimation} for real-valued signals, courtesy of \cite{2007_Blu_SURELET}, is as follows:

\begin{lemma}[Stein’s lemma for real-valued signals]
	\label{stein:lemma1}
	Let $\bf{g}(\bf z')$ be a mapping from  $\mathbb{R}^N$ to $ \mathbb{R}^N$ such that $\E \cbr{|\partial g_\k(\bf z')/\partial \bf z'_\k |}< \infty $,  $\bf{z' = x' + n'} \in \mathbb{R}^N$, ${\bf x'}$ is a fixed parameter, and $\bf n'$ is a zero-mean i.i.d Gaussian random vector  with variance $\sigma^2$. Then we have \footnote{For the lemma to be rigorously precise, it is to be assumed that $f_\k(\bf z')$ does not explode at infinity; typically such that $|f_\k(\bf z')| \leq ce^{\frac{\norm{\bf z'}^2}{2t^2}}$, where $c$ is a constant and $t > \sigma$ \cite{2007_Blu_SURELET}.}
	\begin{align}
	\mathbb{E} \left\{\bf{g}(\bf{z'})^T \bf{x'}\right\}&=\mathbb{E}\left\{\bf{g}( \bf{z'})^T  \bf z'-\sigma^2 \nabla.\,\bf{g}(\bf{z'})\right\}, \label{lemma1}
	\end{align}
\end{lemma}
where $T$ denotes the transpose operator and $\nabla$ is the divergence operator, such that
\begin{align}
\nabla.\,\bf{g}(\bf{z'}) =\sum_{\k \in G}\frac{\partial  g_\k(\bf{z'})}{\partial \bf z'_\k }. \label{eqn:div0}
\end{align}
It is to be noted that Lemma \ref{stein:lemma1} assumes $  g_\k(\bf{z'})$ is differentiable. This is a fundamental result, as it do not make any assumption regarding ${\bf x}$ that it is treated as a fixed parameter, and the source of randomness comes only from the additive  Gaussian i.i.d. noise. We use Lemma \ref{stein:lemma1} to derive SURE for complex-valued signal.
\subsubsection{SURE for complex-valued estimators}
Stein's Lemma demands the estimating function $\bf g$ to be differentiable. But the WFT-based estimator $\bf f:\mathbb{C}^N \mapsto \mathbb{C}^N$ operates on complex-valued signals. Hence we adopt a general framework of differentiation, termed as Wirtinger calculus \cite{1927_Wirtinger,2011_Adali_Cvalued}, which can handle the differentiation involving complex-valued signals. Following this framework, for any \textit{real differentiable}\footnote{A function $f:\mathbb{C} \rightarrow \mathbb{C}$ is real differentiable means that $f$ is differentiable when expressed as a function $f:\mathbb{R}^2 \mapsto \mathbb{R}^2$.} function $f_k(\bf{z})$, the derivative can be defined as:
\begin{align}
\frac{\partial f_\k(\bf{z})}{\partial \bf{z}_\k}:=\frac{1}{2}\nbr{\frac{\partial f_\k(\bf{z})}{\partial \bf z_\k^\Re}-j\frac{\partial f_\k(\bf{z})}{\partial \bf z_{\k}^\Im}}, \label{eqn:wirt}
\end{align}
where the superscript $\Re$ and $\Im$ are used to denote the real and imaginary parts respectively, i.e., $\bf z_\k=\bf z_{\k}^\Re+j\bf z_\k^\Im$. In order to derive SURE, we consider the expectation of $\bf{mse}$ defined in \eqref{eqn:mse_def}:
\begin{align}
  \mathbb{E} \cbr{\bf{mse}(\widehat{\bf x})} = & \frac{1}{N}\mathbb{E} \cbr{ \|\bf{f}(\bf{z}) - {\bf x}\|^2} \nonumber\\
=&\frac{1}{N}\mathbb{E} \cbr{ \|\bf{f}(\bf{z})^\Re - \bf x^\Re\|^2}+\frac{1}{N}\mathbb{E} \cbr{ \|\bf{f}(\bf{z})^\Im - \bf x^\Im\|^2}\nonumber \\
=&\frac{1}{N}\mathbb{E} \sbr{\norm{\bf f\nbr{\bf z}}^2+\norm{\bf x}^2-2\cbr{\bf f\nbr{\bf z}^\Re}^T \bf x^\Re-2\cbr{\bf f\nbr{\bf z}^\Im}^T \bf x^\Im}.
\intertext{Lemma \ref{stein:lemma1} is applied to real and imaginary parts seperately. This is possible since the observation model \eqref{obsmodel} assumes that the real and imaginary noise components ($\bf n_\k^\Re$ and $\bf n_\k^\Im$) are zero-mean independent Gaussian random variables with variance $\sigma^2/2$.}
\mathbb{E} \cbr{\bf{mse}(\widehat{\bf x})} =&\frac{1}{N}\mathbb{E} \sbr{\norm{\bf f\nbr{\bf z}}^2+\norm{\bf x}^2-2\cbr{\bf f\nbr{\bf z}^\Re}^T \bf z^\Re-\sigma^2 \nabla f\nbr{\bf z}^\Re - 2 \cbr{\bf f\nbr{\bf z}^\Im}^T \bf z^\Im-\sigma^2 \nabla f\nbr{\bf z}^\Im} \nonumber\\
=&\frac{1}{N}\mathbb{E} \sbr{\norm{\bf f\nbr{\bf z}}^2+\norm{\bf x}^2-2\Re \cbr{\bf f\nbr{\bf z}^H \bf z}+\sigma^2  \sum_{\k \in G} \cbr{ \frac{\partial  f_{ \k}(\bf z)^\Re}{\partial \bf z^\Re_{ \k} }+\frac{\partial  f_{\k}(\bf z)^\Im}{\partial \bf z^\Im_{\k}  } }} \nonumber\\
=&\frac{1}{N}\mathbb{E} \sbr{\norm{\bf f\nbr{\bf z}}^2+\norm{\bf x}^2-2\Re \cbr{\bf f\nbr{\bf z}^H \bf z}+\sigma^2  \sum_{\k \in G} \Re \cbr{ \frac{\partial  f_{\k}(\bf z)}{\partial \bf z^\Re_{\k} }-j\frac{\partial  f_{\k}(\bf z)}{\partial \bf z^\Im_{\k} }}}. \label{eqn:sure0}\\
\intertext{\eqref{eqn:wirt} in \eqref{eqn:sure0} implies,}
\mathbb{E} \cbr{\bf{mse}(\widehat{\bf x})} =&\frac{1}{N}\mathbb{E} \sbr{\norm{\bf f\nbr{\bf z}}^2+\norm{\bf x}^2-2\Re \cbr{\bf f\nbr{\bf z}^H \bf z}+2\sigma^2 \Re  \cbr{ \sum_{\k \in G}  \frac{\partial  f_{\k}(\bf z)}{\partial \bf z_{\k} }}} \nonumber\\
=&\mathbb{E}  \sbr{\frac{1}{N}\cbr{\norm{\bf f\nbr{\bf z}}^2+\norm{\bf x}^2-2\Re \cbr{\bf f\nbr{\bf z}^H \bf z}+2\sigma^2 \Re \cbr{ \nabla.\,\bf{f}(\bf{z})}}}.\label{eqn:sure1}
\end{align}
In \eqref{eqn:sure1}, the only term depending the true variable $\bf x$ is $\norm{\bf x}^2$. This energy term can be easily replaced as $\norm{\bf x}^2=\norm{\bf z}^2-\sigma^2$. Using \eqref{eqn:sure1}, we define the SURE, an unbiased estimate of  $\bf{mse}$, for complex-valued signal as:
\begin{align}
\bs \varepsilon^{\textbf{sure}}&:=\frac{1}{N}\sbr{|| f(\bf z) || ^2+|| \bf z|| ^2 -N\sigma^2-2\Re \{\bf{f}({\bf z})^H \bf{z}\}+2\sigma^2\Re \cbr{\nabla.\,\bf{f}( \bf z) }}.  \label{sure}
\end{align}
\noindent It is quite evident from \eqref{eqn:sure1} that $\bs \varepsilon^{\textbf{sure}}$ is an unbiased estimate of $\bf{mse}(\widehat{\bf x})$, i.e., $\mathbb{E}\cbr{\bs \varepsilon^{\textbf{sure}}}=\mathbb{E}\cbr{\bf{mse}(\widehat{\bf x})}.$ If the standard deviation of $\bf{mse}(\widehat{\bf x})$ is ``small'', then  $\bs \varepsilon^{\textbf{sure}}$ is an extremely useful estimate of the $\bf{mse}(\widehat{\bf x})$, as it depends only on the observations ${\bf z}$,  the estimator $\bf{f}$, and the noise variance $\sigma^2$.
\subsection{Threshold function for denoising}	
\begin{figure}[h!]
	\centering
	\begin{subfigure}{0.49\textwidth}
		\includegraphics[width=\textwidth]{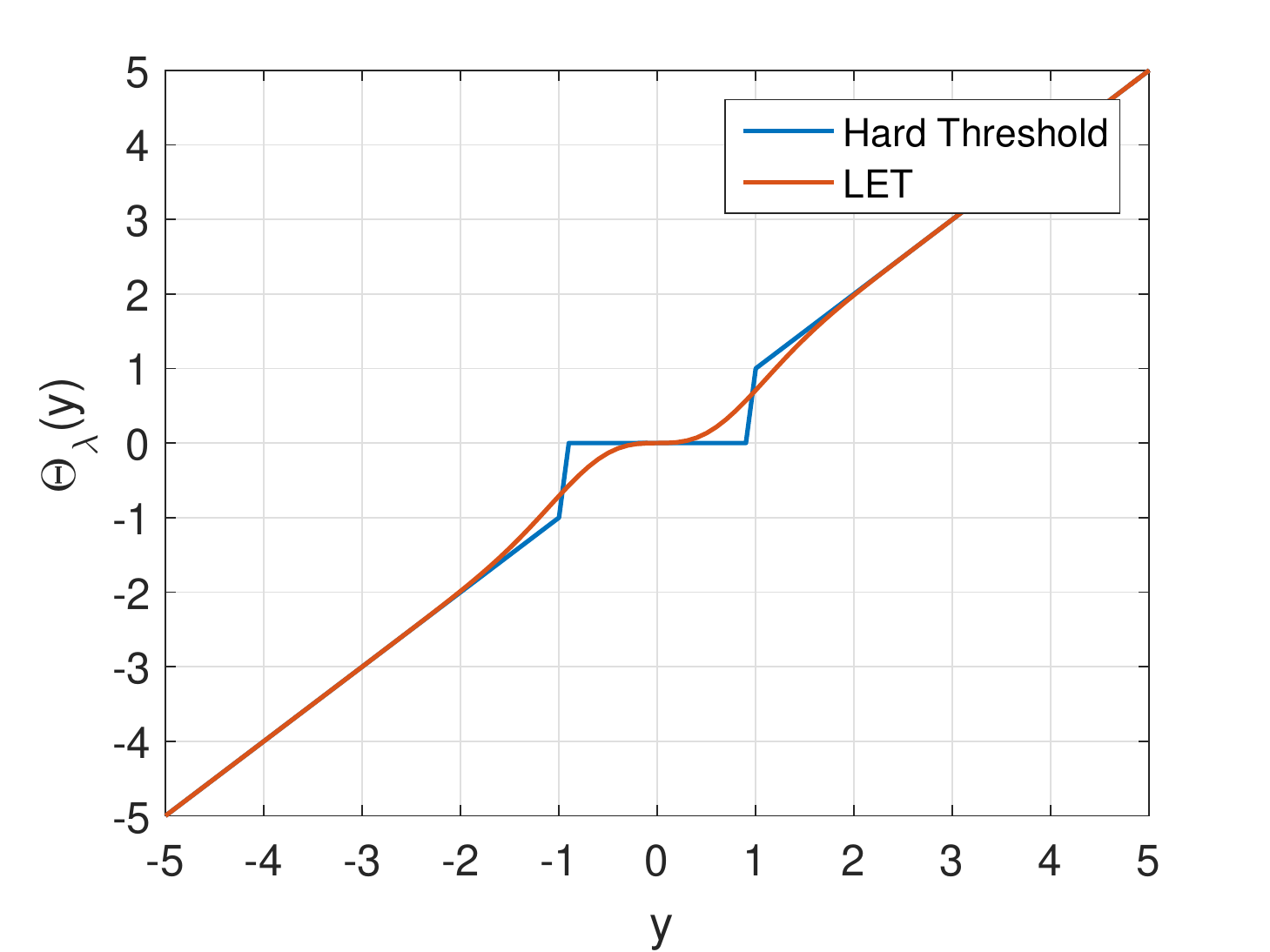}
		\caption{$\sigma=0.3$}
		\label{let1}
	\end{subfigure}
	\hfill
	\begin{subfigure}{0.49\textwidth}
		\includegraphics[width=\textwidth]{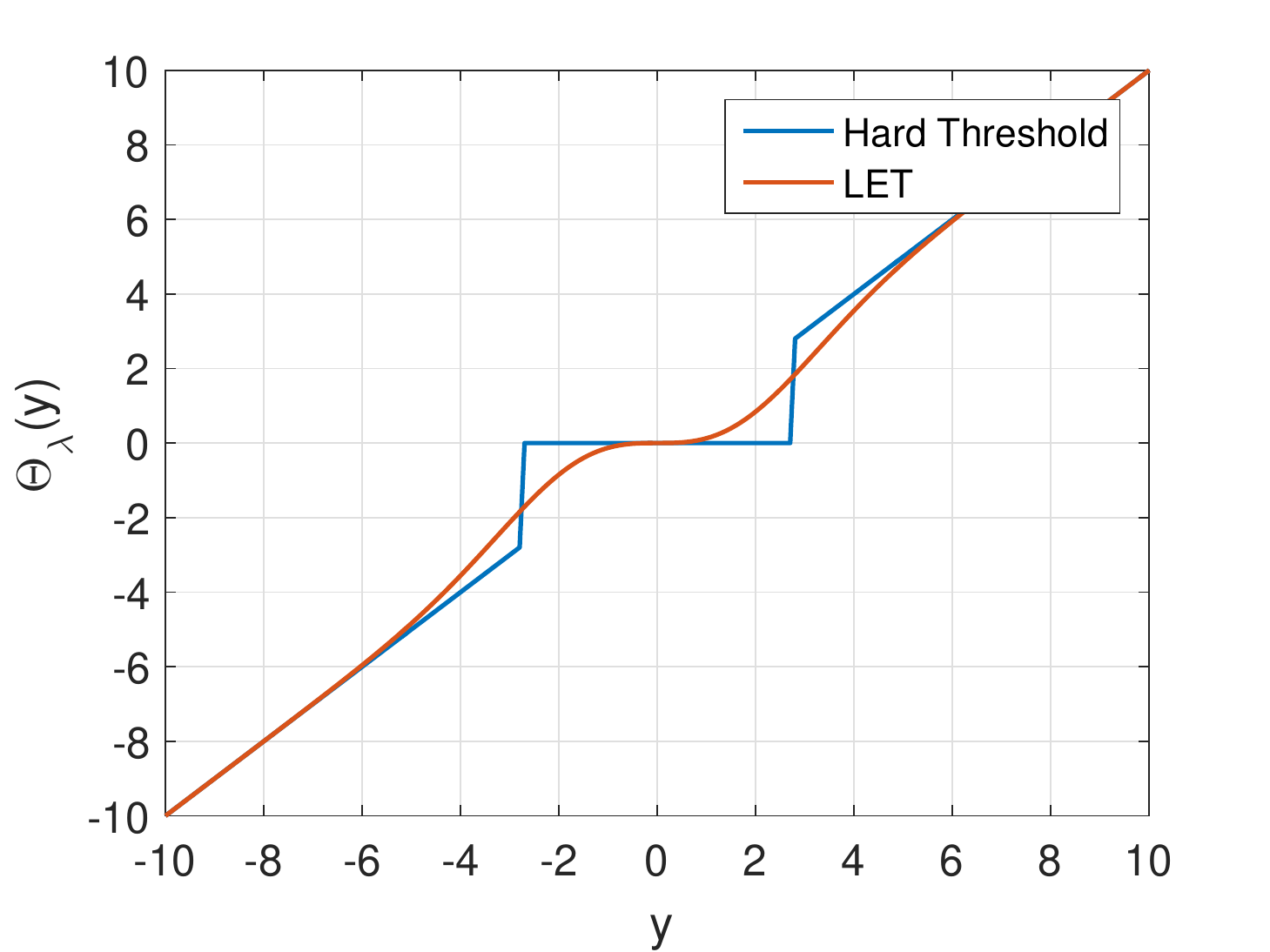}
		\caption{$\sigma=0.9$}
		\label{let2}
	\end{subfigure}
	\caption{Shapes of LET and Hard Threshold function for $\lambda=3\sigma$}
	\label{letplot}
\end{figure}

Stein's lemma \ref{stein:lemma1} assumes differentiable functions- However, the WFF estimator  defined in \eqref{eqn:wftestmn} contains the hard threshold function  $\Theta_H^\lambda \nbr{.}$, which is not differentiable.  As in \cite{2007_Luisier_New_sure}, this roadblock is tackled by selecting the following threshold function, known as Linear expansion of threshold (LET)\footnote[1]{A soft thresholding function, which naturally might come to mind, is not applicable,  as it is not differentiable as well.} in literature:
\begin{align}
\Theta_L^\lambda(y) &:= y(1-e^{-\frac{|y|^2}{\l^2}}),\label{eqn:LET}
\end{align}
where $\l$ is the threshold parameter.  In our implementation, $\l$ is fixed as $3\sigma$, where $\sigma$ is the noise standard deviation. This value is obtained by an empirical tuning, specific to the SURE-fuse algorithm, which is explained in detail in \cref{sec:paramset}.  \Cref{letplot} shows the shape of an LET-based threshold function $\Theta_L^\lambda$ and a hard threshold function $\Theta_H^\lambda$ for $\sigma=0.3 ~\text{and}~ 0.9$, representing a low and a high noise level respectively.
The rationale behind LET is the following: \textbf{i}) it has \textit{almost}  shrinking capability, since $\Theta_L^\lambda(y) \simeq 0$ for $|y|\ll \l $; \textbf{ii}) LET is differentiable and thus can be applied with mathematical formulation involving  Stein's lemma \ref{stein:lemma1}; \textbf{iii}) LET exhibit anti-symmetry and thus avoids sign preferences; and  \textbf{iv}) LET does not make changes to large valued $y$, i.e., $\Theta_L^\lambda(y) $ tends to $y$ for large values of $y$. 
Equation \eqref{eqn:wftestmn} is modified using LET function to define the WFF$^{\textrm{let}}$ as
\begin{align}
\widehat{\bf x}_\k=f_\k\nbr{\bf z}=\frac{1}{E_{\bf h^s}|W|}\sumk{\k''}\sumw  \Theta_L^\lambda \nbr{\bf Z _{\k'',\w}} \bf h^s_{\k''-\k} \expp{\k}. \label{eqn:wftestmn2}
\end{align}
\subsection{Finding the divergence term, i.e., $\nabla.\,\bf{f}( \bf z)$}
\label{sec:derv}
Calculation of $\nabla.\,\bf{f}( \bf z)$ involves the complex derivative $\frac{\partial  f_\k(\bf z)}{\partial \bf z_\k }$ in Wirtinger calculus sense \cite{2011_Adali_Cvalued}. Expanding \eqref{eqn:wftestmn2} and using the LET function \eqref{eqn:LET} yields:
\begin{align}
f_\k\nbr{\bf z}&=\frac{1}{E_\bf {h^s}|W|}\sumk{\k''}\sumw  \bf Z _{\k'',\w} \bf h^s_{\k''-\k} \expp{\k} 
- \frac{1}{E_\bf {h^s}|W|}\sumk{\k''}\sumw  \bf Z _{\k'',\w}e^{\frac{-|| \bf Z _{\k'',\w}||^2}{\l^2}}\bf h^s_{\k''-\k} \expp{\k} \\
&=\bf z_{\k}- \frac{1}{E_\bf {h^s}|W|}\sumk{\k''}\sumw  \bf Z _{\k'',\w}e^{\frac{-|| \bf Z _{\k'',\w}||^2}{\l^2}}\bf h^s_{\k''-\k} \expp{\k},\label{eqn:dive1}
\end{align}
From \cref{eqn:dive1},  the derivative  $\frac{\partial  f_\k(\bf z)}{\partial \bf z_\k }$ may be written as (the details are developed in \cref{app:div})
\begin{align}
\frac{\partial f_\k\nbr{\bf z}}{\partial \bf z_{\k}}&=1-\frac{1}{E_\bf {h^s}|W|}\sumk{\k''}\sumw  \underbrace { e^{\frac{-|| \bf Z _{\k'',\w}||^2}{\l^2}}}_{\upPsi_{\k'',\w}}\underbrace {\nbr{ {\bf h^s_{\k''-\k}}}^{2}}_{\upGamma_{\k''-\k}}
+\frac{1}{E_\bf {h^s}|W|\l^2}\sumk{\k''}\sumw  \underbrace {|| \bf Z_{\k'',\w}||^2 e^{\frac{-|| \bf Z_{\k'',\w}||^2}{\l^2}}}_{\upOmega_{\k'',\w}}\underbrace {\nbr{ \bf h^s_{\k''-\k}}^2}_{\upGamma_{\k''-\k}} \nonumber \\
&=1-\frac{1}{E_\bf {h^s}|W|}\sumk{\k''} \sumw { \upPsi_{\k'',\w}}  {\upGamma_{\k''-\k}} +\frac{1}{E_\bf {h^s}|W|\l^2}\sumk{\k''} \sumw {\upOmega_{\k'',\w}}  {\upGamma_{\k''-\k}}\nonumber \ \\
&=1-\frac{1}{E_\bf {h^s}|W|}\sumw  \upPsi_{\k,\w}\circledast {\upGamma_{-\k}} +\frac{1}{E_\bf {h^s}|W|\l^2}\sumw  \upOmega_{\k,\w}\circledast {\upGamma_{-\k}}. \label{derv0}
\end{align}
	\begin{figure}[h!]
	\centering
	\begin{subfigure}{0.49\textwidth}
		\includegraphics[width=\textwidth]{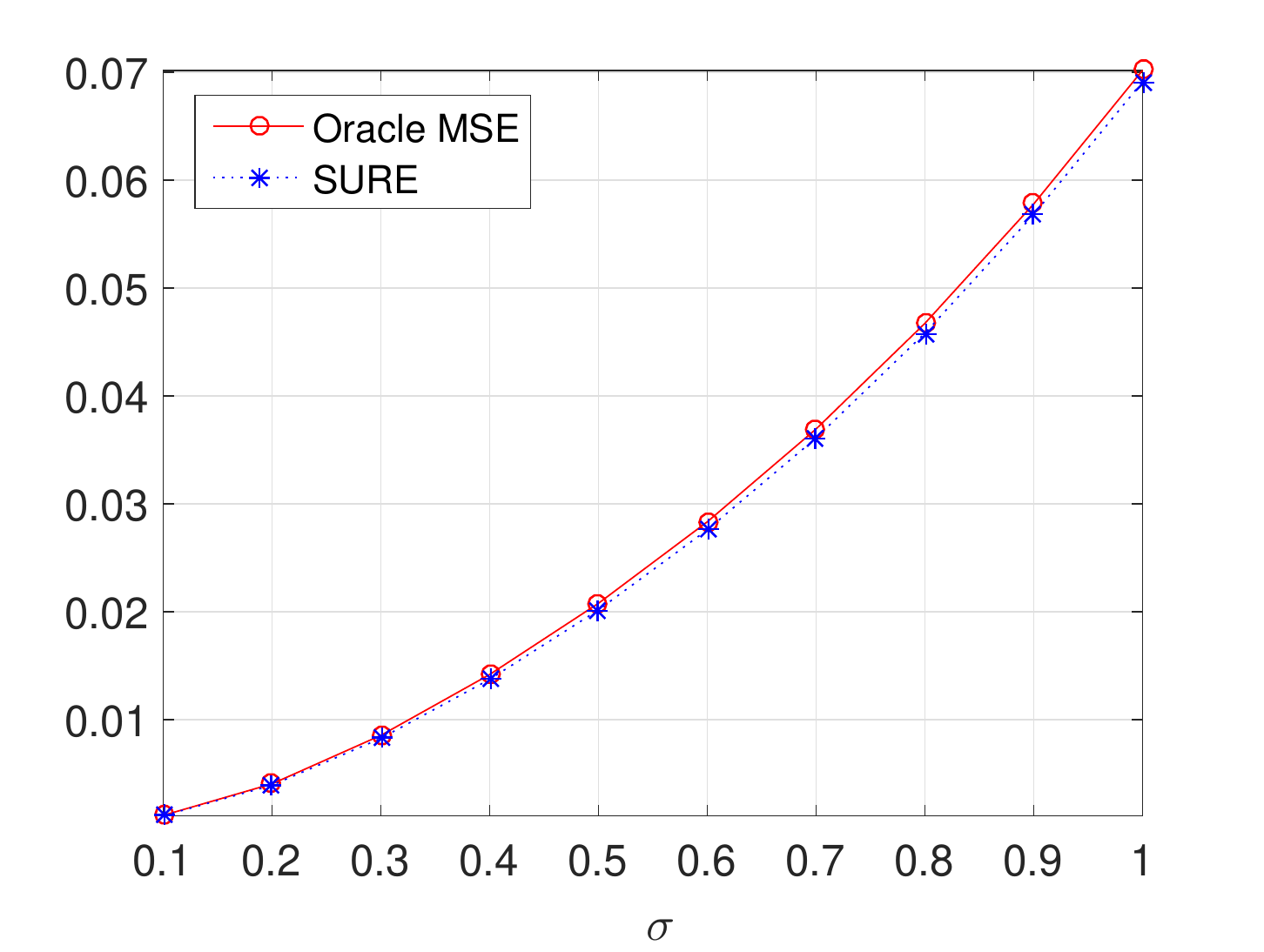}
		\caption{Truncated Gaussian}
		\label{sureMSEa}
	\end{subfigure}
	\hfill
	\begin{subfigure}{0.49\textwidth}
		\includegraphics[width=\textwidth]{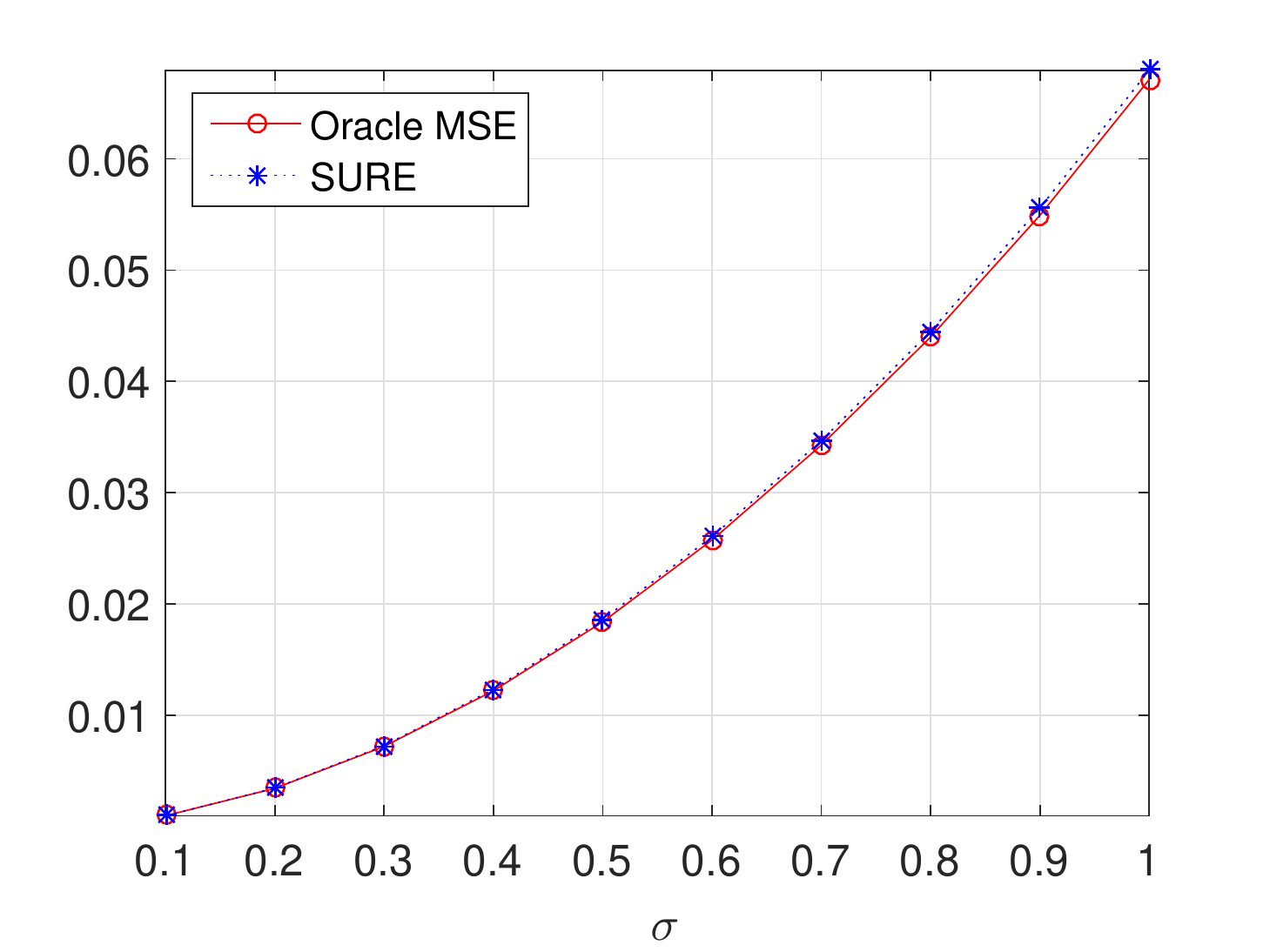}
		\caption{Smooth mountains}
		\label{sureMSEb}
	\end{subfigure}
    \hfill
	\begin{subfigure}{0.49\textwidth}
	\includegraphics[width=\textwidth]{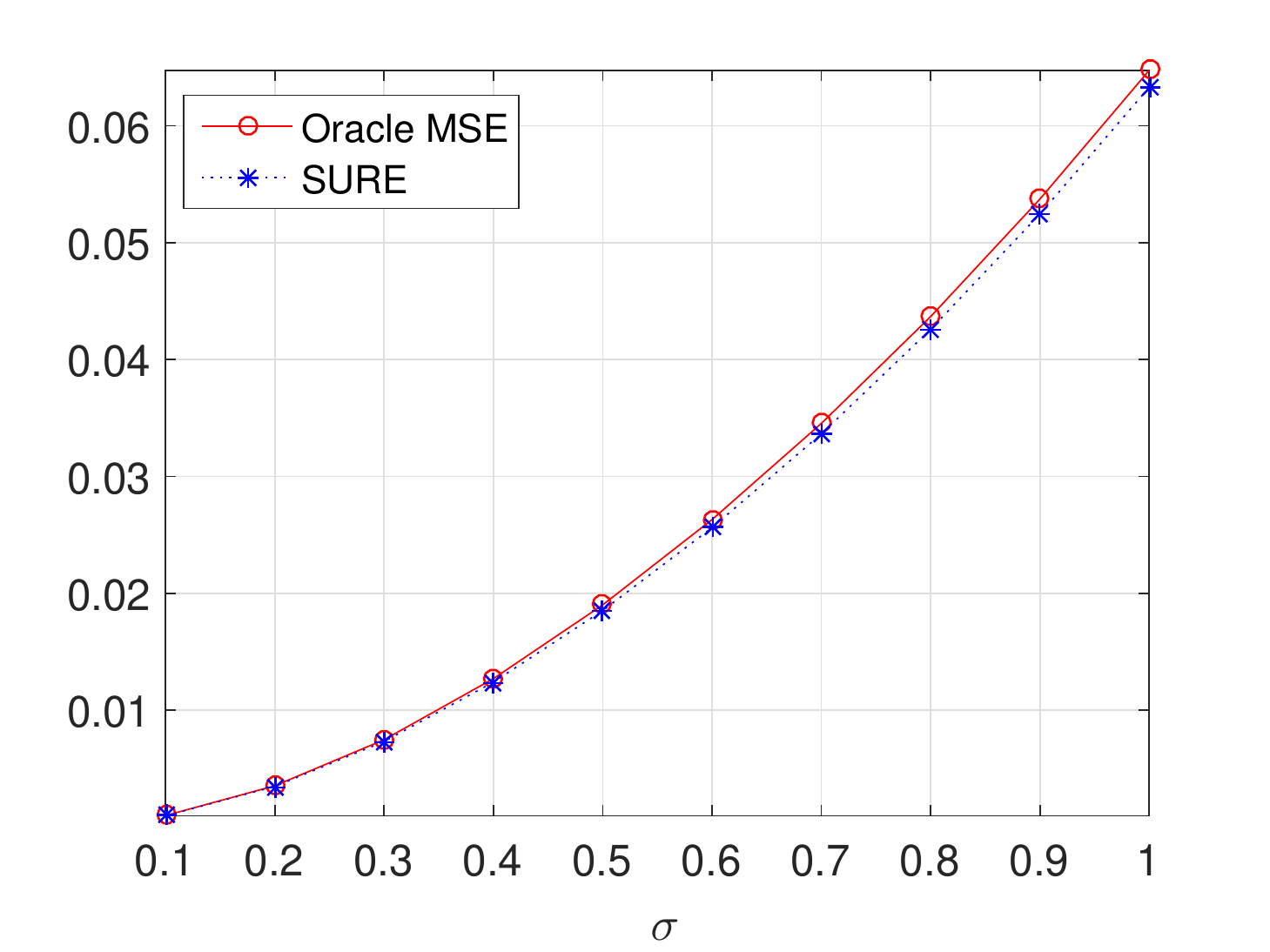}
	\caption{Shear Plane}
	\label{sureMSEc}
    \end{subfigure}
     \hfill
     \begin{subfigure}{0.49\textwidth}
	\includegraphics[width=\textwidth]{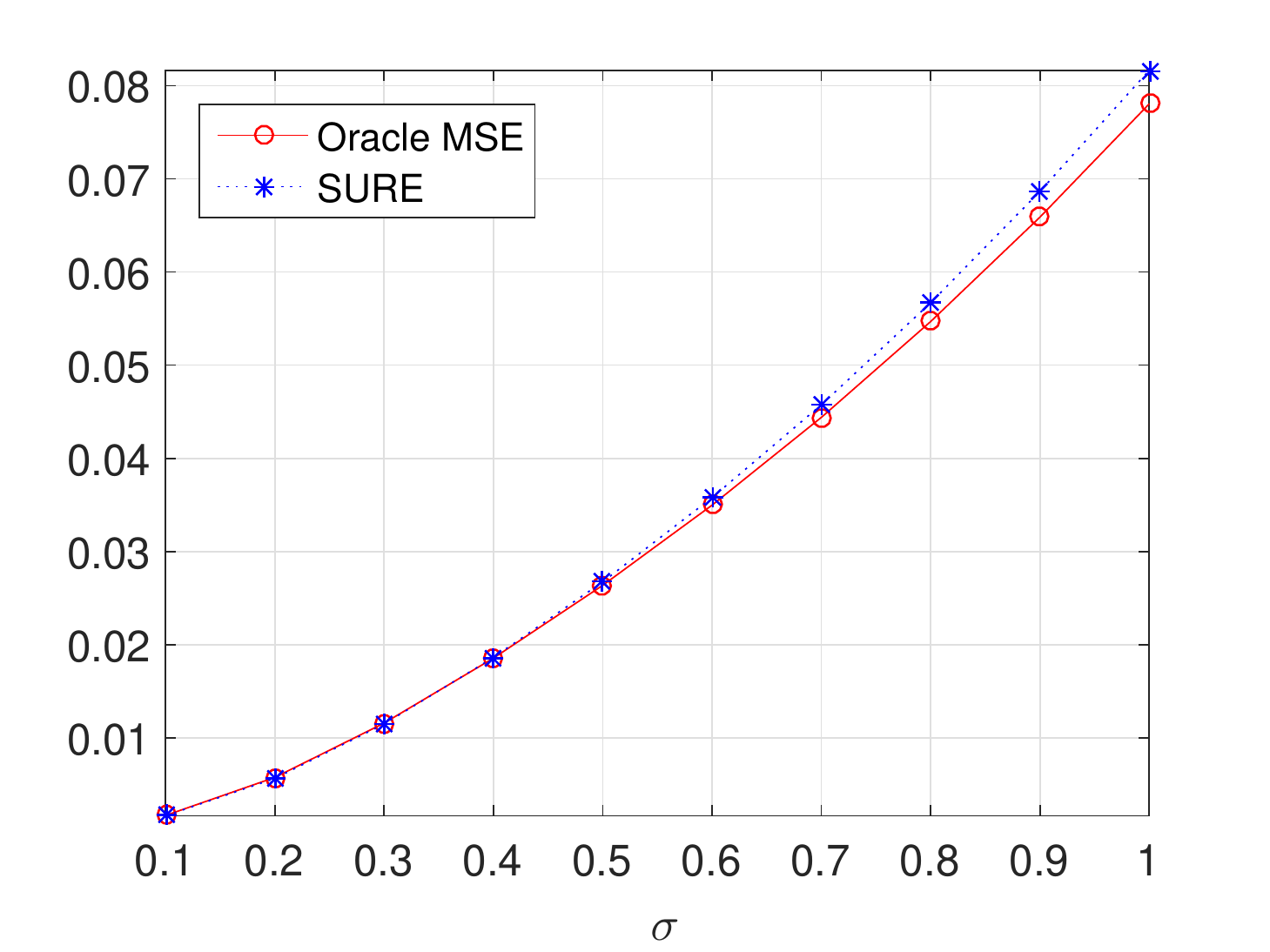}
	\caption{Peak-valley surface}
	\label{sureMSEd}
    \end{subfigure}
	\caption{SURE vs oracle $\bf{mse}$. WFF window corresponding to the scale $s=4$ is used}
	\label{sureMSE}
    \end{figure}
The complex derivative and the derived risk estimate (SURE) is assessed in the following experiment. Here four different surfaces given in \cref{TG,mnt,sp,pvsub} are considered as the true absolute phase $ \upPhi$ with unit amplitude $\bf a$. Observations following model  \eqref{obsmodel} are generated for low to high level of noise ($\sigma \in \cbr{0.3, 0.5, 0.7, 0.9}$). The values of SURE \eqref{sure} and of oracle $\bf {mse}$ are plotted in \cref{sureMSE} and all of them show that SURE is a very good approximation of the oracle $\bf {mse}$.
\subsection{SURE-Based fusing}
Let us consider $S$ different WFF$^{\textrm{let}}$ denoisers according to \eqref{eqn:wftestmn2} with scales $s=1,2,\cdots,S$. Let the windows be $\cbr{\bf h^i}_{i=1}^{i=S}$ and the corresponding WFF$^{\textrm{let}}$ estimates at pixel $\k$, arranged in vector form, be $\overline{ f}_\k(\bf z) :=\sbr{f^1_\k\nbr{\bf z},~ f^2_\k\nbr{\bf z},\cdots,f^S_\k\nbr{\bf z} } \in \mathbb{C}^S$.  For each pixel $\k \in G$, a fused estimate $f^{\textrm{fuse}}_{\k}(\bf z)$ is obtained by the following fusion formula:
\begin{align}
f^{\textrm{fuse}}_{\k}(\bf z)&:=\overline{\bf {a}}_\k^T   \overline{ f}_\k(\bf z)  \label{F},
\shortintertext {where}
\overline{\bf  a}_\k&:=\sbr{\bf a_\k^1 \ \bf a_\k^2 .... \bf a_\k^S}^T.
\end{align}
We design an optimization frame work, aiming at finding best fusion coefficient $\overline{\bf  a}_\k$, by minimizing SURE of $f^{\textrm{fuse}}_{\k}(\bf z)$. To design an efficient multi-resolution WFF, $\overline{ f}_\k(\bf z) $ should contain WFF estimates having resolutions in all possible ranges. To accomplish this, $S$ different windows are chosen in such a way that they cover a wide range of window sizes yielding estimates with low, medium and high resolutions. In such a choice of windows, the optimization is expected to pick up the best window out of the $S$ available choices. Mathematically, this is possible only if the fusion coefficient for the best window is positive and real-valued and the coefficients for the other windows are close to zero. This indicates that even if we use complex-valued $\overline{\bf  a}_\k$'s, the preferred feasible region of the SURE-fuse optimisation is a subspace that accommodates only real valued $\overline{\bf  a}_\k$'s. Keeping this in mind, we design SURE-fusion using real-valued non-negative fusion coefficients, i.e., $\overline{\bf  a}_\k \in \mathbb{R}^S_{\geq 0}$

We first define the \textbf{mse} of $f^{\textrm{fuse}}_{\k}(\bf z)$ and then derive an optimizing frame work to minimize its SURE. It is to be noted that in \eqref{eqn:mse_def}, a global \textbf{mse} is defined. Hence the sample mean of the squared error is computed by considering all the samples from the image, i.e., $\k \in G$. But for a pixel-wise fusion process, \textbf{mse} for a pixel $\k$ is to be defined. This is taken care by defining a neighbourhood, $ \xi_\k$, for each $\k$ and by assuming that the fusion coefficient is same within $ \xi_\k$. We define $ \xi_\k$  to be the set of pixels that falls within a square window centred at $\k$.\footnote{Implementation settings of $\xi_\k$ is explained in \cref{sec:paramset}.}  With this assumption in hand, the \textbf{mse} of $f^{\textrm{fuse}}_{\k}(\bf z)$ is defined as:
\begin{align}
\label{eqn:mse_def2}
\bf{mse}\cbr{f^{\textrm{fuse}}_{\k}(\bf z)} := \frac{1}{|\xi_\k|}\sum_{\n\in \xi_\k}\|f^{\textrm{fuse}}_{\n}(\bf z) - {\bf x}\|^2=\frac{1}{|\xi_\k|}\sum_{\n\in \xi_\k}\|\overline{\bf {a}}_\k^T   \overline{ f}_\n(\bf z) - {\bf x}\|^2.
\end{align}
In the above definition, we used the result $\overline{\bf {a}}_\n=\overline{\bf {a}}_\k,~\forall \n\in \xi_\k$, since we assumed that fusion coefficient is same in the neighbourhood $\xi_\k$. The unbiased estimate, SURE, of this \textbf{mse} can easily be define with the help of \eqref{sure} as:
\begin{align}
	\bs \varepsilon^{\textbf{sure}}_\k&=\frac{1}{| \xi_\k|}\sum_{\bf n\in  \xi_\k}\sbr{|| f^{\textrm{fuse}}_{\n}(\bf z) || ^2-2\Re \cbr{ \nbr{f^{\textrm{fuse}}_{\n}(\bf z)}^H \bf z_\n}+2\sigma^2\Re \cbr{\frac{\partial f^{\textrm{fuse}}_{\n}(\bf z)}{\partial \bf z_{\n}}} } + c, \label{opt_1}\\
	&=\frac{1}{| \xi_\k|} \sum_{\n\in  \xi_\k}\sbr{[||\a_\k^T \f|| ^2-2\Re \cbr{\f^H \a_\k \bf z_\n} +2\sigma^2\Re \cbr{\frac{\partial \a_\k^T \f}{\partial \bf  z_{\n}}}} + c,\\
	 &=\frac{1}{| \xi_\k|}\sum_{\n\in  \xi_\k}\sbr{\a_\k^T \f \f^H \a_\k-2  \Re \cbr{\a_\k^T \f^{^*} \bf z_\n }+2\sigma^2  \Re \cbr{\frac{\partial \a_\k^T \f}{\partial \bf z_{\n}}}} + c,\\
	 \intertext{(where * denotes the complex conjugate operator and $c$ is a conxtant not depending on ${\bf a}_{\bf k}$)}
	&=\frac{2}{| \xi_\k|}\sbr{ \frac{1}{2}\a_\k^T \cbr{\sum_{\n\in  \xi_\k}\f \f^H}\a_\k+ \a_\k^T \sum_{\n\in  \xi_\k} \Re\cbr{  -\cbr{\f^{^*} \bf z_\n  }+\sigma^2   \cbr{\frac{\partial  \f}{\partial \bf z_{\n}}}}}+ c,\\
	&= \frac{2}{| \xi_\k|} \sbr{\frac{1}{2} \a_\k^T \bf{H}_\k \a_\k+ \boldsymbol{\gamma}_\k^T \a_\k} +c, \label{eqn:qp0}
\shortintertext{where}
\bf{H}_\k&:= \sum_{\n\in  \xi_\k}\f \f^H,\\
\bs{\gamma}_\k&:=\Re \sbr{\sum_{\n\in  \xi_\k}\cbr{ - \cbr{ \f^{^*} \bf z_\n  }+\sigma^2   \cbr{\frac{\partial  \f}{\partial \bf z_{\n}}}}}.
\end{align}
The objective is to obtain $\bf a_\k$ such that it minimizes $\bs \varepsilon^{\textbf{sure}}_\k$. From \eqref{eqn:qp0}, the following optimization problem is formulated:
\begin{equation}
\begin{matrix}
\displaystyle \a_\k^{\mathrm{sure}} &= \underset{\a_\k}{\mathrm{argmin}}& \frac{1}{2}\a_\k^T \bf{H}_\k\a_\k+ \bs{\gamma}_\k^T \a_\k,\\
\textrm{s.t.} & \a_\k  \geq  0. \label{opt_end}\\
\end{matrix}
\end{equation}
This is a quadratic program with non-negative constraints and can easily be solved.

\begin{figure}[h!]
	\centering
	\begin{subfigure}{0.48\textwidth}
		\includegraphics[width=\textwidth]{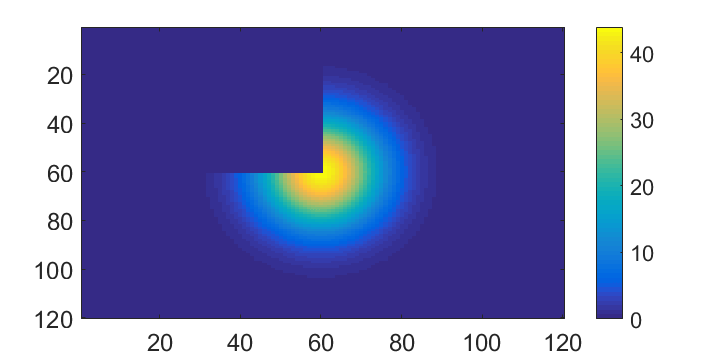}
		\caption{Truncated Gaussian ($\bs \upPhi_{}$)}
		\label{a11}
	\end{subfigure}
	\hfill
	\begin{subfigure}{0.48\textwidth}
		\includegraphics[width=\textwidth]{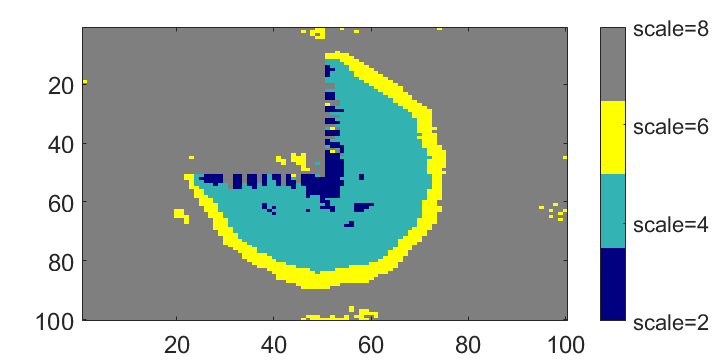}
		\caption{Point-wise minim. error indicator}
		\label{a12}
	\end{subfigure}
	\hfill
	\begin{subfigure}{0.23\textwidth}
		\includegraphics[width=\textwidth]{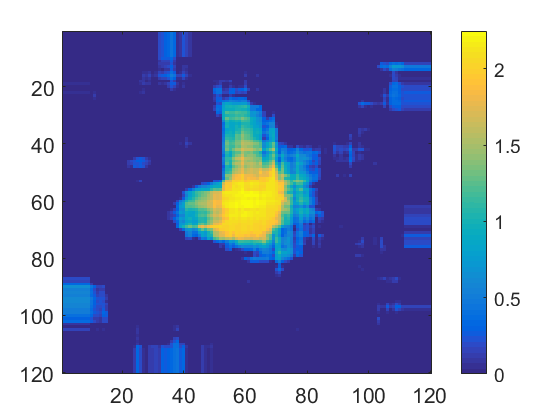}
		\caption{\centering Fusing coef. $\bf a$  for $scale=2$}
		\label{a13}
	\end{subfigure}
	\hfill
	\begin{subfigure}{0.23\textwidth}
	\includegraphics[width=\textwidth]{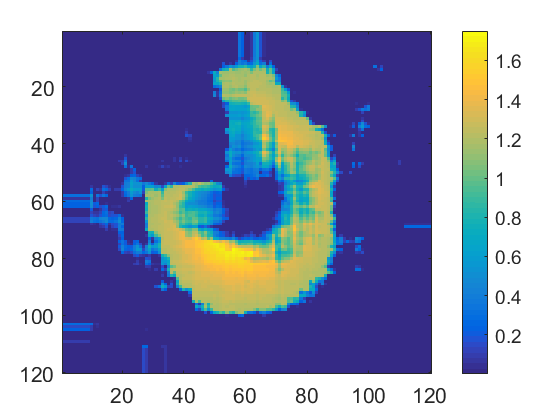}
	\caption{\centering Fusing coef. $\bf a$  for $scale=4$}
	\label{a21}
\end{subfigure}
\hfill
\begin{subfigure}{0.23\textwidth}
	\includegraphics[width=\textwidth]{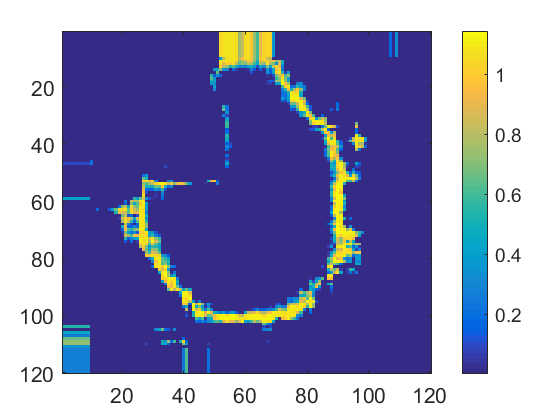}
		\caption{\centering Fusing coef. $\bf a$  for $scale=6$}
	\label{a22}
\end{subfigure}
\hfill
\begin{subfigure}{0.23\textwidth}
	\includegraphics[width=\textwidth]{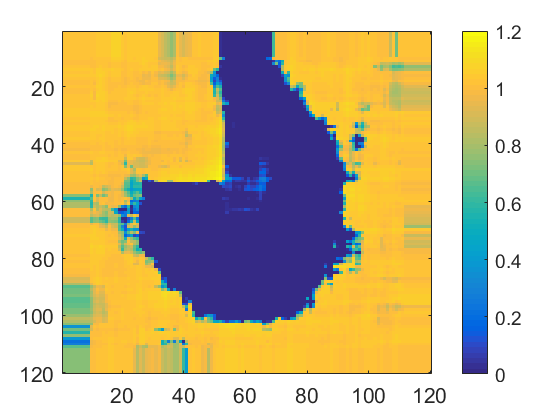}
		\caption{\centering Fusing coef. $\bf a$  for $scale=8$}
	\label{a23}
\end{subfigure}
	\caption{Image representation of SURE-fusion coefficients for a Truncated Gaussian surface of size:$120\times 120$ and phase range: 0 to 44 radians.}
	\label{TGcoef}
\end{figure}

In the following experiment,  SURE-fusion is analysed using an image representation of the fusion coefficients $\cbr{\a_\k}_{\k \in  G}$. A \textit{Truncated Gaussian} phase surface is used for the experiment. This is a Gaussian surface in which one quarter is removed to introduce sharp discontinuity. A highly noisy complex-valued observation is simulated as per model \eqref{obsmodel} ($a=1, \sigma=0.9$). SURE-fusion is done using four different scales $s \in \cbr{2, 4, 6, 8}$. The best scale, for each pixel, computed using pixel-wise minimum error (oracle), is indicated in \cref{a12}. From the image representation of the fusion coefficients (\cref{a13,a21,a22,a23}), it is observed that the SURE-fusion is in agreement with the pixel-wise minimum error and adapts to the terrain topography.

%
%
\subsection{Parameter settings of SURE-fuse WFF}
\label{sec:paramset}
\textit{\textbf{WFT window:}} For the windowed Fourier analysis, we choose Gaussian function, i.e., $\bf h_{\k}=\bf h_{k_1,k_2}=e^{-\frac{k_1^2+k_2^2}{s^2}}$. It is to be noted that the support of a Gaussian function  should be larger than 6 times the standard deviation ($6s/\sqrt{2}$), to accommodate most of the significant non-zero values of the function. Accordingly, the 2D-support of $\bf h$, which we denote as $\bf{n_{h}} $, is chosen as:
\begin{align}
\bf{n_{h}} =\nbr{n_{\bf h_1},n_{\bf h_2}}:=\nbr{ \lceil 6s \rceil_{odd}, \lceil 6s \rceil_{odd}},
\end{align}
where the function $\lceil y \rceil_{odd}$ finds the nearest odd integer greater than or equal $y$. We prefer an odd-valued support to place the window symmetrically around the central pixel.

\textit{\textbf{PRC settings:}} For a window function with scale $s$ and support $\bf {n_h}$, the values of $D_1$ and $D_2$ are carefully chosen to to ensure PRC. We recall the PRC criteria discussed in \cref{sec:WFT}: \textbf{i}) The support of $\bf h_{k_1,k_2}$ w.r.t. $k_1$ and $k_2$ are not greater than $\frac{m}{D_1}$ and $\frac{n}{D_2}$ respectively. Accordingly, we may set  $\frac{m}{D_1}=n_{\bf h_1}$ and $\frac{n}{D_2}=n_{\bf h_2}$. \textbf{ii}) $D_1$ and $D_2$ should be sub-multiples of $m$ and $n$ respectively. In order to guarantee \textbf{i} and \textbf{ii} together, me modify image size $\nbr{m,n}$ to the multiple of $\nbr{n_{\bf h_1},n_{\bf h_2}}$, in the respective dimensions, by padding zeros. The size of the new zero-padded image, denoted as $\nbr{\overbar{m},\overbar{n}}$, and the value of $\nbr{D_1,D_2}$ for perfect reconstruction are as follows: 
\begin{alignat}{2}
\overbar{m}&:=n_{\bf h_1}\lceil \frac{m}{n_{\bf h_1}}\rceil;  &&\overbar{n}:=n_{\bf h_2}\lceil \frac{n}{n_{\bf h_2}}\rceil \label{eqn:padd},\\
D_1&:=\frac{\overbar{m}}{n_{\bf h_1}}=\lceil \frac{m}{n_{\bf h_1}}\rceil;~~~~~&&D_2:=\frac{\overbar{n}}{n_{\bf h_2}}=\lceil \frac{n}{n_{\bf h_2}}\rceil, \label{d1d2}
\end{alignat}
where $\lceil \rceil$ is the ceiling function that performs `rounds to the nearest higher integer' operation. The updated 2D-frequency resolution of our WFT analysis is $\nbr{\frac{2\pi}{\overbar{m}/D_1},\frac{2\pi}{\overbar{n}/D_2}}$.

\begin{figure}[h!]
	\centering
	\begin{subfigure}{0.31\textwidth}
		\includegraphics[width=\textwidth]{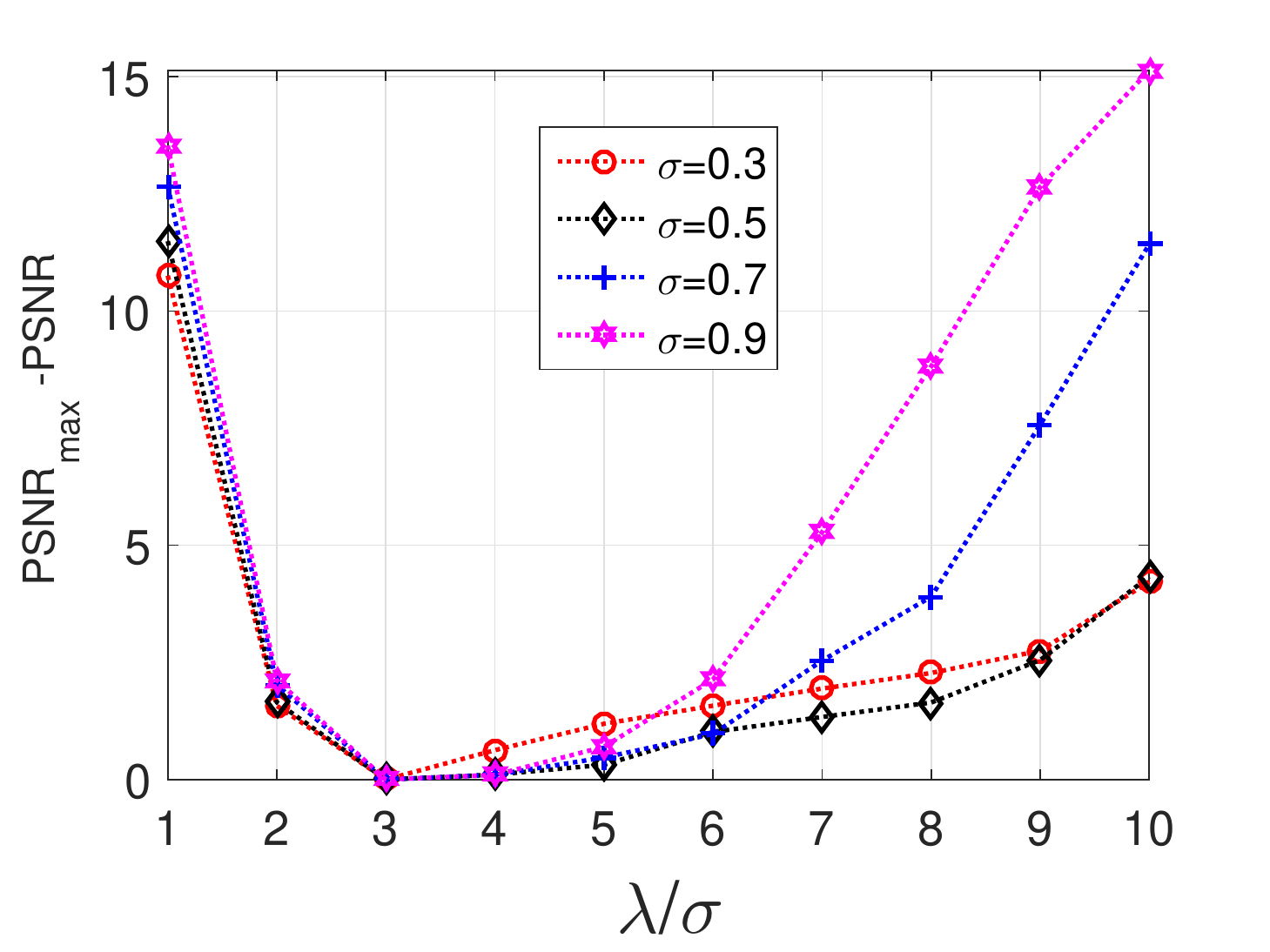}
		\caption{Truncated Gaussian }
		\label{TG_lam}
	\end{subfigure}
	\hfill
	\begin{subfigure}{0.31\textwidth}
		\includegraphics[width=\textwidth]{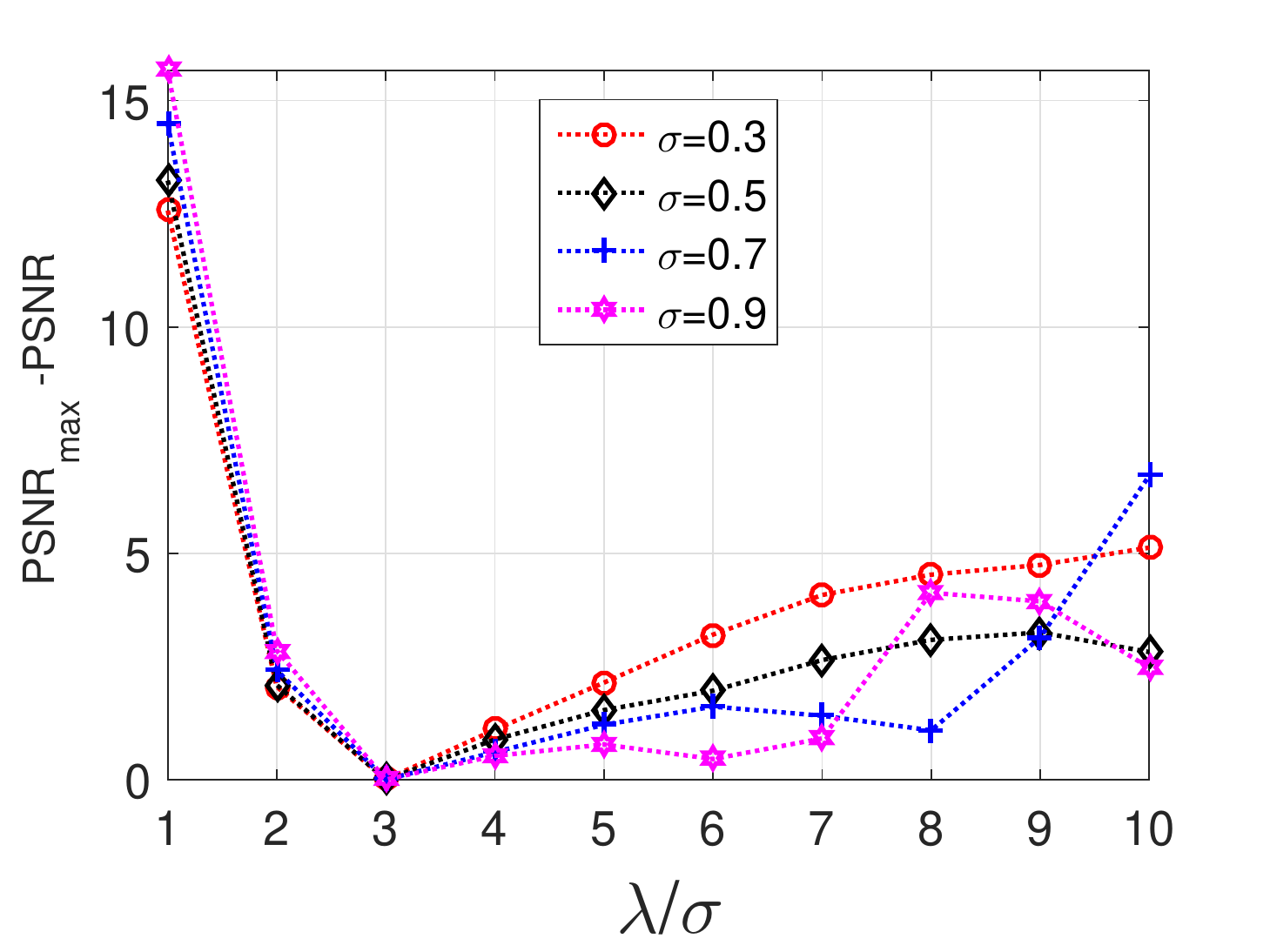}
		\caption{Spear Plane}
		\label{SP_lam}
	\end{subfigure}
	\hfill
	\begin{subfigure}{0.31\textwidth}
		\includegraphics[width=\textwidth]{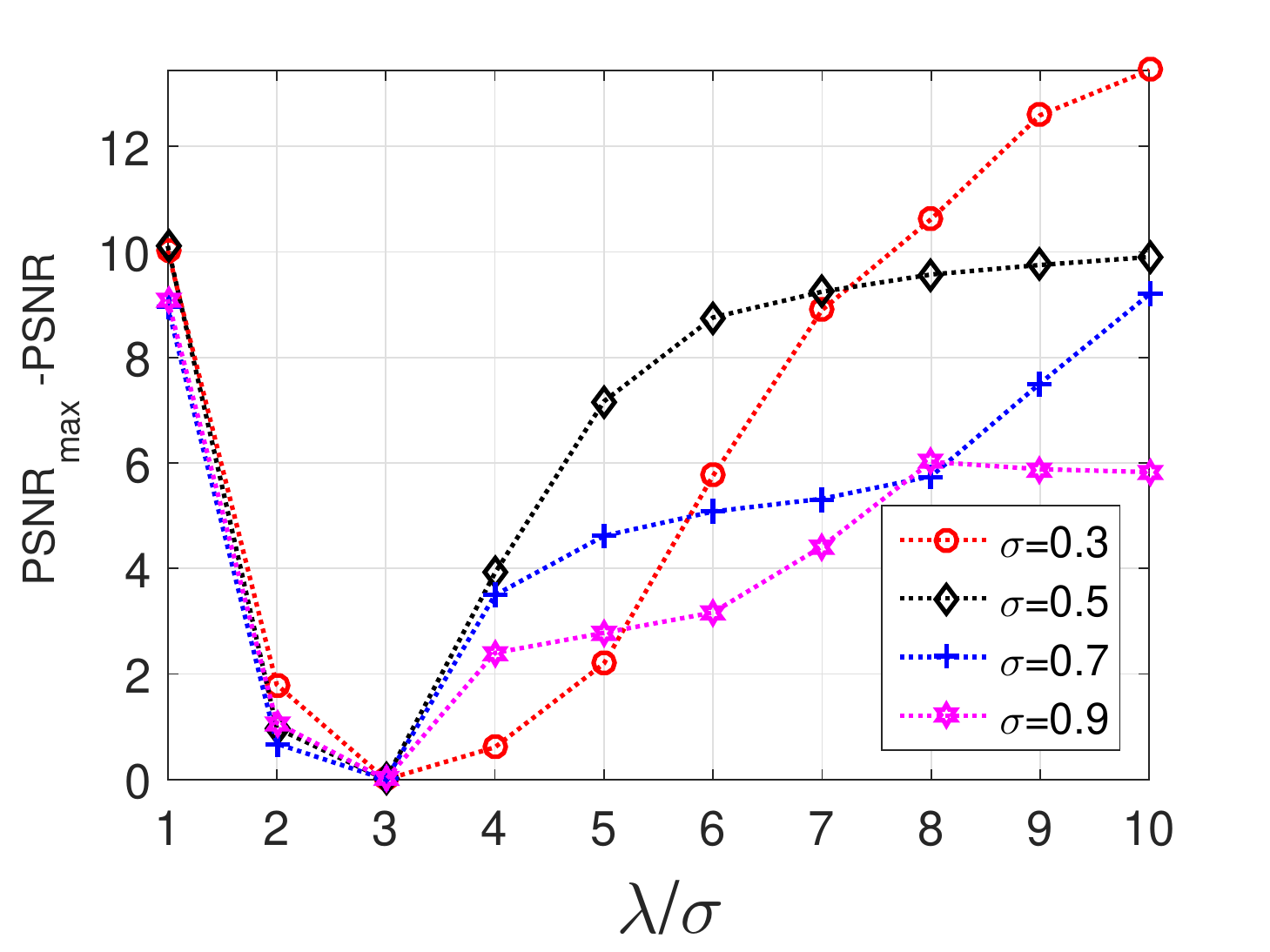}
		\caption{Peak-valley Surface}
		\label{PVS_lam}
	\end{subfigure}
	
	\caption{Emperical tuning for $\lambda$ using 3 different test images. (The absolute phase surface are given in \cref{TG,sp,pvsub} respectively)}
	\label{lambda}
\end{figure}

\textit{\textbf{LET parameter $\lambda$:}} The parameter $\lambda$ of the LET-based thresholding is heuristically tuned. It has been observed from the literature that this threshold performs well when expressed as a function of the noise level \cite{2007_Blu_SURELET,2007_Luisier_New_sure}. In our implementation we perform an empirical tuning  by observing the performance of SURE-fuse WFF, in terms of PSNR \eqref{psnr}, for a wide range of of $\lambda$, expressed as a function of the noise level $\sigma$. The experiment is conducted with different test images and it is observed that the best value of $\lambda$ in majority of cases is $\lambda=3\sigma$. The tuning of $\lambda$ done for 3 different test images are shown in \cref{lambda}

\textit{\textbf{Neighbourhood  $\xi_\k$ for \textbf{mse} computation:}} Another important parameter is the size of the square neighbourhood $\xi_\k$ which is used to compute \textbf{mse}. In general, for a better \textbf{mse} estimation, large number of samples are preferred. But on the other hand, since the non-local regions of the phase surface can be structurally very different, a very large neighbourhood is also not preferred. In our implementation we use $7\times 7$ square neighbourhood centred at the pixel of interest. This is a heuristically tuned trade off. 
\section{Experiments \& Results}
\label{sec:exp}
In this section, we present a series of experiments to evaluate the performance of the developed algorithm and to compare it with the state-of-the-art. The SURE-fuse WFF is expected to beat the best resolution WFF or to show a result close to it. This is demonstrated by comparing the performance of SURE-fuse WFF with the individual WFFs that are being fused. 
We use the implementation by Kemao\cite{2007_Kemao_Twodimensional} for the individual WFFs. In addition to this, two algorithms, namely SpInPhase\cite{2015_Hongxing_Interferometric} and MoGInPhase\cite{2017_Joshin_MoGInpahse} are also considered as the competitors, which are the state-of-the-art in InPhase denoising to the best of our knowledge. All these algorithms are tuned for the optimum performance as indicated in the respective papers. To evaluate the quality of the estimates, given that our main objective is InPhase denoising, we adopt \textit{peak signal-to-noise ratio} (PSNR), defined as:
\begin{equation}
\text{PSNR}:=10\log_{10}\frac{4N\pi^2}{ \norm{\mathcal{W}(\widehat{\boldsymbol{\upPhi}}_{2\pi}-\boldsymbol{\upPhi}_{})}_F^2} \ \  [\text{dB}], \label{psnr}
\end{equation}
where $\boldsymbol{\upPhi}_{}$ is the true phase (unwrapped), $\widehat{\boldsymbol{\upPhi}}_{2\pi}$ , is the estimated wrapped phase and $\mathcal{W} $ is the wrapping operator defined in \eqref{eqn:wrap}.
In the initial parts, we deal with simulated data set that are carefully selected to represent various structurally different phase surfaces. In the final parts, we also use phase data from InSAR and MRI to illustrate the effectiveness of the developed algorithm in real world  remote sensing and medical imaging scenarios.  
\subsection{Comparative analysis of the algorithms using Simulated data}
\subsubsection{SURE-fuse WFF versus WFF}
\begin{figure}[h!]
	\centering
	\begin{subfigure}{0.49\textwidth}
		\includegraphics[width=\textwidth]{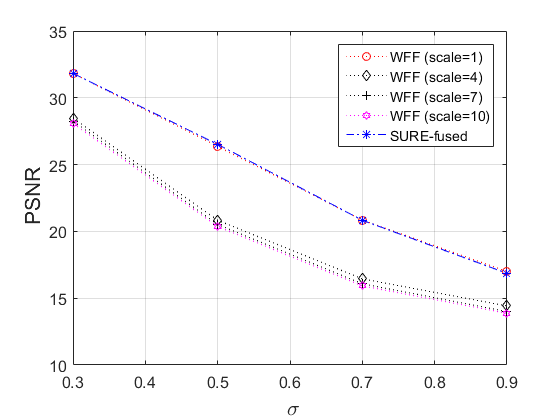}
		\caption{Smooth mountains}
		\label{mnt_lf2}
	\end{subfigure}
	\hfill
	\begin{subfigure}{0.49\textwidth}
		\includegraphics[width=\textwidth]{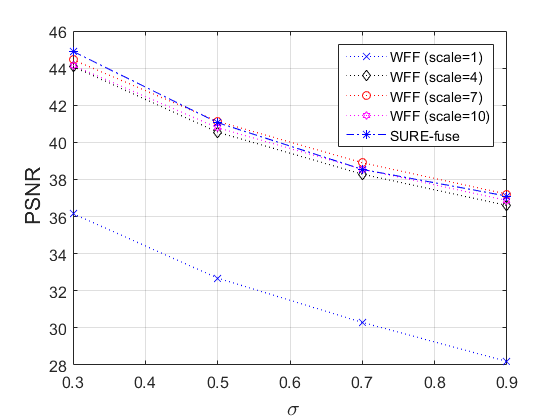}
	\caption{Non-smooth mountains}
    \label{mnt_hf2}
	\end{subfigure}
	\caption{PSNR comparisons }
			\label{hfmain}
    \end{figure}

\begin{figure}[h!]
	\centering
	\begin{subfigure}{0.33\textwidth}
		\includegraphics[width=\textwidth]{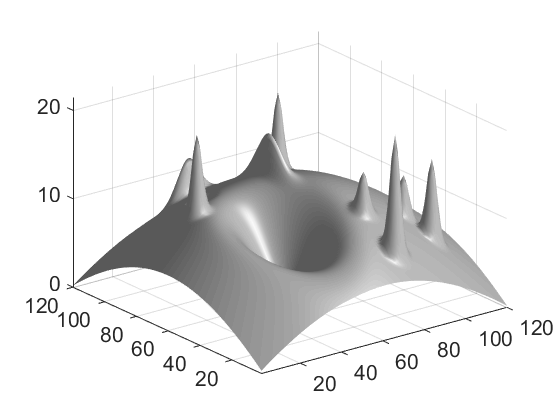}
		\caption{peak-valley surface}
		\label{pvs_1}
	\end{subfigure}
	\hfill
	\begin{subfigure}{0.33\textwidth}
		\includegraphics[width=\textwidth]{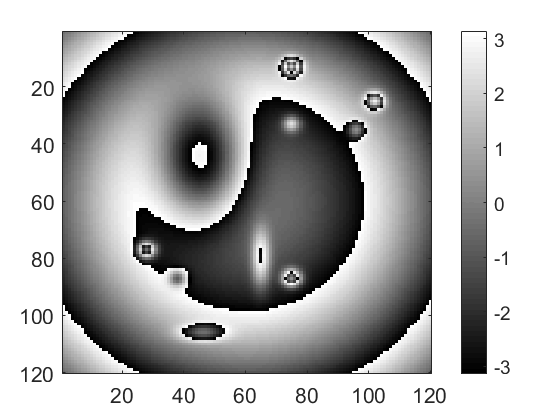}
		\caption{interferogram}
		\label{pvs_2}
	\end{subfigure}
	\hfill
\begin{subfigure}{0.32\textwidth}
	\includegraphics[width=\textwidth]{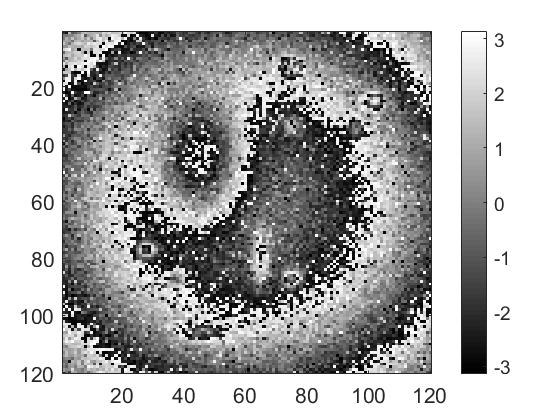}
	\centering \caption{ noisy interferogram ($\sigma=0.9$)}
	\label{pvs_3}
\end{subfigure}
	\caption{Peak-valley surface. Simulated phase surface of size $120 \times 120$ and phase range from 0 to 22 radians}
	\label{pvsmain}
\end{figure}

Initially, the same experiments conducted with smooth (\cref{pre_lf}) and non-smooth (\cref{pre_hf}) mountains in \cref{sec:sure} are repeated here using SURE-fuse WFFs. \Cref{hfmain} shows the PSNR plots of SURE-fuse WFF and the individual WFFs as a function of noise variance. For  better readability of the plots, WFFs with representative scales $s=1,4,7,10$ are shown;  the performance of other WFFs lies in between. It can be observed from \cref{mnt_lf2,mnt_hf2} that SURE-fusion's shows very competitive results, which are close to or better than the best WFF being fused.
In the above experiment, although a smooth and a non-smooth topology are considered, the structural properties of a particular topology is spatially similar throughout the image. In such cases, the optimum resolution does not much vary from pixel to pixel and the SURE-fusion will have close performance with the best WFF. Now we consider more complex structures where a single phase image has regions with both smooth and sharp phase variations. In such cases, SURE-fusion is expected to beat the best WFF, as fusion mechanism can pick-up the best resolution in a pixel-wise manner. To demonstrate this, we simulate a special surface, which we term as \textit{peak-valley surface} (\cref{pvs_1}), having pit like dips and sudden rising peaks. The corresponding interferogram is shown in \cref{pvs_2}. A heavily noisy version of the image with noise level $\sigma=0.9$ as shown in \cref{pvs_3} is denoised by considering WFF with eight different window sizes ($S=8$) corresponding to the scales $s \in \lbrace 1, 2, ..., 8\rbrace$. 

	\begin{figure}[h!]
		\captionsetup{justification=centering}
		\hspace{-2.5cm}
		\includegraphics[scale=0.4]{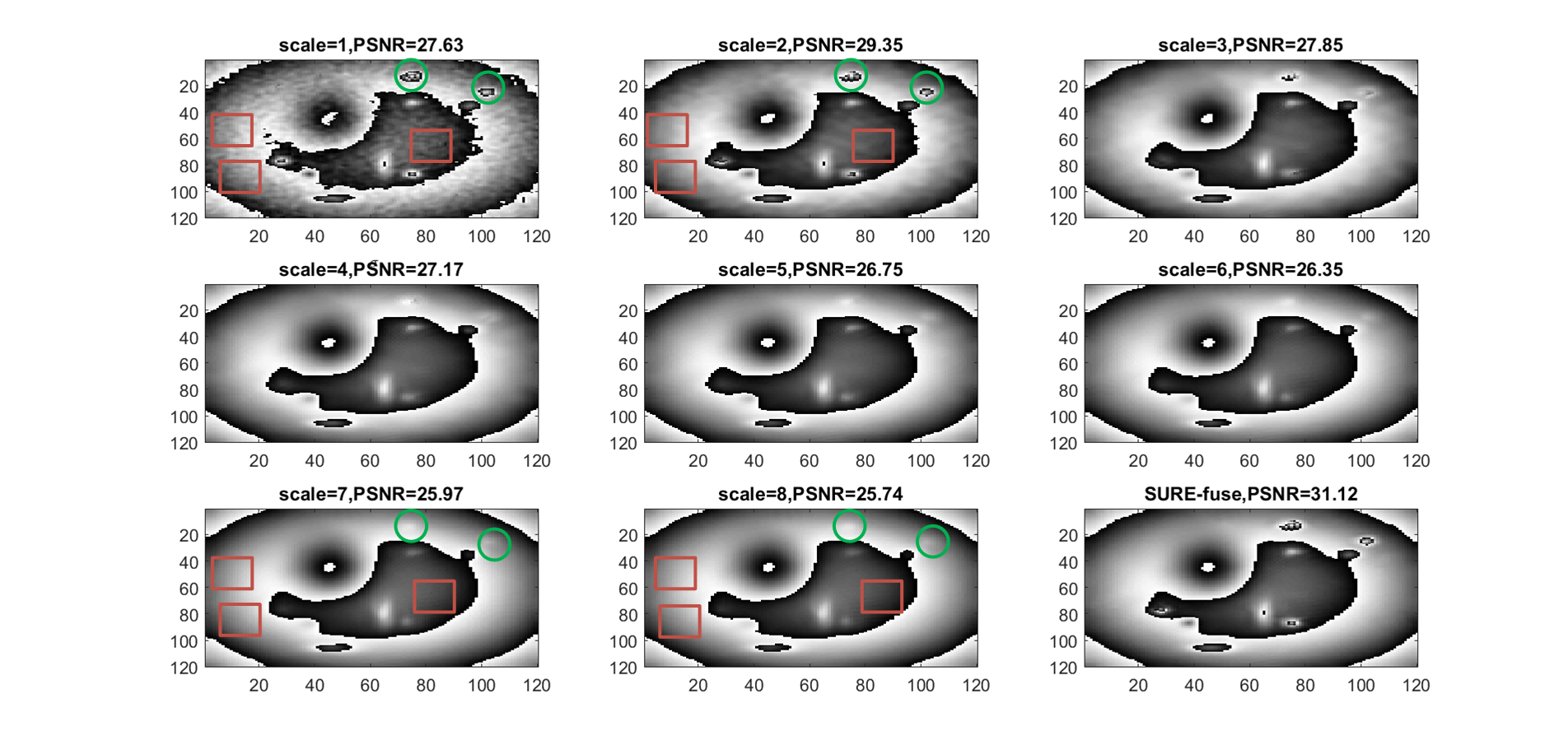}
		\caption{Performance indicators of SURE-fusing algorithm in comparison with WFF ($scale=1,2,...8.$). Green circles indicate sharp peaks and red squares indicate smooth regions.}
		\label{pvs_4}
\end{figure}
 The estimated interferograms along with the PSNR values are shown in \cref{pvs_4} and the following conclusions are drawn: (\textbf{i}) The sharp peaks (or valley), indicated by green circles, are well estimated by small window-WFF, i.e., $s= 1 \ \& \ 2$. These peaks are almost absent for large  window-WFF, i.e., $s= 7 \  \& \ 8$.  (\textbf{ii}) The smooth regions, indicated by red squares, are well estimated by large window-WFF. These regions are of poor quality in small window-WFF estimates.  (\textbf{iii}) SURE-fusion combines the advantages of the individual WFFs in a data dependent manner and estimates high quality interferogram with clear sharp peaks and smooth regions. These visual evidences are very well supported by the PSNR value (31.12 dB) compared to its WFF counter parts. The performance indicators for peak-valley surface for other noise levels are listed in the next section.
\subsubsection{SURE-fuse WFF versus state-of-the-art competitors}
\begin{figure}[h!]
	\centering
	\begin{subfigure}{0.19\textwidth}
		\includegraphics[width=\textwidth]{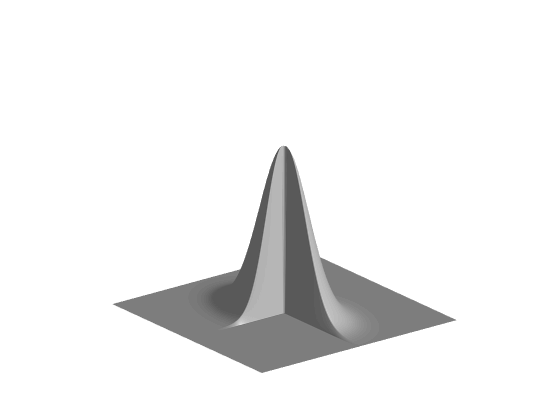}
		\caption{\centering Truncated Gaussian}
		\label{TG}
	\end{subfigure}
	\hfill
	\begin{subfigure}{0.19\textwidth}
	\includegraphics[width=\textwidth]{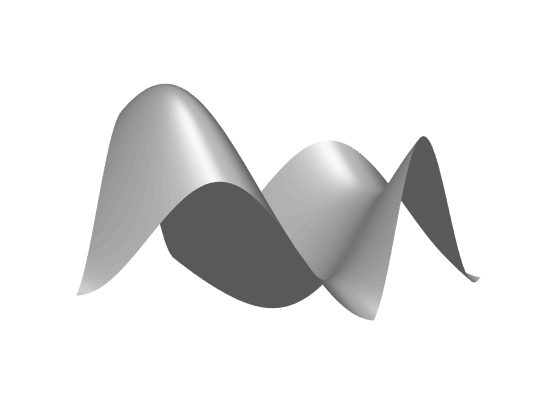}
	 \caption{\centering Smooth mountain}
	\label{mnt}
\end{subfigure}
	\hfill
\begin{subfigure}{0.19\textwidth}
	\includegraphics[width=\textwidth]{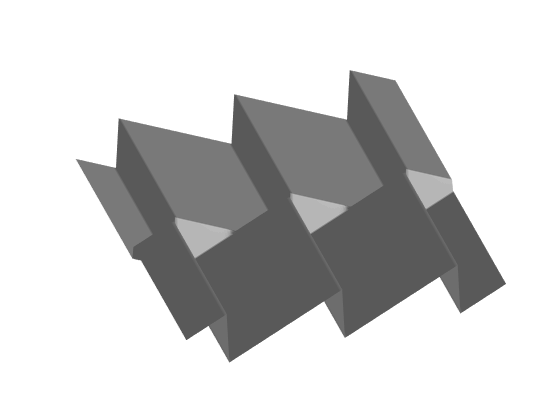}
	\caption{Spear Plane}
	\label{sp}
\end{subfigure}
	\hfill
\begin{subfigure}{0.19\textwidth}
	\includegraphics[width=\textwidth]{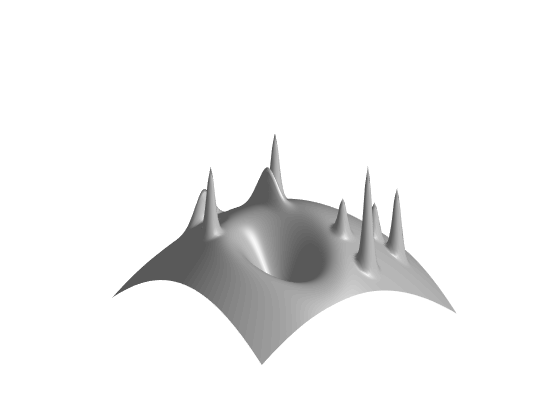}
	\caption{\centering Peak-valley surface}
	\label{pvsub}
\end{subfigure}
	\hfill
\begin{subfigure}{0.19\textwidth}
	\includegraphics[width=\textwidth]{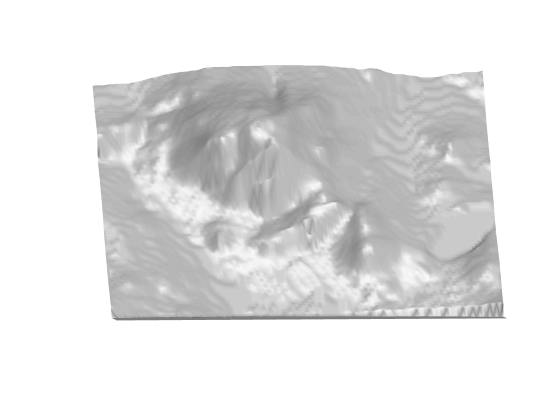}
	\caption{Longs Peak}
	\label{longs}
\end{subfigure}
	\caption{Data set}
\label{simdataset}
\end{figure}

The multi-resolution ability of SURE-fuse WFF, in comparison with the normal WFF, was demonstrated in the previous section. In this section, we present more experiments and bring the comparisons with the state-of-the-art competitors. To consider a wide range of phase topologies, we use a data set which is shown in \cref{simdataset}. Out of these surfaces, \cref{TG,mnt,sp,pvsub} are synthetic data. But we also use data from Interferometric SAR (InSAR),  which is shown in \cref{longs}. This  is the digital elevation model of a mountainous terrain around Longs Peak, Colorado, distributed with book \citep{1998_Ghiglia_Two}\footnote{For the performance evaluation of the InSAR data, only the valid image pixels, as indicated by the quality masks supplied with the data set, are considered.}. All these surfaces are of size $120\times120$ with spatial phase variations in the following range: Truncated Gaussian (\cref{TG})- 0 to 44 radians, Smooth mountains (\cref{mnt})- 0 to 44 radians, Spear plane (\cref{sp})- 0 to 7 radians, Peal-valley surface (\cref{pvsub})- 0 to 22 radians, Long Peaks (\cref{longs})- 0 to 90 radians. A complex-valued data is created with each of these surfaces as the absolute phase ($\bs \upPhi$) and with an invariant amplitude equal to 1. These is a very tough data set for the phase denoising algorithms as the underlying absolute phases contain wide range of structural properties that include smooth regions, planes, sharp peaks and valleys, sharp discontinuities, etc. Observations as per model \eqref{obsmodel} are generated for low to high level of noise ($\sigma \in \lbrace0.3, 0.5, 0.7, 0.9 \rbrace$).

Ten different windows ($S=10$), corresponding to the scales $s \in \lbrace 1, 2, 3,...., 10 \rbrace$, are considered. \Cref{table1} shows performance indicators of the developed algorithms in comparison with the state-of-the-art competitors. In the same table, we have also tabulated the results for the normal WFF, for each of the scales, using the implementation by Kemao \cite{2007_Kemao_Twodimensional}. In \cref{table1}, the WFF results corresponding to the best window sizes are highlighted using rectangle boxes. It is evident from the table that in most of the cases, SURE-fusing yields the best result, even considering the  best WFF estimate. Also in majority of the test cases, except for the Spear Plane, SURE-fusion beats SpInPhase and MoGInPhase with considerable margins. We would like to remark that the Spear Plane considered here has a periodic structure with high level of non-local self-similarity. For such structures, algorithms like SpInPhase and MoGInPhase are expected to perform very well as they are designed to exploit the non-local self-similarity. However, such synthetic images, a sinusoidal surface for another example, are structurally very different from the real life data.

\begin{table}[h!]
	\begin{center}
		\scalebox{0.85}{
			\begin{tabular}{r||c||cccccccccc||c||c||c}
				\multicolumn{1}{r||}{\multirow{4}[1]{*}{Surf.}} & \multirow{3}[1]{*}{$\sigma$} & \multicolumn{13}{c}{{\normalsize PSNR (dB) }} \\
				\multicolumn{1}{r||}{\multirow{2}[1]{*}{}} & \multirow{2}[1]{*}{} & \multicolumn{10}{c||}{\footnotesize WFF (Kemao)}  & \multicolumn{1}{c||}{} & \multicolumn{1}{c||}{} & \multicolumn{1}{c}{SURE-}\\
				\multicolumn{1}{r||}{} &       & \multicolumn{1}{c}{s=1} & \multicolumn{1}{c}{s=2} & \multicolumn{1}{c}{s=3} & \multicolumn{1}{c}{s=4} & \multicolumn{1}{c}{s=5} & \multicolumn{1}{c}{s=6}& \multicolumn{1}{c}{s=7}& \multicolumn{1}{c}{s=8}& \multicolumn{1}{c}{s=9} & \multicolumn{1}{c||}{s=10} & \multicolumn{1}{c||}{Sp} & \multicolumn{1}{c||}{MoG} & \multicolumn{1}{c}{fuse}\\
				
				\midrule
				\midrule
				\clineB{6-6}{2.5}
				\multirow{2}[1]{*}{Trunc. }      & 0.3   &36.28	&40.14	&41.37	&\LR{41.43}	&41.09	&40.62	&40.11	&39.57	&39.06	&38.64	&42.47	&41.66	&\B{42.60}\\
											 	 & 0.5   &32.75	&36.44	&37.48	&\LR{37.62}	&37.29	&36.74	&36.25	&35.64	&35.12	&34.72	&\B{39.26}	&37.46	&39.01\\
				\multirow{2}[1]{*}{Gauss.}       & 0.7   &30.36	&34.15	&35.24	&\LR{35.25}	&34.85	&34.08	&33.39	&32.87	&32.50	&32.09	&35.99	     &34.90&\B{36.20}\\
												 & 0.9   &28.25	&32.42	&33.59	&\LR{33.60}	&32.96	&32.22	&31.62	&31.04	&30.51	&29.97	&33.79	    &34.30	&\B{34.46}\\
				\clineB{6-6}{2.5}
				\midrule
				\midrule
				\clineB{8-8}{2.5}
                \multirow{2}[1]{*}{Mountain}     & 0.3   &36.15	&40.74	&42.95	&44.08	&44.59	&\LR{44.78}	&44.45	&44.40	&44.39	&44.14  &42.82 &44.40  &\B{44.90}\\
                                                 & 0.5   &32.68	&37.15	&39.41	&40.54	&41.08	&\LR{41.32}	&41.13	&41.09	&41.02	&40.78  &39.41 &40.92  &\B{41.06}\\
                \multirow{2}[1]{*}{(smooth)}     & 0.7   &30.29	&34.82	&37.13	&38.28	&38.81	&\LR{39.04}	&38.90	&38.85	&38.79	&38.55  &36.30 &38.15  &\B{38.54}\\
                                                 & 0.9   &28.19	&33.09	&35.44	&36.59	&37.13	&\LR{37.34}	&37.21	&37.15	&37.11	&36.86  &34.48 &37.01  &\B{37.10}\\ 
				\clineB{8-8}{2.5}
				\midrule
				\midrule
				\clineB{8-8}{2.5}
				\multirow{2}[1]{*}{Spear}        & 0.3   &35.85	&40.05	&42.06	&43.01	&43.35	&\LR{43.36}	&43.06	&42.76	&42.48	&42.24  &\B{48.33}	&47.58      &43.41\\
											     & 0.5   &32.23	&36.13	&37.82	&38.55	&38.73	&\LR{38.74}	&38.43	&38.29	&38.13	&38.02  &\B{43.81}	&43.20      &39.42\\
				\multirow{2}[1]{*}{plane}        & 0.7   &29.77	&33.59	&35.28	&36.08	&36.30	&\LR{36.32}	&36.18	&36.22	&36.07	&36.00  &36.72	    &\B{38.85}  &36.95\\
											     & 0.9   &27.61	&31.83	&33.59	&34.36	&34.64	&\LR{34.71}	&34.65	&34.56	&34.41	&34.37  &34.17	    &\B{36.18}  &35.23\\
				\clineB{8-8}{2.5}
				\midrule
				\midrule
				\clineB{4-4}{2.5}
				\multirow{2}[1]{*}{Peak-}       & 0.3   &35.98	&\LR{39.38}	&38.88	&36.30	&35.80	&33.44	&31.29	&29.79	&29.14	&28.68  &40.59	&40.43  &\B{41.90}\\
											    & 0.5   &32.45	&\LR{35.29}	&33.22	&30.49	&28.63	&27.78	&27.39	&27.12	&26.94	&26.84  &36.36	&29.99  &\B{38.22}\\
				\multirow{2}[1]{*}{Valley}      & 0.7   &29.97	&\LR{31.14}	&29.00	&27.63	&27.11	&26.68	&26.50	&26.35	&26.21	&26.09  &32.23	&26.29  &\B{34.86}\\
												& 0.9   &27.61	&\LR{29.31}	&27.76	&27.29	&26.77	&26.39	&25.80	&25.77	&25.41	&25.30  &29.26	&25.35  &\B{31.88}\\
				\clineB{4-4}{2.5}
				\midrule
				\midrule
				\clineB{4-4}{2.5}
				\multirow{2}[1]{*}{Longs}        & 0.3   &33.71	&\LR{33.18}	&32.46	&32.09	&31.89	&31.59	&31.35	&31.19	&31.04	&30.88	&33.56	&32.98 &\B{35.08}\\
											   	 & 0.5   &30.13	&\LR{30.51}	&29.57	&28.98	&28.77	&28.55	&28.37	&28.21	&28.06	&27.91	&30.86	&28.71 &\B{32.20}\\
				\multirow{2}[1]{*}{Peak}         & 0.7   &27.66	&\LR{28.29}	&28.03	&27.75	&27.41	&27.11	&26.87	&26.67	&26.54	&26.44	&28.90	&27.35 &\B{30.40}\\
												 & 0.9   &25.51	&\LR{27.16}	&27.16	&26.89	&26.54	&26.19	&25.85	&25.60	&25.44	&25.27	&27.57	&26.15 &\B{29.12}\\
				\clineB{4-4}{2.5}
				\midrule
				\midrule
		\end{tabular}}%
		\caption{Performance indicators for surfaces shown in \cref{simdataset}. Sp:SpInPhase, MoG: MoGInPhase, s: scale of WFF. Best values among WFF estimates are shown in boxes. Best among Sp, MoG and SURE-fuse are shown in bold.}
		\label{table1}
	\end{center}
\end{table}
\subsection{Real MRI data}
\begin{figure}[htbp!]
	\centering
	\begin{subfigure}{0.33\textwidth}
		\includegraphics[width=\textwidth]{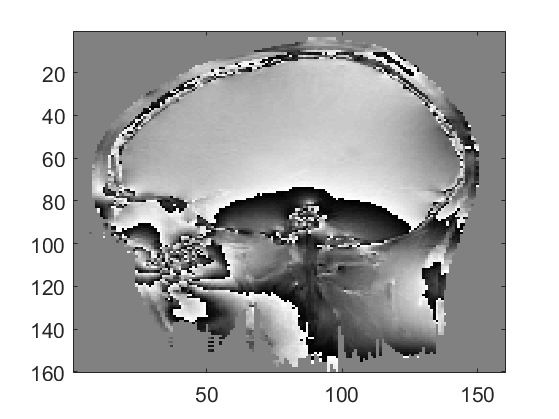}
		\caption{Side view}
		\label{mri_a1}
	\end{subfigure}
	\hfill
	\begin{subfigure}{0.33\textwidth}
		\includegraphics[width=\textwidth]{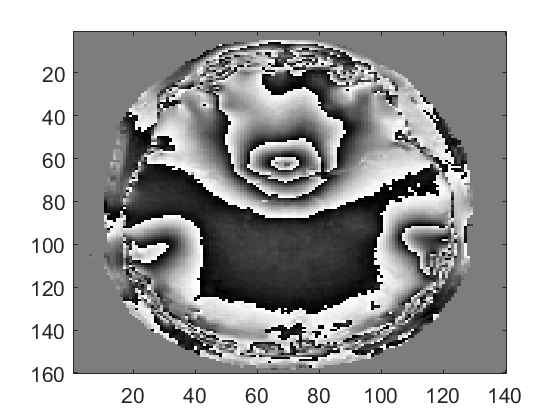}
		\caption{Top view}
		\label{mri_a2}
	\end{subfigure}
	\hfill
	\begin{subfigure}{0.32\textwidth}
		\includegraphics[width=\textwidth]{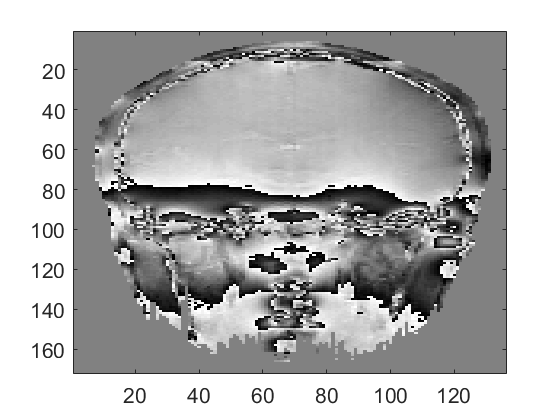}
		\centering \caption{ Front view}
		\label{mri_a3}
	\end{subfigure}
	\caption{Real MRI interferograms of a human head obtained by scanning along 3 different directions}
	\label{mri_a}
\end{figure}

\begin{figure}[htbp!]
	\centering
	\begin{subfigure}{0.33\textwidth}
		\includegraphics[width=\textwidth]{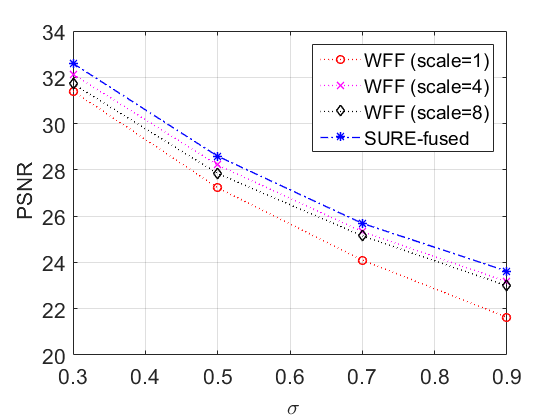}
		\caption{PSNR plots: Side view}
		\label{mri_b1}
	\end{subfigure}
	\hfill
	\begin{subfigure}{0.33\textwidth}
		\includegraphics[width=\textwidth]{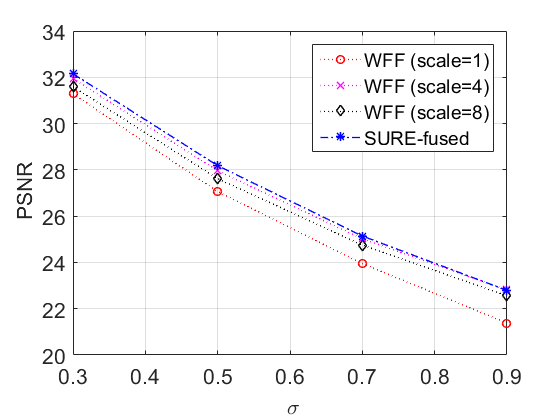}
		\caption{PSNR plots: Top view}
		\label{mri_b2}
	\end{subfigure}
	\hfill
	\begin{subfigure}{0.32\textwidth}
		\includegraphics[width=\textwidth]{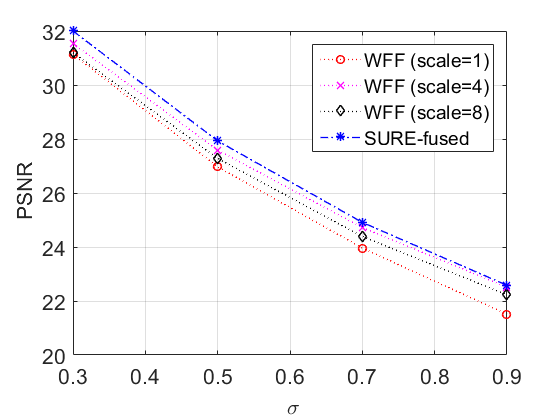}
		\centering \caption{PSNR plots: Front view}
		\label{mri_b3}
	\end{subfigure}
	\caption{SURE-fusing vs WFF for MRI interferograms}
	\label{mri_b}
\end{figure}

\begin{figure*}[htbp!]
	\hskip-1.5cm
	\begin{tabular}{ccccc}
		\includegraphics[scale=0.23]{mri_side.png}&\includegraphics[scale=0.23]{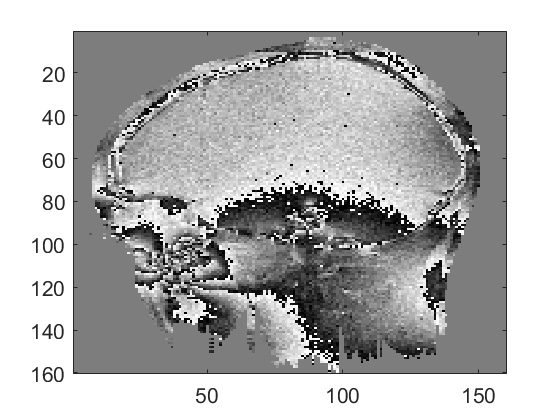}&\includegraphics[scale=0.23]{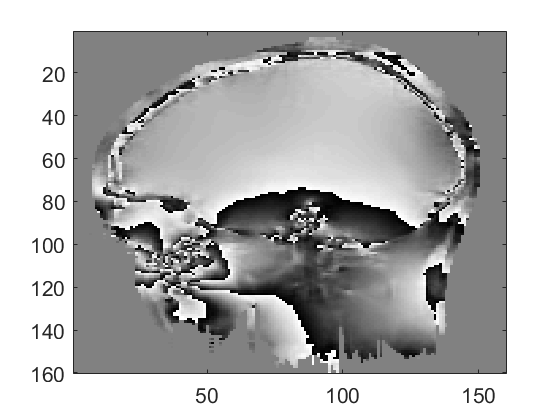}
		&\includegraphics[scale=0.23]{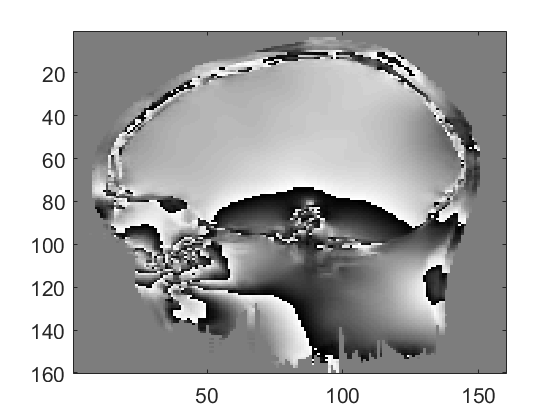}&\includegraphics[scale=0.23]{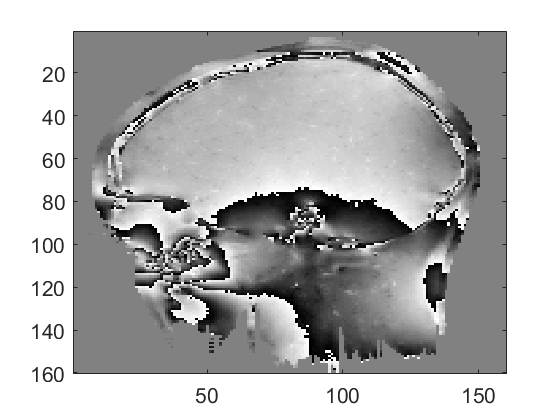}\\
		\scriptsize	Original phase &  \scriptsize	Noisy phase    & \scriptsize SpInPhase estimate   & \scriptsize MoGInPhase estimate & \scriptsize SURE-fuse estimate\\
	    \scriptsize (Side view)    &  \scriptsize	($\sigma=0.5$) & \scriptsize $PSNR=28.34$         & \scriptsize $PSNR=26.76$        & \scriptsize $\bf{PSNR=28.59}$ \\
		\includegraphics[scale=0.23]{mri_top.png}&\includegraphics[scale=0.23]{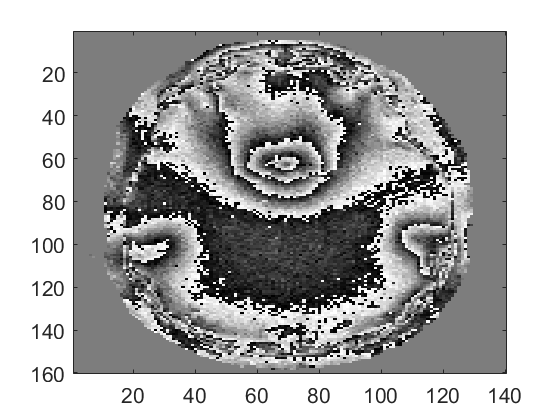}&\includegraphics[scale=0.23]{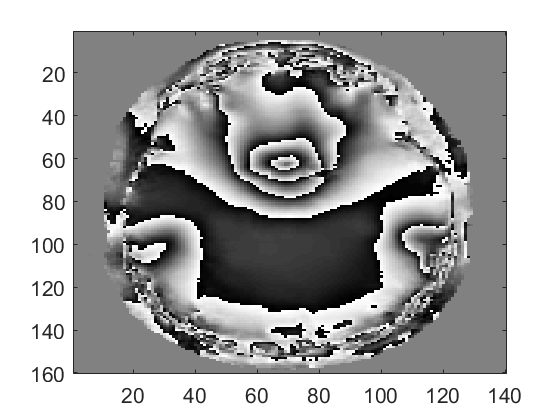}
		&\includegraphics[scale=0.23]{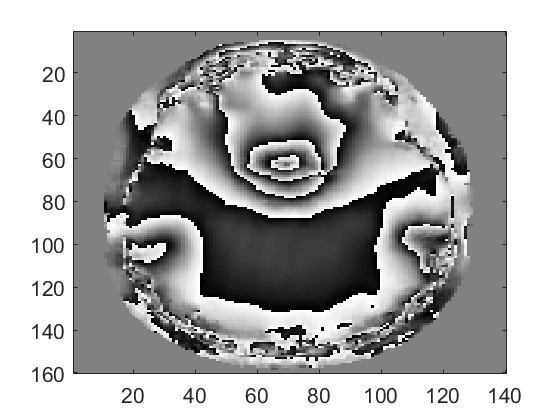}&\includegraphics[scale=0.23]{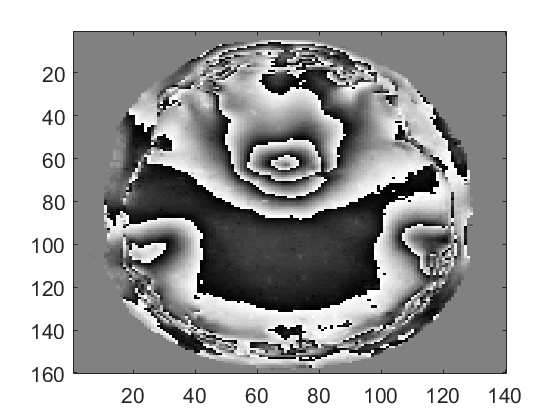}\\
		\scriptsize	Original phase&  \scriptsize	Noisy phase    & \scriptsize SpInPhase estimate   & \scriptsize MoGInPhase estimate & \scriptsize SURE-fuse estimate \\
		\scriptsize (Top view)    &  \scriptsize	($\sigma=0.5$) & \scriptsize $PSNR=27.91$         & \scriptsize $PSNR=27.60$        & \scriptsize $\bf{PSNR=28.18}$ \\
		\includegraphics[scale=0.23]{mri_front.png}&\includegraphics[scale=0.23]{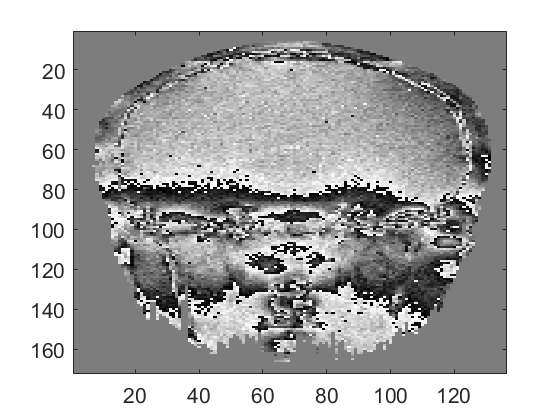}&\includegraphics[scale=0.23]{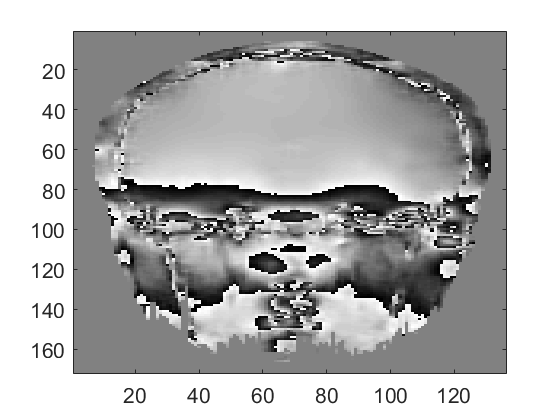}
		&\includegraphics[scale=0.23]{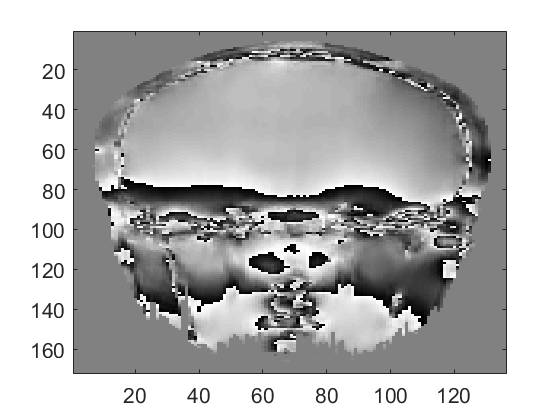}&\includegraphics[scale=0.23]{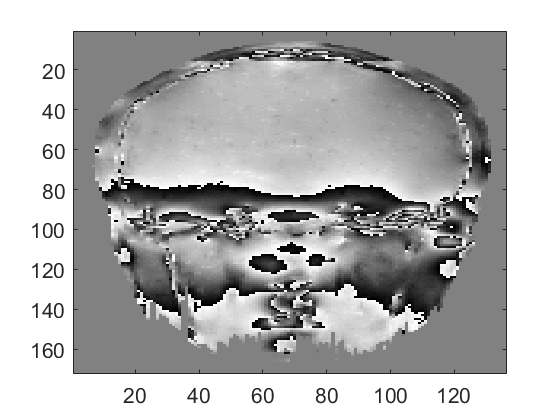}\\
		\scriptsize	Original phase&  \scriptsize	Noisy phase    & \scriptsize SpInPhase estimate   & \scriptsize MoGInPhase estimate & \scriptsize SURE-fuse estimate \\
		\scriptsize (Front view) &  \scriptsize	($\sigma=0.5$) & \scriptsize $PSNR=26.83$         & \scriptsize $PSNR=26.95$        & \scriptsize $\bf{PSNR=28.05}$ \\
\end{tabular}
	 \caption{ \footnotesize Clean, noisy and estimated MRI phase images. Results for SURE-fuse WFF versus state-of-the-art algorithms are shown. Top, middle and bottom rows contain side, top and front views respectively.}
	\label{mri_c}
\end{figure*}

Experiments using real MRI interferograms are presented in this section. We remark that the MRI InPhase data shows same statistical properties as any other interferograms and is well modelled by \eqref{obsmodel} (please refer \cite{1998_Ghiglia_Two} for more details). \Cref{mri_a} shows the InPhase images collected from real MRI scan \footnote[2]{The work was carried out on a 1.5 T GE Signa clinical scanner operating within Western General Hospital (WGH), University of Edinburgh.} from a human head along side, top and front orientations. These three phase images are used as clean InPhase images and iid white Gaussian noise with variance corresponds to $\sigma \in \lbrace0.3, 0.5, 0.7, 0.9\rbrace$ is added, as per \eqref{obsmodel} to get the noisy data.  
Phase denoising is done for all these noise levels using SURE-fuse WFF with $scale \in \lbrace1, 2, ...,8 \rbrace$. In \cref{mri_b}, SURE-fusion is comared with WFF. For better readability of the graphs, only the WFF plots corresponding to $scale=1, 4 \ \& \ 8$, representing low, medium and large sized windows, are shown.  \Cref{mri_b1,mri_b2,mri_b3} show that SURE-fuse WFF always outperforms the best scale WFF. The comparison with the state-of-the-art competitors, using the same experiments is also done and the resulting estimates  along with PSNR values are shown in \cref{mri_c}. In this experiment, a moderate noise variance ($\sigma=0.5$) is considered. In addition to this, the same experiment is repeated for other noise levels and the results are tabulated in \cref{mri_table}. From \Cref{mri_c} and \cref{mri_table} it can be concluded that the developed algorithm is very competitive with the state-of-the-art, providing experimental evidence to its effectiveness and competitiveness to perform InPhase denoising in real data. 

\begin{table}[h!]
	\begin{center}
		\begin{tabular}{r|c|ccc}
			\multicolumn{1}{r|}{\multirow{2}[1]{*}{Surf.}} & \multirow{2}[1]{*}{$\sigma$} & \multicolumn{3}{c}{PSNR (dB)} \\
			\multicolumn{1}{r|}{} &       & \multicolumn{1}{c}{SpInPhase} & \multicolumn{1}{c}{MoG} & \multicolumn{1}{c}{SURE-fuse} \\
			
			\midrule
			\midrule
			\multirow{2}[1]{*}{Side }         & 0.3   &32.49	    &32.27  &\B{32.60}  \\
										      & 0.5   &28.34	    &26.76	&\B{28.59}  \\
			\multirow{2}[1]{*}{view}          & 0.7   &25.60	    &23.21	&\B{25.69}  \\
											  & 0.9   &\B{23.87}	&21.19	&23.62      \\
			\midrule
			\midrule
			\multirow{2}[1]{*}{Top}           & 0.3   &31.99	    &31.94	&\B{32.15}   \\
			                                  & 0.5   &27.91	    &27.60	&\B{28.18}   \\
			\multirow{2}[1]{*}{view}          & 0.7   &25.05	    &23.45	&\B{25.14}   \\
			                                  & 0.9   &23.05  	    &20.81	&\B{23.10}   \\
			\midrule
			\midrule
			\multirow{2}[1]{*}{Front}         & 0.3   &31.29  	    &31.60	&\B{32.18}  \\
			                                  & 0.5   &26.89	    &26.95	&\B{28.05}  \\
			\multirow{2}[1]{*}{view}          & 0.7   &24.27	    &22.86	&\B{25.25}  \\
			                                  & 0.9   &22.62	    &20.22	&\B{23.13}  \\
			\midrule
			\midrule
		\end{tabular}%
	\end{center}
	\caption{Performance indicators for surfaces shown in \cref{mri_a}. \\ Best values are shown in bold.}
	\label{mri_table}
\end{table}

\subsection{Interferometric SAR data (InSAR)}
\label{sec:insarexp1}
\begin{figure}[h!]
			\captionsetup{justification=centering}
	\centering
	\begin{subfigure}{0.15\textwidth}
		\includegraphics[width=\textwidth]{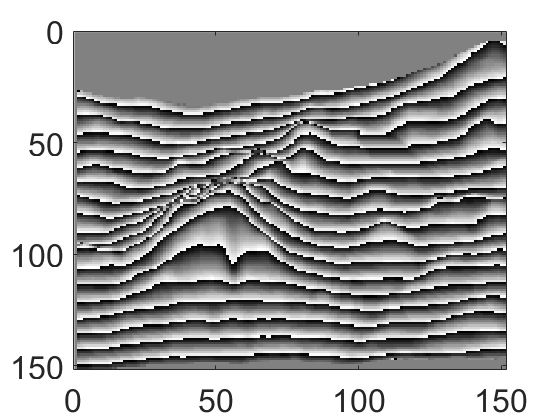}
		\caption{LP-1}
		\label{long_a1}
	\end{subfigure}
	\hfill
	\begin{subfigure}{0.15\textwidth}
		\includegraphics[width=\textwidth]{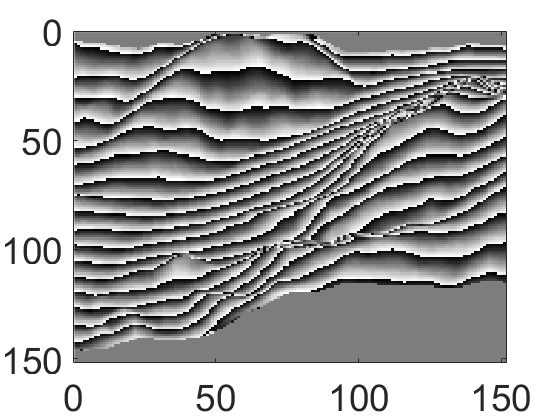}
		\caption{LP-2}
		\label{long_a2}
	\end{subfigure}
	\hfill
	\begin{subfigure}{0.15\textwidth}
		\includegraphics[width=\textwidth]{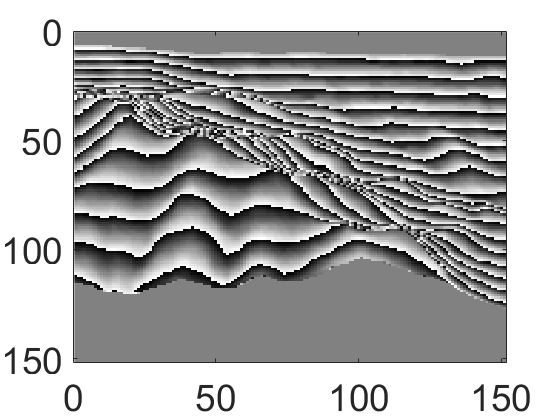}
		\centering \caption{LP-3}
		\label{long_a3}
	\end{subfigure}
    \hfill
    	\begin{subfigure}{0.15\textwidth}
    	\includegraphics[width=\textwidth]{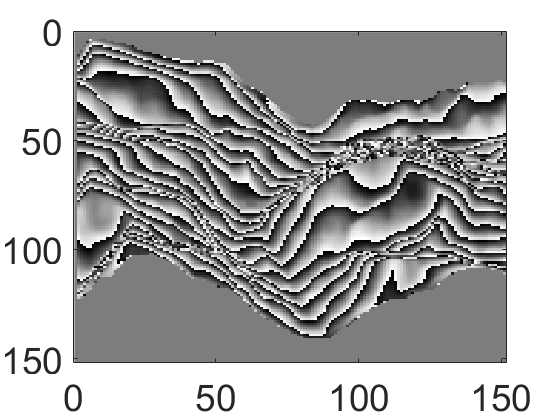}
    	\caption{ISP-1}
    	\label{isola_a1}
    \end{subfigure}
    \hfill
    \begin{subfigure}{0.15\textwidth}
    	\includegraphics[width=\textwidth]{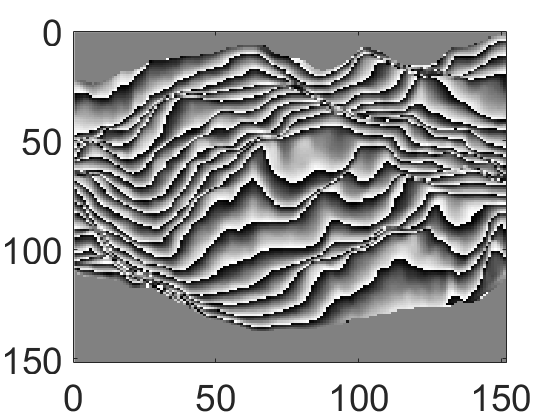}
    	\caption{ISP-2}
    	\label{isola_a2}
    \end{subfigure}
    \hfill
    \begin{subfigure}{0.15\textwidth}
    	\includegraphics[width=\textwidth]{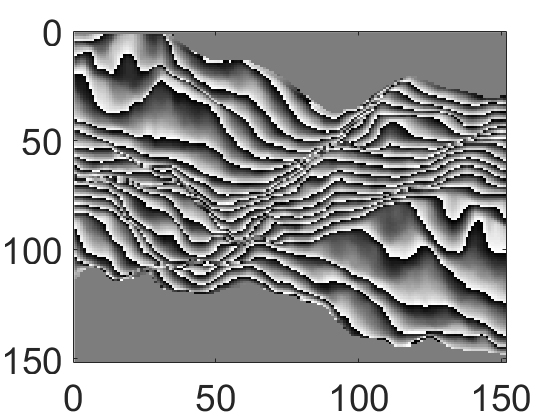}
    	\centering \caption{ISP-3}
    	\label{isola_a3}
    \end{subfigure}
    \hfill
	\begin{subfigure}{0.15\textwidth}
	\includegraphics[width=\textwidth]{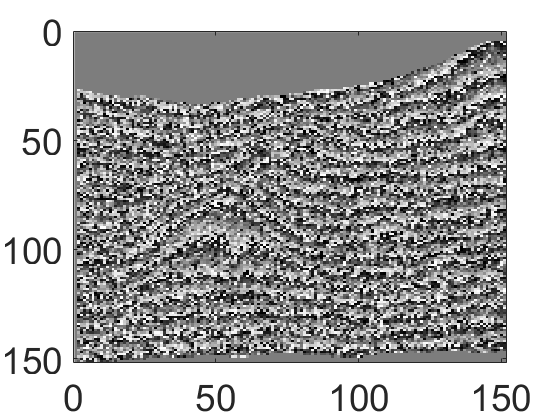}
	\caption{LP-1 (noisy)}
	\label{long_b1}
     \end{subfigure}
    \hfill
    \begin{subfigure}{0.15\textwidth}
	\includegraphics[width=\textwidth]{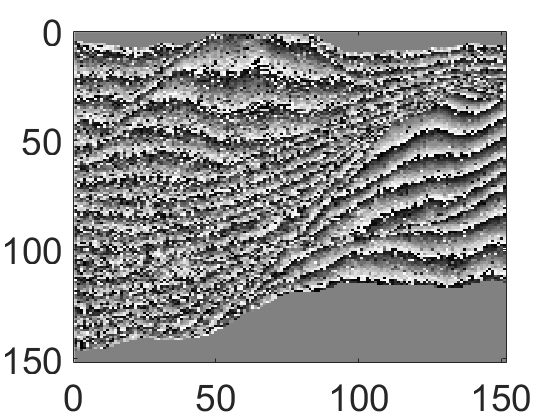}
	\caption{LP-2 (noisy)}
	\label{long_b2}
   \end{subfigure}
    \hfill
    \begin{subfigure}{0.15\textwidth}
	\includegraphics[width=\textwidth]{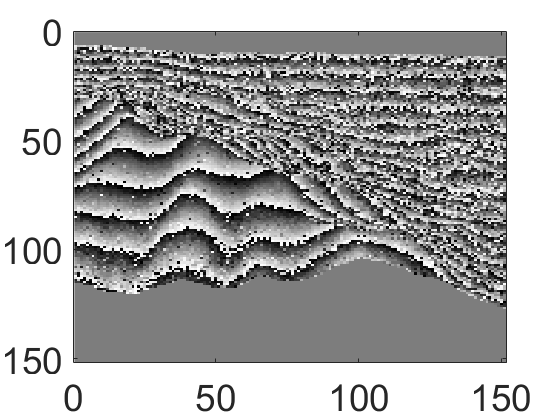}
	\caption{LP-3 (noisy)}
	\label{long_b3}
    \end{subfigure}
\hfill
	\begin{subfigure}{0.15\textwidth}
	\includegraphics[width=\textwidth]{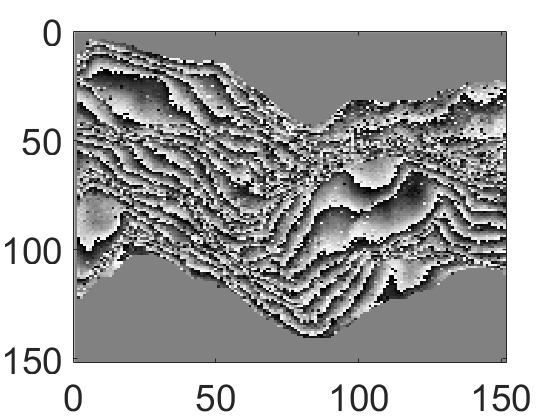}
	\caption{ISP-1 (noisy)}
	\label{isola_b1}
\end{subfigure}
\hfill
\begin{subfigure}{0.15\textwidth}
	\includegraphics[width=\textwidth]{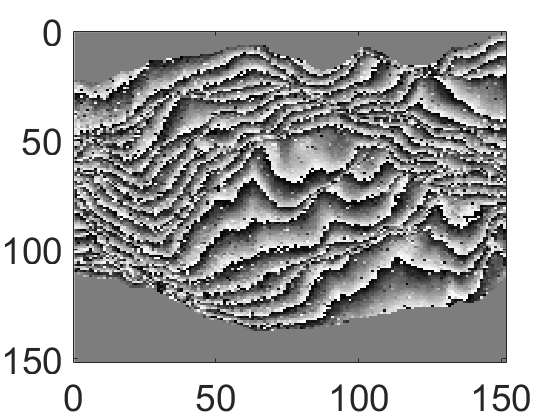}
	\caption{ISP-2 (noisy)}
	\label{isola_b2}
\end{subfigure}
\hfill
\begin{subfigure}{0.15\textwidth}
	\includegraphics[width=\textwidth]{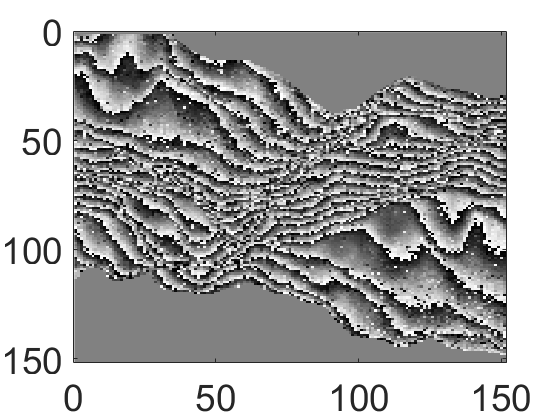}
	\caption{ISP-3 (noisy)}
	\label{isola_b3}
\end{subfigure}
	\caption{Real digital elevation model data from Longs Peak (LP) and Isolation Peak (ISP). \\ Top row : Clean interferogram, Bottom row: Corresponding noisy data with InSAR noise.}
	\label{long_ab}
\end{figure}

In this section, we present strong experimental evidences to demonstrate the robustness of SURE-fuse WFF in denoising real InSAR data. In SAR interferometry, the terrain topography is reconstructed from the InPhase data which are extracted from the complex-valued noisy SAR images collected at slightly displaced antennas. Let $\bf x_1$ and $\bf x_2$ be the acquired SAR images. Our focus is to denoise the noisy interferogram given by $\boldsymbol{\upPhi}_{2\pi}=\arg \nbr{\bf x_1 \odot \bf x_2^*}$, where $\bf x_2^*$ denotes the complex conjugate of $\bf x_2$ and $\odot$ represents the Hadamard product \citep{1998_Bamler_Synthetic}. The InSAR data used in the following experiments are distributed with the book \citep{1998_Ghiglia_Two}. These data sets were generated based on real digital elevation models of mountainous terrains around Longs Peak and Isolation Peak, Colorado, using a high-fidelity InSAR simulator. We use six different SAR data, each having pixel size $152 \times 152$. The first three interferograms are collected from adjacent areas around Longs Pleak (LP), which we term as LP-1 (\cref{long_a1}), LP-2 (\cref{long_a2}), LP-3 (\cref{long_a3}). The latter three are from Isolation peak (ISP), and we term them as ISP-1 (\cref{isola_a1}), ISP-2 (\cref{isola_a2}), ISP-3 (\cref{isola_a3}). The corresponding interferograms, corrupted with InSAR noise, are shown in  \cref{long_b1,long_b2,long_b3,isola_b1,isola_b2,isola_b3}. The detailed description of the InSAR simulators are beyond the focus of this discussion and the readers are suggested to refer \cite[Chapter~3]{1998_Ghiglia_Two}.  Also, for the additive and uniform noise variance modelling of the InSAR noise, we adopt the methodologies described in \cite{2015_Hongxing_Interferometric}.

\begin{figure}[h!]
	\captionsetup{justification=centering}
	\centering
	\begin{subfigure}{0.32\textwidth}
		\includegraphics[width=\textwidth]{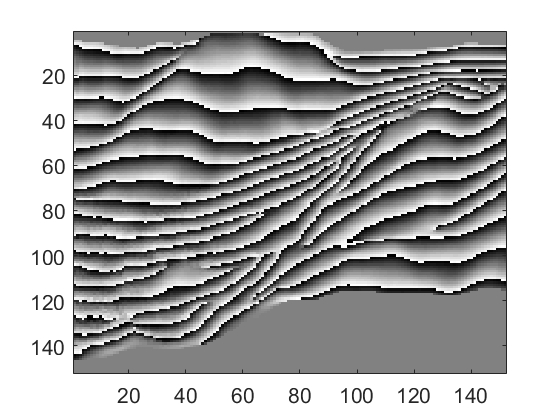}
		\caption{SpInPhase estimate\\ 	$PSNR=26.09$}
		\label{l_est_1}
	\end{subfigure}
	\hfill
	\begin{subfigure}{0.32\textwidth}
		\includegraphics[width=\textwidth]{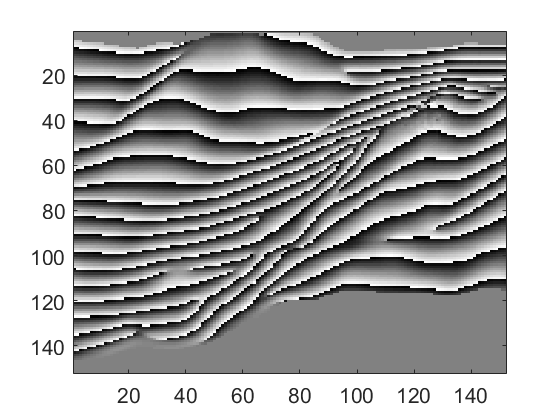}
		\caption{ MoGInPhase estimate \\$PSNR=24.39$}
		\label{l_est_2}
	\end{subfigure}
	\hfill
	\begin{subfigure}{0.32\textwidth}
		\includegraphics[width=\textwidth]{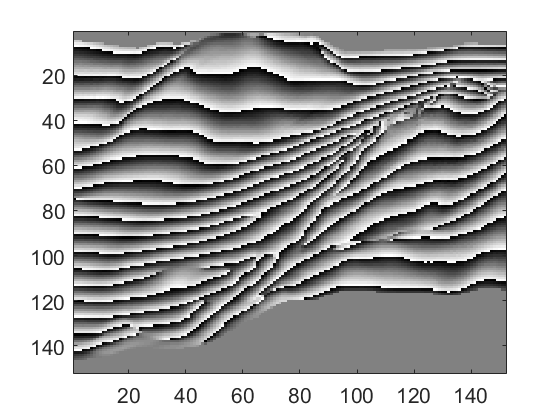}
		 \caption{SURE-fuse estimate\\ $\bf{PSNR=26.59}$}
		\label{l_est_3}
	\end{subfigure}
	\hfill
	\begin{subfigure}{0.32\textwidth}
		\includegraphics[width=\textwidth]{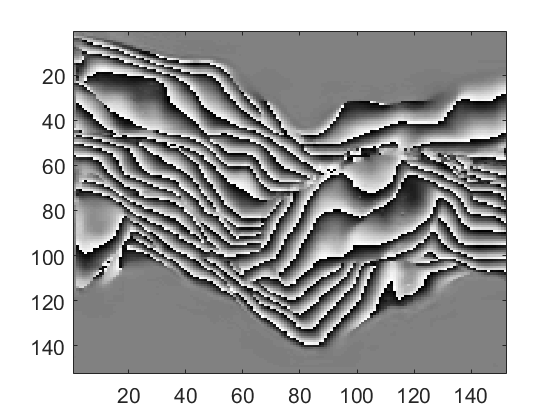}
		\caption{SpInPhase estimate\\ 	$PSNR=25.14$}
		\label{i_est_1}
	\end{subfigure}
	\hfill
	\begin{subfigure}{0.32\textwidth}
		\includegraphics[width=\textwidth]{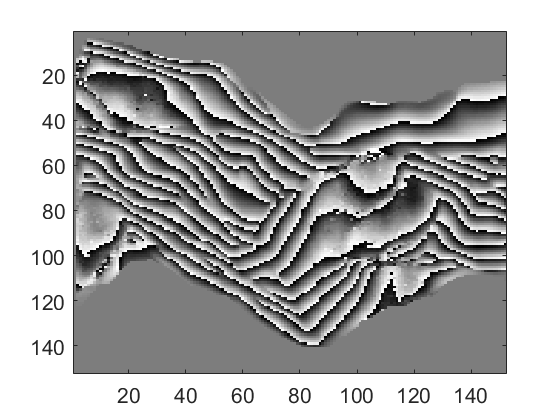}
		\caption{ MoGInPhase estimate \\ $PSNR=24.03$}
		\label{i_est_2}
	\end{subfigure}
	\hfill
	\begin{subfigure}{0.32\textwidth}
		\includegraphics[width=\textwidth]{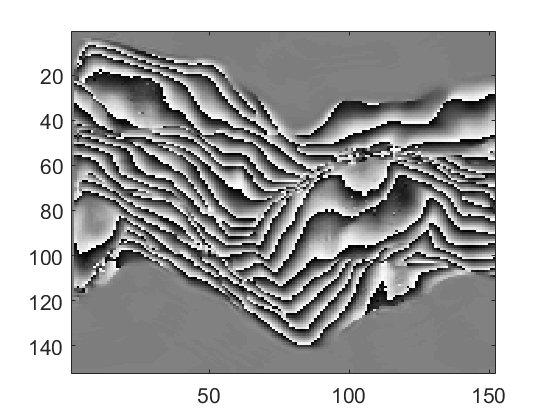}
		\caption{SURE-fuse estimate\\ $\bf{PSNR=26.75}$ }
		\label{i_est_3}
	\end{subfigure}
	\caption{InPhase estimates: SURE-fusing vs state-of-the-arts. Top row: Long Peak-2. Bottom row: Isolation Peak-1}
	\label{isola_ab}
\end{figure}

The estimation is carried out by neglecting the `bad' pixels as indicated by the quality maps supplied along with the data. In \cref{isola_ab}, the InPhase estimates for two selected interferograms [Long Peak-1 (\cref{long_b1}) and Isolation Peak-2 (\cref{isola_b2})] are given for visual comparison. Careful examination of \cref{isola_ab} shows ability of SURE-fusion to retain the minute details, which are over smoothened for its competitors. This is clearly supported by the PSNR values. The detailed results for the experiments conducted on these data sets are concluded in \cref{tableinsar}. The displayed values indicates that SURE-fuse WFF brings remarkable improvements in quality of the estimates. This results provide strong experimental evidence of the advantages of SURE-fusion and leads to the conclusion that SURE-fusion is very well suited to the real and challenging InSAR data.

\begin{figure}[h!]
	\centering
		\includegraphics[width=0.6\textwidth]{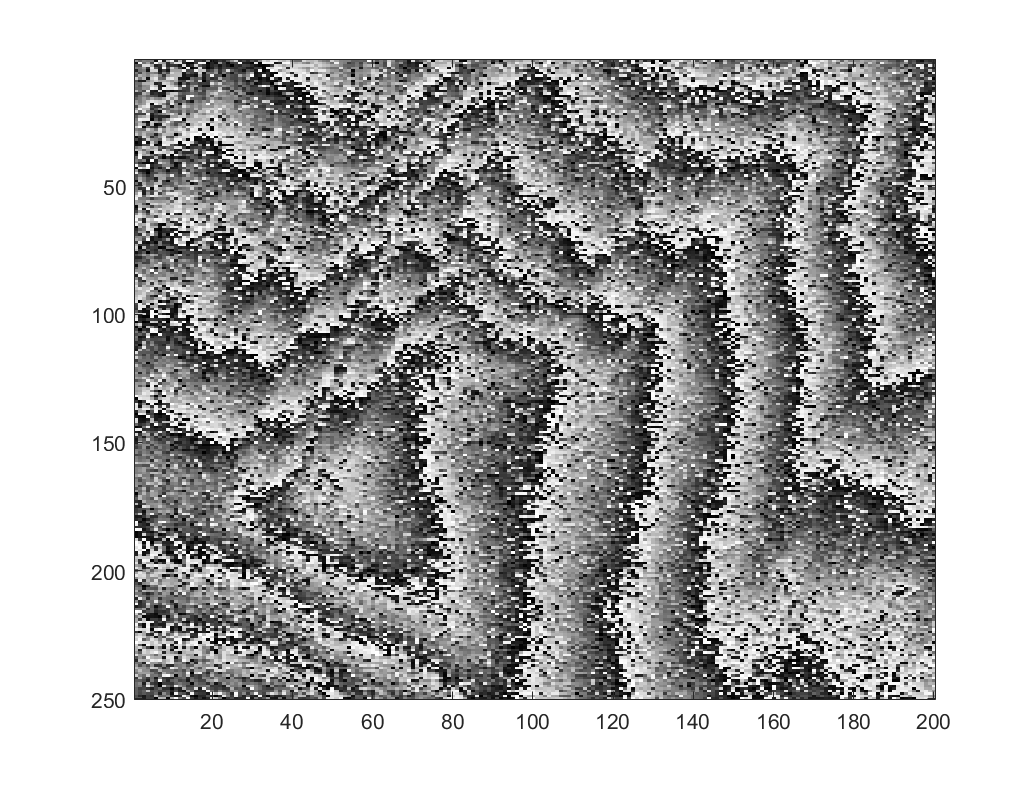}
	\caption{Real interferometric SLC data collected from Evaggelistria, Greece.}
	\label{slc}
\end{figure}

\begin{table}[h!]
	\begin{center}
		\begin{tabular}{c|ccc}
			\multirow{2}[1]{*}{Surface} & \multicolumn{3}{c}{PSNR (dB)} \\
			&\multicolumn{1}{c}{SpInPhase} & \multicolumn{1}{c}{MoG} & \multicolumn{1}{c}{SURE-fuse} \\
			
			\midrule
			\midrule
			Longs Peak-1   		&22.27	&24.20  &\B{25.28}  \\
			Longs Peak-2   		&26.09	&24.39  &\B{26.59}  \\
			Longs Peak-3   		&26.72	&23.74  &\B{27.10}  \\
			\midrule
			\midrule
			Isolation Peak-1   	&25.14	&24.03  &\B{26.75}  \\
			Isolation Peak-2   	&27.25	&25.45  &\B{28.60}  \\
			Isolation Peak-3   	&26.86	&25.14  &\B{28.46}  \\
			\midrule
			\midrule			
		\end{tabular}%
	\end{center}
	\caption{ PSNR comparison for InSAR data}
	\label{tableinsar}
\end{table}
Next we present experiments based on an InSAR Single Look Complex (SLC) data set, having pixel size $200\times250$, collected from an area near Evaggelistria, Greece (from 38$^\circ$18'30''N to 38$^\circ$19'32''N and from 23$^\circ$01'15''E to 23$^\circ$04'24''E), distributed by European Space Agency (ESA)\footnote[3]{http://eo-virtual-archive4.esa.int}. As it is evident from \cref{slc}, this is a highly challenging Interferometric data set which is corrupted with heavy noise.

	\begin{figure}[h!]
	\hskip -0.5cm
	\centering
	\begin{subfigure}{0.33\textwidth}
		\includegraphics[width=1.1\textwidth]{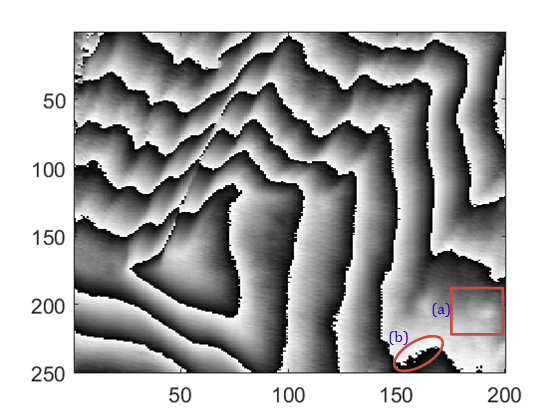}
		\caption{SpInPhase Estimate}
		\label{slc_1}
	\end{subfigure}
	\hfill
	\begin{subfigure}{0.32\textwidth}
		\includegraphics[width=1\textwidth]{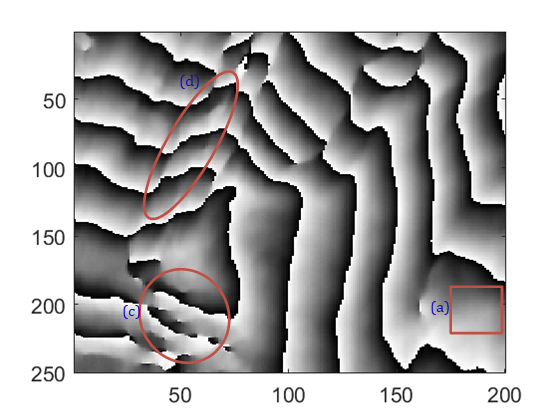}
		\caption{MoGInPhase Estimate}
		\label{slc_2}
	\end{subfigure}
	\hfill
	\begin{subfigure}{0.32\textwidth}
		\includegraphics[width=1\textwidth]{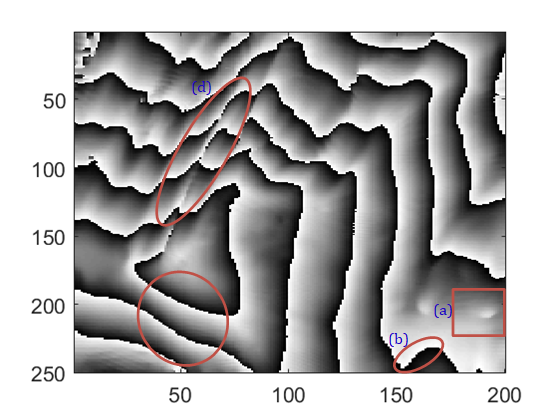}
		\centering \caption{ SURE-fuse estimate}
		\label{slc_3}
	\end{subfigure}
	\caption{Qualitative comparison of the denoised results for the InSAR SLC data }
	\label{slcmain}
\end{figure}
The estimates produced by SpInPhase, MoGInPhase and SURE-fuse are shown in \cref{slcmain}. For this data set, since the true image is not available for PSNR calculation, we present a qualitative analysis. After careful examination of the estimates, we arrive at the following remarks: (\textbf{i}) For the SpInPhase (\cref{slc_1}), the peak at region (a) is not clearly estimated. Also, region (b) is not very sharp and it contains ambiguities. (\textbf{ii}) For the MoGInPhase (\cref{slc_2}), peak at region (a) is almost smoothen out and almost invisible. The discontinuities at region (d) is washed out here. Also, region (c) is badly estimated with lot of missing areas. (\textbf{iii}) For the SURE-fuse estimate, as evident from \cref{slc_3}, all these region from (a) to (d) are having remarkable improvements in quality compared to other two estimates. Also, in general, the interferometric fringes are more sharp and clear for SURE-fuse estimate.

\subsection{Absolute phase imaging with phase unwrapping}
The absolute phase estimation from noisy interferograms, is usually accomplished in two stages- Phase Denosing and Phase Unwrapping.  This paper focuses on phase denoising part. But we remark that phase denoising is a crucial step in absolute phase estimation as the quality of the denoised InPhase images plays a major role in the success of the proceeding stage, i.e., phase unrapping. The preservation of sharp fringes and discontinuities of the interferogram is an important attribute of any phase denoising algorithm.  In this section, the SURE-fuse WFF estimates are unwrapped using the state-of-the-art phase unwrapping algorithm, termed as PUMA\cite{2007_Bioucas_Phase}. The section contains both qualitative and quantitative evaluation of SURE-fuse WFF in comparison with its competitors. The quality of the unwrapped denoised phase ($\widehat{\bs{\upPhi}}$) is measured based on the number of estimated pixels having error less than $\pi$ compared to the true phase image ($\bs{\upPhi}$). We define \textit{number of error larger than $\pi$} ($NELP$ \cite{2015_Hongxing_Interferometric}) as:
\begin{align}
\text{NELP} &:=N-|J|,
\end{align}
where  $J:=\cbr{j:|\widehat{\bs \upPhi}_j - \bs \upPhi_j|\leq \pi, j=1,\cdots,N}$. High quality InPhase estimates should have low $NELP$ values and vice versa. Also, based on the set $J$ we define a new \textit{peak signal-to-noise ratio}, denoted as $\text{PSNR}_a$ as follows:
 \begin{align}
 PSNR_a &:=10\log_{10}\frac{4N\pi^2}{ \norm{(\widehat{{\bs \upPhi}}_{J}-{\bs \upPhi}_{J})}_F^2} \ \  [\text{dB}],
 \end{align}
where the notation ${\bs \upPhi}_{J}$ stands for the restriction of ${\bs \upPhi}$ to the set $J$ \cite{2015_Hongxing_Interferometric}. 

\subsubsection{Phase unwrapping of simulated data set}
\begin{figure}[h!]
			\captionsetup{justification=centering }
	\begin{subfigure}{0.24\textwidth}
		\includegraphics[width=\textwidth]{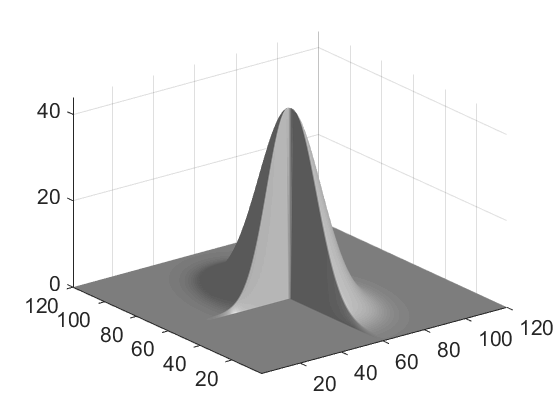}
		\caption{Truncated Gaussian \\ \scriptsize(Original Phase) }
		\label{pumaTG1}
	\end{subfigure}
	\hfill
		\begin{subfigure}{0.24\textwidth}
		\includegraphics[width=\textwidth]{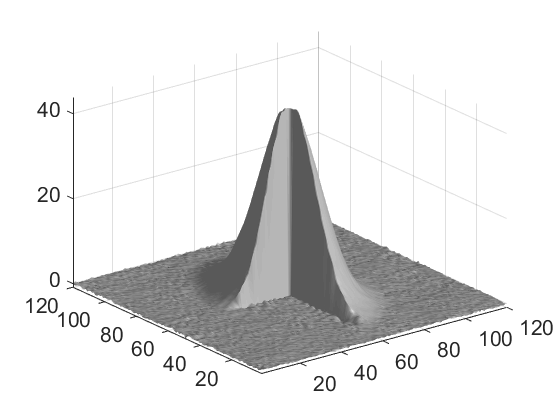}
		\caption{ SpInPhase \\ \scriptsize $NELP=18,\ PSNR_a=33.16 \ dB$}
		\label{pumaTG2}
	\end{subfigure}
	\hfill
			\begin{subfigure}{0.24\textwidth}
		\includegraphics[width=\textwidth]{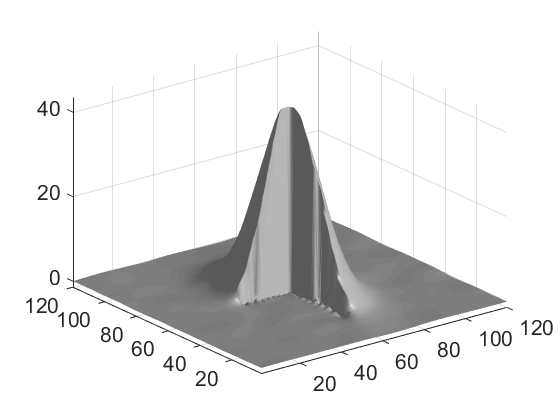}
		\caption{ MoGInPhase \\ \scriptsize $NELP=89,\ PSNR_a=32.44 \ dB$ }
		\label{pumaTG3}
	\end{subfigure}
	\hfill
			\begin{subfigure}{0.24\textwidth}
		\includegraphics[width=\textwidth]{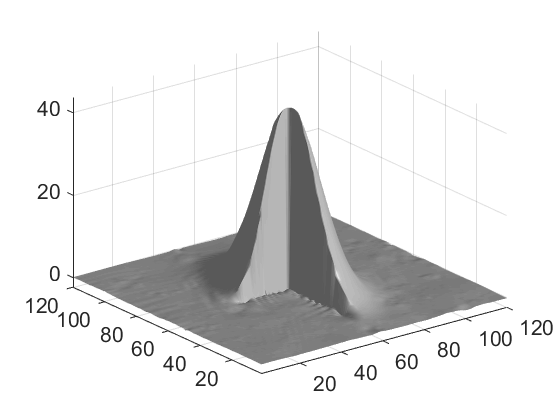}
		\caption{SURE-fuse \\ \scriptsize $NELP=\B{18},\ PSNR_a=\B{35.25 \ dB}$ }
		\label{pumaTG4}
	\end{subfigure}
	\hfill
		\begin{subfigure}{0.24\textwidth}
		\includegraphics[width=\textwidth]{PVS3d.png}
		\caption{Peak-valley Surface\\ \scriptsize (Original Phase)}
		\label{pumaPVS1}
	\end{subfigure}
	\hfill
	\begin{subfigure}{0.24\textwidth}
		\includegraphics[width=\textwidth]{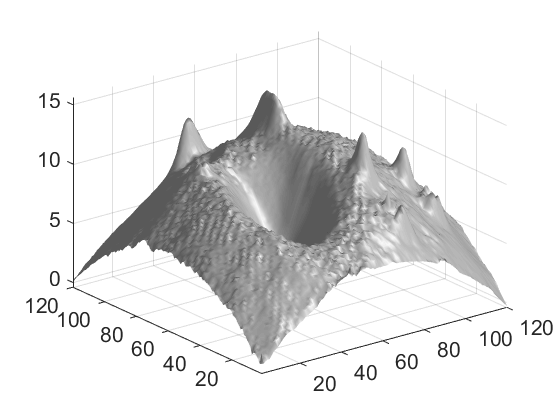}
		\caption{SpInPhase \\ \scriptsize $NELP=56,\ PSNR_a=30.06 \ dB$}
		\label{pumaPVS2}
	\end{subfigure}
	\hfill
	\begin{subfigure}{0.24\textwidth}
		\includegraphics[width=\textwidth]{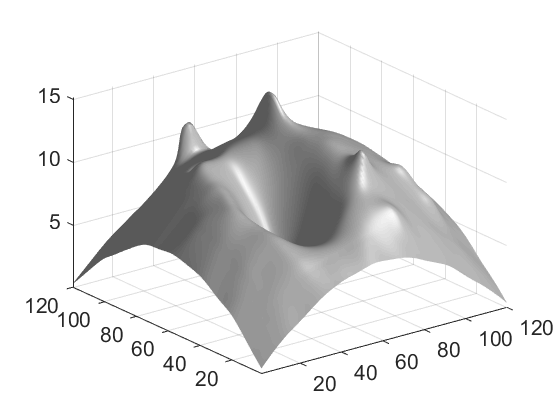}
		\caption{MoGInPhase\\ \scriptsize $NELP=92, \ PSNR_a=27.06 \ dB$}
		\label{pumaPVS3}
	\end{subfigure}
	\hfill
	\begin{subfigure}{0.24\textwidth}
		\includegraphics[width=\textwidth]{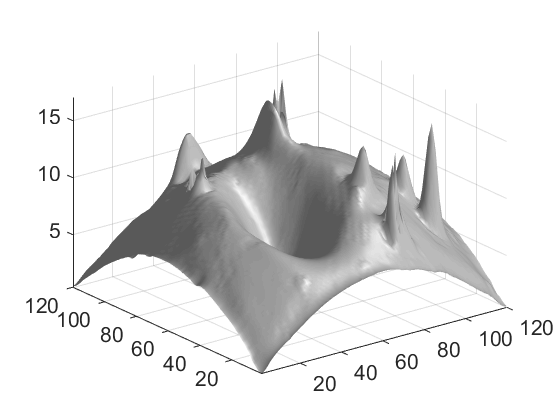}
		\caption{ SURE-fuse \\ \scriptsize $NELP=\B{19}, \ PSNR_a=\B{32.17 \ dB}$}
		\label{pumaPVS4}
	\end{subfigure}

	\caption{Phase Unwrapping Results: Top row: Truncated Gaussian, Bottom row: Peak-valley surface.}
\label{puma_sim}	
\end{figure}

Phase unwrapping is demonstrated using two different surfaces- Truncated Gaussian (\cref{{pumaTG1}}) and peak-valley (\cref{{pumaPVS1}}) surfaces. Heavily noisy interferograms are generated ($\sigma=0.9,\bf a=1$ in model \eqref{obsmodel}) using these surfaces as the absolute phases ($\bs \upPhi$). The underlying absolute phase estimation is quite challenging as the interferograms are highly noisy and they contain sharp irregularities. These noisy wrapped phases are denoised using SURE-fuse WFF and its competitors and then unwrapped using PUMA \cite{2007_Bioucas_Phase}. \Cref{pumaTG2,pumaTG3,pumaTG4} show the results for truncated Gaussian. Form a careful examination, the competitive nature of the SURE-fuse WFF in retaining the sharp discontinuities as well as the smooth and flat regions can be observed. This is supported by the $NELP$ and $PSNR_a$ values.

The second surface in \cref{pumaPVS1} is more challenging as it contains narrow peaks and pits. The SpInPhase estimate (\cref{pumaPVS2}) has very poor quality on the smooth areas. The MoGInPhase estimate (\cref{pumaPVS3}) has better smooth areas but many narrow peaks are missing as a result of the smoothening. The SURE-fuse estimate (\cref{pumaPVS4}) is much better, compared to its competitors, in retaining the narrow peaks as well as the smooth regions. These qualitative observations are supported by $NELP$ and $PSNR_a$ values with more than 2 dB improvement.  We remark that, even in SURE-fuse estimate, some of the narrow peaks are missing or badly estimated; but this is a very challenging data with a heavily noisy interferogram.
\subsubsection{Phase unwrapping of InSAR images}
\begin{figure}[h!]
	\centering
	\begin{subfigure}{0.33\textwidth}
		\includegraphics[width=\textwidth]{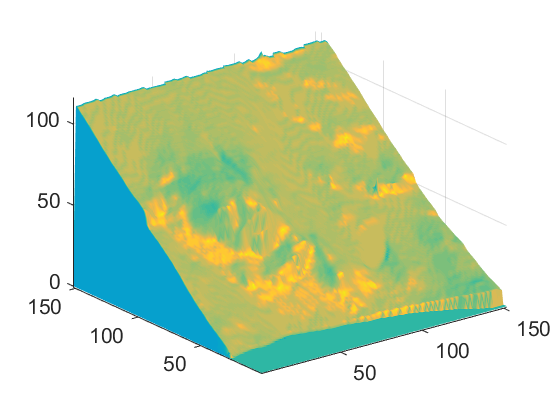}
		\caption{Longs Peak-1}
		\label{LP1_3d}
	\end{subfigure}
	\hfill
	\begin{subfigure}{0.32\textwidth}
		\includegraphics[width=\textwidth]{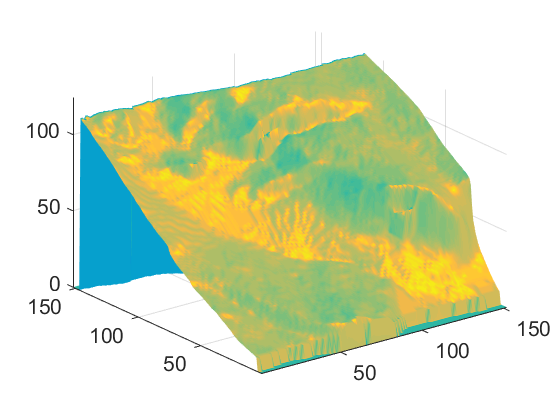}
		\caption{Longs Peak-2}
		\label{LP2_3d}
	\end{subfigure}
	\hfill
	\begin{subfigure}{0.32\textwidth}
		\includegraphics[width=\textwidth]{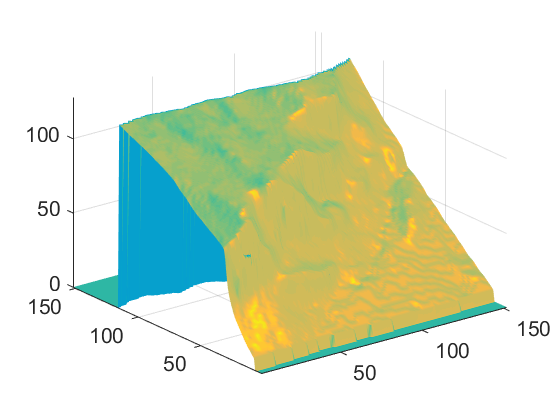}
		\caption{Longs Peak-3}
		\label{LP3_3d}
	\end{subfigure}
	\hfill
	\begin{subfigure}{0.32\textwidth}
		\includegraphics[width=\textwidth]{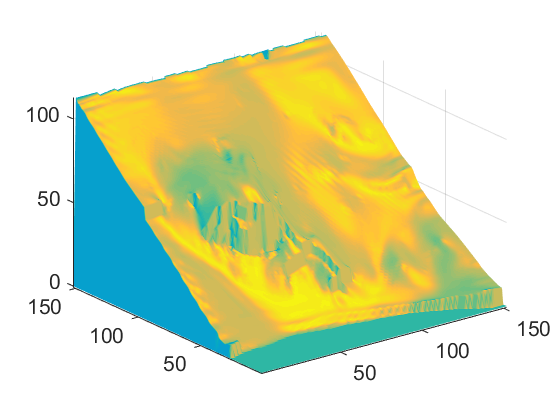}
		\caption{$NELP=13, \ PSNR_a=24.93 \ dB$}
		\label{LP1_sure}
	\end{subfigure}
	\hfill
	\begin{subfigure}{0.32\textwidth}
		\includegraphics[width=\textwidth]{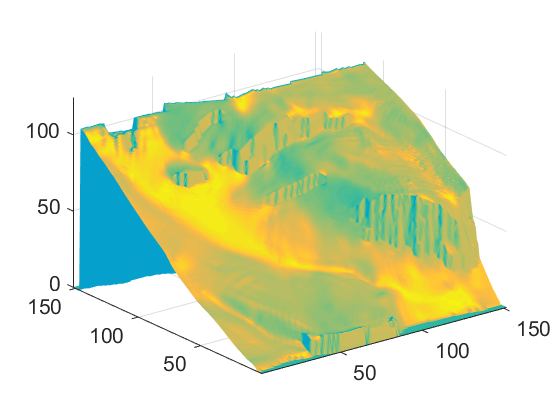}
		\caption{$NELP=44, \ PSNR_a=26.48 \ dB$}
		\label{LP2_sure}
	\end{subfigure}
	\hfill
	\begin{subfigure}{0.32\textwidth}
		\includegraphics[width=\textwidth]{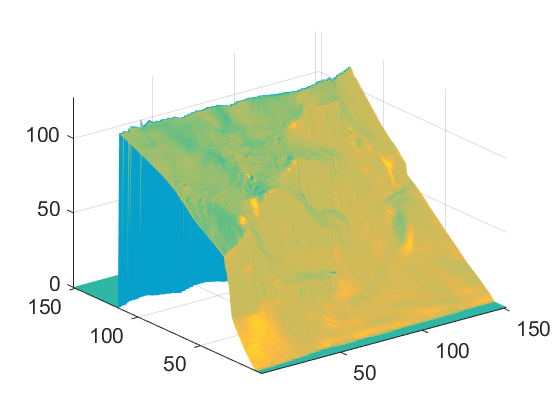}
		\caption{$NELP=16, \ PSNR_a=27.94 \ dB$}
		\label{LP3_sure}
	\end{subfigure}
	\caption{Top row: DEM of the clean InSAR data collected from Longs Peak. Bottom row: The absolute phase estimates obtained from the noisy interferograms of the top row (please see \cref{long_b1,long_b2,long_b3} for the noisy interferograms) using SURE-fuse WFF and PUMA unwrapping \cite{2007_Bioucas_Phase}.}
	\label{LP_sure}
\end{figure}

The absolute phase estimation is performed using real InSAR data described in \cref{sec:insarexp1} (\cref{long_a1,long_a2,long_a3,isola_a1,isola_a2,isola_a3}). The original digital elevation model (DEM) of the clean InSAR data collected from Longs Peak is shown in \cref{LP1_3d,LP2_3d,LP3_3d}. The noisy interferograms (refer \cref{long_b1,long_b2,long_b3}) are denoised using SURE-fuse algorithm and unwrapped. The absolute phase estimates, shown in \cref{LP1_sure,LP2_sure,LP3_sure}, are of very good quality and preserve most of the informations in the original DEM. The corresponding $NELP$ values of 13, 44, and 16 support these observations.

\begin{figure}[h!]
	\begin{subfigure}{0.33\textwidth}
		\includegraphics[width=\textwidth]{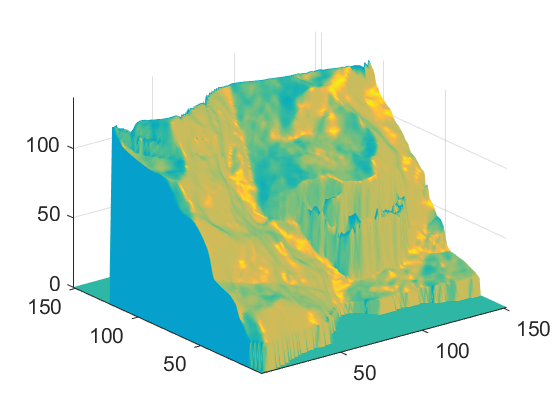}
		\caption{Isolation Peak-1}
		\label{IS1_3d}
	\end{subfigure}
	\hfill
	\begin{subfigure}{0.32\textwidth}
		\includegraphics[width=\textwidth]{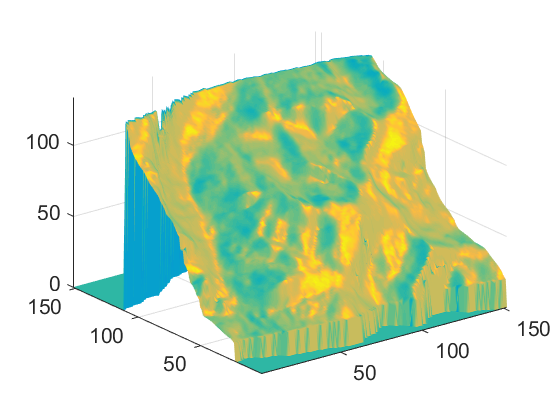}
		\caption{Isolation Peak-2}
		\label{IS2_3d}
	\end{subfigure}
	\hfill
	\begin{subfigure}{0.32\textwidth}
		\includegraphics[width=\textwidth]{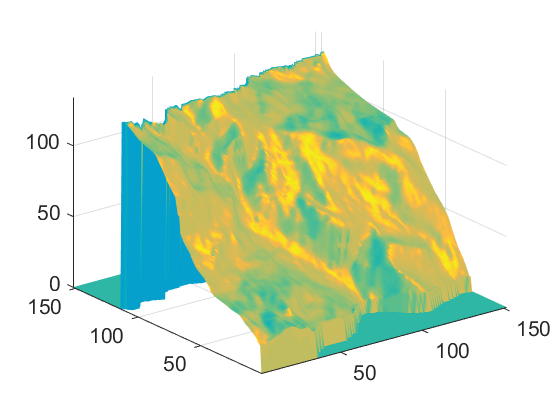}
		 \caption{Isolation Peak-3}
		\label{IS3_3d}
	\end{subfigure}
    \hfill
\begin{subfigure}{0.32\textwidth}
	\includegraphics[width=\textwidth]{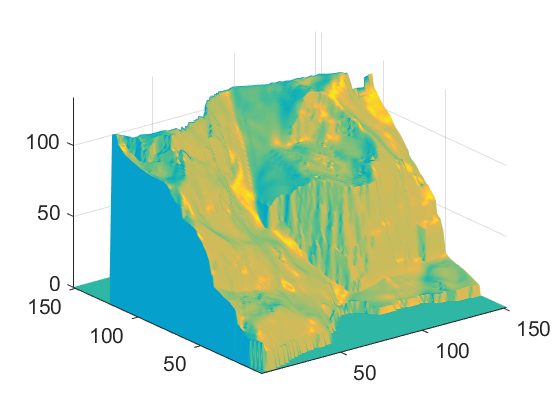}
	\caption{$\scriptsize NELP=349,PSNR_a=27.30 dB$}
	\label{IS1_sure}
\end{subfigure}
    \hfill
\begin{subfigure}{0.32\textwidth}
	\includegraphics[width=\textwidth]{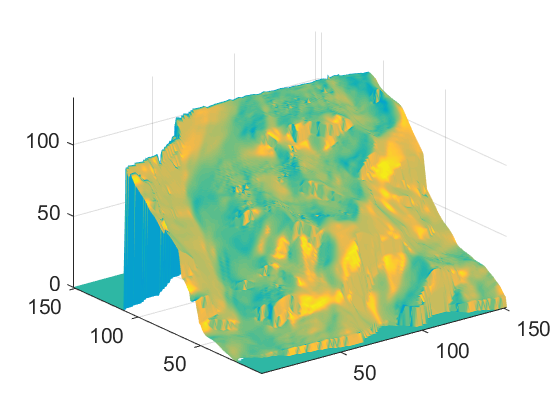}
	\caption{$\scriptsize NELP=240,PSNR_a=28.42 \ dB$}
	\label{IS2_sure}
\end{subfigure}
    \hfill
\begin{subfigure}{0.32\textwidth}
	\includegraphics[width=\textwidth]{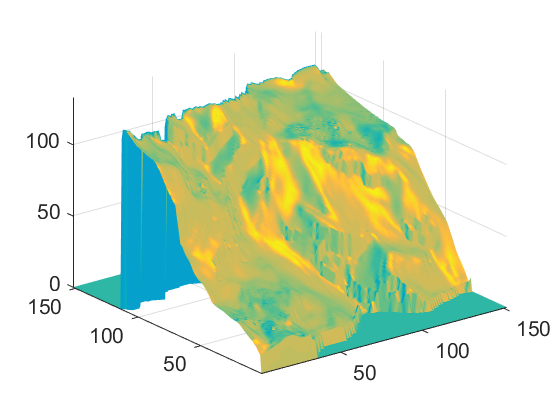}
	\caption{$\scriptsize NELP=35, \ PSNR_a=28.31 \ dB$}
	\label{IS3_sure}
\end{subfigure}
	\caption{ Top row: DEM of the clean InSAR data collected from Isolation Peak. Bottom row: The absolute phase estimates obtained from the noisy interferograms of the top row (please see \cref{isola_b1,isola_b2,isola_b3} for the noisy interferograms) using SURE-fuse WFF and PUMA unwrapping \cite{2007_Bioucas_Phase}.}
	\label{IS_3d}
\end{figure}

The same experiment is repeated using the InSAR data from Isolation peaks. The  DEMs of the clean data (\cref{IS1_3d,IS2_3d,IS3_3d}) indicate that the topology is bit more challenging, compared to Longs Peak, due to the presence of non-smooth and non-structured discontinuities. Despite these difficulties, the visual quality of the SURE-fuse estimates  \cref{IS1_sure,IS2_sure,IS3_sure} and the corresponding $NELP$ values of 349, 240 and 39 are in agreement with the previous conclusions.
  
Finally, we repeat the previous experiments with real InSAR data using other phase denoising algorithms. The results tabulated in \cref{tableinsar2} show that SURE-fuse WFF has around 1 dB (or more) improvement in $PSNR_a$ for most of the test cases. Also the $NELP$ values of SURE-fuse are much better compared to that of its competitors. This is a very strong empirical proof that makes the SURE-fuse WFF a very good choice in real InSAR phase imaging applications.
\begin{table}[h!]
	\begin{center}
		\begin{tabular}{c|ccc|ccc}
			\multirow{2}[1]{*}{Surface} & \multicolumn{3}{c|}{NELP}  & \multicolumn{3}{c}{$PSNR_a$ (dB)} \\
			&\multicolumn{1}{c}{SpInPhase} & \multicolumn{1}{c}{MoG} & \multicolumn{1}{c|}{SURE-fuse}&\multicolumn{1}{c}{SpInPhase} & \multicolumn{1}{c}{MoG} & \multicolumn{1}{c}{SURE-fuse}  \\
			
			\midrule
			\midrule
			Longs Peak-1   		&48	&67	    &\B{13}    &22.31	&24.50	&\B{24.92}\\
			Longs Peak-2   		&90	&141	&\B{44}    &26.31	&24.61	&\B{26.48}\\
			Longs Peak-3   		&89	&1218	&\B{16}    &26.85	&23.88	&\B{27.93}\\
			\midrule
			\midrule
			Isolation Peak-1   	&1590	&2852	&\B{349}   &25.85	&24.55	&\B{27.30}\\
			Isolation Peak-2   	&248	&1911	&\B{240}   &27.49	&25.61	&\B{28.42}\\
			Isolation Peak-3   	&790	&2212	&\B{35}    &27.01	&25.2	&\B{28.31}\\
			\midrule
			\midrule			
		\end{tabular}%
	\end{center}
	\caption{Performance Comparison for Phase Unwrapping with real InSAR data}
	\label{tableinsar2}
\end{table}
\section{Conclusion}
\label{sec:concl}
This paper introduced SURE-fuse WFF, an algorithm based on multi-resolution windowed Fourier analysis, for interferometric phase image denoising. InPhase images, being natural images, show high level of sparsity in the frequency domain and WFT is a powerful tool to be exploited to denoise such images. Through a brief discussion on the pros and cons of WFT-based representations, we arrived at the conclusion that the fixed resolution (window size ) sets a major bottle neck in the performance of WFT. This issue has been addressed by proposing a new multi-resolution WFT, termed as SURE-fuse WFF, aiming at fusing the WFF-estimates having different resolutions to achieve multi-resolution. A linear, pixel-wise and data-dependent fusion mechanism in the complex domain is introduced. An optimization frame work that minimizes the unbiased estimate of mean square error, known as SURE, is developed to obtain the fusion weights. 

Through a series of demonstrative experiments, we show the data dependent fusing ability of SURE-fuse WFF to achieve multi-resolution. A very detailed experiment section is provided to prove that SURE-fusion outperforms the best hand tuned WFF. Also strong qualitative and quantitative experimental evidences are provided using a set of synthetic, real InSAR and real MRI data to show that the SURE-fuse WFF outperforms the state-of-the-art InPhase denoising algorithms.

\newpage
\appendixtitles{} 
\appendixsections{multiple} 
\appendix
\section{: Derivation of divergent $\nabla.f( \bf z)$ }
\label{app:div}
As per \eqref{eqn:div0}, calculation of $\nabla.f( \bf z)$ involves the complex derivative $\frac{\partial  f_\k(\bf z)}{\partial \bf z_\k }$. Here the derivative w.r.t. complex-valued variable are not straight forward and we adopt Wirtinger calculus \cite{2011_Adali_Cvalued} which is explained in \eqref{eqn:wirt}. According to this framework, $\frac{\partial \bf z_\k }{\partial \bf z_{\bf k}}=1$ and $\frac{\partial \bf z_\k }{\partial \bf z_{\bf k}^*}=0$. Expanding \eqref{eqn:wftestmn2} using the LET function \eqref{eqn:LET} yields:
\begin{align}
f_\k(\bf z) &=\frac{1}{E_{\bf h^s}|W|}\sumk{\k''}\sumw  \bf Z _{\k'',\w}\bf h^s_{\k''-\k} \expp{\k} - \frac{1}{E_{\bf h^s}|W|}\sumk{\k''}\sumw  \bf Z _{\k'',\w}e^{\frac{-|| \bf Z _{\k'',\w}||^2}{\l^2}}\bf h^s_{\k''-\k} \expp{\k} \nonumber\\
&=\bf z_{\bf k}- \frac{1}{E_{\bf h^s}|W|}\sumk{\k''}\sumw  \bf Z _{\k'',\w}e^{\frac{-|| \bf Z _{\k'',\w}||^2}{\l^2}}\bf h^s_{\k''-\k} \expp{\k} \nonumber \\
 \frac{\partial f_\k(\bf z)}{\partial \bf z_{\bf k}}&=1 - \frac{1}{E_{\bf h^s}|W|}\sumk{\k''}\sumw  \frac{\partial \bf Z_{\k'',\w}}{\partial \bf z_{\bf k}}e^{\frac{-|| \bf Z _{\k'',\w}||^2}{\l^2}}\bf h^{s}_{\k''-\k} \expp{\k} -\frac{1}{E_{\bf h^s}|W|}\sumk{\k''}\sumw  \bf Z_{\k'',\w}\frac{\partial e^{\frac{-|| \bf Z _{\k'',\w}||^2}{\l^2}}}{\partial \bf z_{\bf k}}\bf h^{s}_{\k''-\k} \expp{\k} \label{pre_derv}
\end{align}
Now from \eqref{eqn:wft} we have,
\begin{align}
\bf Z_{\k'',\w} &= \sumk{\k'}\bf z_{\k'}\bf h^{s}_{\k''-\k'}\expn{\k'} \nonumber \\ 
\Rightarrow \frac{\partial \bf Z_{\k'',\w}}{\partial \bf z_{\bf k}}&= \bf h^{s}_{\k''-\k}\expn{\k} \label{d1}
\shortintertext{Also,} \frac{\partial e^{\frac{-|| \bf Z _{\k'',\w}||^2}{\l^2}}}{\partial \bf z_{\bf k}}&=-\frac{1}{\l^2}e^{\frac{-|| \bf Z _{\k'',\w}||^2}{\l^2}}\bf Z^* _{\k'',\w} \frac{\partial {\bf Z _{\k'',\w}}}{\partial \bf z_{\bf k}} \label{d2}
\shortintertext{\eqref{d1} in \eqref{d2} }
\frac{\partial e^{\frac{-|| \bf Z _{\k'',\w}||^2}{\l^2}}}{\partial \bf z_{\bf k}}&=-\frac{1}{\l^2}e^{\frac{-|| \bf Z _{\k'',\w}||^2}{\l^2}}\bf Z^* _{\k'',\w} \bf h^s_{\k''-\k}\expn{\k} \label{d4}
\end{align}
\begin{align}
\shortintertext{\eqref{d1} and \eqref{d4} in \eqref{pre_derv}}
\frac{\partial f_\k(\bf z)}{\partial \bf z_{\bf k}}&=1 -\frac{1}{E_{\bf h^s}|W|}\sumk{\k''}\sumw  e^{\frac{-|| \bf Z _{\k'',\w}||^2}{\l^2}}\bf h^s_{\k''-\k}\bf h^{s}_{\k''-\k} 
+\frac{1}{E_{\bf h^s}|W|\l^2}\sumk{\k''}\sumw  ||\bf Z _{\k'',\w}||^2 e^{\frac{-|| \bf Z _{\k'',\w}||^2}{\l^2}}  \bf h^s_{\k''-\k}\bf h^{s}_{\k''-\k}\nonumber \\
&=1-\frac{1}{E_{\bf h^s}|W|}\sumk{\k''}\sumw  \underbrace { e^{\frac{-|| \bf Z _{\k'',\w}||^2}{\l^2}}}_{\bs  \upPsi_{\k'',\w}}\underbrace {\nbr{ \bf h^s_{\k''-\k}}^2}_{\bs \upGamma_{\k''-\k}}
+\frac{1}{E_{\bf h^s}|W|\l^2}\sumk{\k''}\sumw  \underbrace {|| \bf Z _{\k'',\w}||^2 e^{\frac{-|| \bf Z _{\k'',\w}||^2}{\l^2}}}_{\bs \upOmega_{\k'',\w}}\underbrace {\nbr{ \bf h^s_{\k''-\k}}^2}_{\bs \upGamma_{\k''-\k}} \nonumber\\
&=1-\frac{1}{E_{\bf h^s}|W|}\sumk{\k''} \sumw {\bs \upPsi_{\k'',\w}}  {\bs \upGamma_{\k''-\k}} +\frac{1}{E_{\bf h^s}|W|\l^2}\sumk{\k''} \sumw {\bs \upOmega_{\k'',\w}}  {\bs \upGamma_{\k''-\k}}\nonumber \\
&=1-\frac{1}{E_{\bf h^s} |W|}\sumw  \bs \upPsi_{\k,\w}\circledast {\bs \upGamma_{-\k}} +\frac{1}{E_{\bf h^s}|W|\l^2}\sumw  \bs \upOmega_{\k,\w}\circledast {\bs \upGamma_{-\k}} \label{derv}.
\end{align}
\newpage
\vspace{1cm}
\section*{\hspace{6cm} \large{R}\normalsize EFERENCES}
\vspace{-1.5cm}

\renewcommand\bf{\bfseries}

\bibliographystyle{IEEEbib}

\bibliography{InPhaseRef}
%
\end{document}